\UseRawInputEncoding
\documentclass[12pt,a4paper]{article}

\usepackage{ucs}
\usepackage[usenames,dvipsnames]{xcolor}
\usepackage{tikz}
\usepackage{tkz-tab}
\usepackage{caption}
\usepackage{latexsym}
\usepackage{amssymb}
\usepackage{amsmath}
\usepackage{subcaption}
\newtheorem{theorem}{Theorem}[section]
\newtheorem{corollary}[theorem]{Corollary}
\newtheorem{lemma}[theorem]{Lemma}

\newtheorem{proposition}{Proposition}[section]
\newtheorem{definition}{Definition}[section]
\newtheorem{remark}{Remark}[section]
\usepackage{amsfonts,amssymb,eucal,amsmath}
\title{\bf Finiteness of Stationary Configurations of the Planar Four-vortex Problem }

\author{{ Xiang Yu\footnote{Email:yuxiang@swufe.edu.cn, xiang.zhiy@gmail.com}} \\
\small \it School of Economic and Mathematics, Southwestern
University of Finance and Economics, \\
\small \it Chengdu 611130, China}

\date{}

\begin{document}
\maketitle

\begin{abstract}
For the planar  four-vortex problem, we show that there are finitely many  stationary configurations consisting of equilibria, rigidly translating configurations, relative equilibria (uniformly rotating configurations)
 and collapse configurations. We also provide upper bounds  for these classes of
stationary configurations.  \\

 %provided that all the vorticities  are nonzero.In particular, we give an example which dose not have any stationary configuration.

% We show that there are finitely many  planar stationary configurations (included equilibria, rigidly translating configurations, relative equilibria (uniformly rotating configurations) and collapse configurations) in the  four-vortex problem. The proof is based on the elegant method of  Albouy and  Kaloshin for celestial mechanics.


{\bf 2020AMS Subject Classification} { 76B47	\and 70F10 \and 70F15 \and 	37Nxx}.

\end{abstract}

\section{Introduction}
\ \ \ \

We consider the motion of $N$  point vortices on a plane. This problem has been investigated by many researchers for a
long time and can be dated back to Helmholtz's work on hydrodynamics in 1858 \cite{helmholtz1858integrale}. Let the $n$-th point vortex have vortex strength (or vorticity) $\Gamma_n\in \mathbb{R}\backslash\{0\}$ and position
$\mathbf{r}_n \in {\mathbb{R}}^2$ ($n=1, 2, \cdots, N$), then the motion of \emph{the $N$-vortex problem} is   governed  by
\begin{equation}\label{eq:vortex's equation1}
\Gamma_n \dot{\mathbf{r}}_n =J\nabla_kH=J\sum_{1 \leq j \leq N, j \neq n} \frac{\Gamma_n\Gamma_j(\mathbf{r}_j-\mathbf{r}_n)}{|\mathbf{r}_j-\mathbf{r}_n|^2}, ~~~~~~~~~~~~~~~n=1, 2, \cdots, N.
\end{equation}
where $J=\left(
           \begin{array}{cc}
             0 & 1 \\
             -1 & 0 \\
           \end{array}
         \right)$, $H=-\sum_{1\leq j<k\leq N}\ln|\mathbf{r}_j-\mathbf{r}_k|$, $\nabla_n$ denotes the two-dimensional
partial  gradient vector    with respect to $\mathbf{r}_n$ and  $|\cdot|$  denotes the Euclidean norm in ${\mathbb{R}}^2$.

The system (\ref{eq:vortex's equation1})  (the $N$-vortex problem)  was introduced by Helmholtz \cite{helmholtz1858integrale}, then Kirchhoff \cite{kirchoff1877vorlesungen} first found its Hamiltonian structure.
%It is noteworthy that the system (\ref{eq:vortex's equation1}) has Hamiltonian structure,  this fact is first explicitly found by  in 1876 .

The $N$-vortex problem is formally derived from the planar Euler equations and is a widely used model for
providing finite-dimensional approximations to vorticity evolution in fluid dynamics. It is well known that the analysis
of  the $N$-vortex problem could lead to the understanding of vortex dynamics in
inviscid flows \cite{aref1983integrable}.  For
example, similar to  the  Newtonian $N$-body problem in celestial mechanics, one of the notable features of  the $N$-vortex problem is the existence of self-similar
collapsing    solutions, and the
mechanism of vortex collapse plays an important role to understand fluid phenomena. Indeed it has been pointed out that the vortex collapse is related to the loss of the
uniqueness of solutions to the Euler equations \cite{novikov1979vortex}.

For considerable physical interest, it is natural to study
stationary
configurations  which produce self-similar solutions of the $N$-vortex problem.  In \cite{o1987stationary} it is shown that the only stationary
configurations of vortices are equilibria, rigidly translating configurations, relative equilibria (uniformly rotating configurations)
 and collapse configurations. It turns out that  $L =\sum_{1\leq j<k\leq N}\Gamma_j\Gamma_k =0$ is a necessary condition on the vorticities
for the existence of equilibria, and  $\Gamma =\sum_{j=1}^{N}\Gamma_j =0$ is a necessary condition on the vorticities
for the existence of rigidly translating solutions.

Although a great amount of work on  stationary
configurations of the $N$-vortex problem has been done (a review can be found in \cite{aref2003vortex}), in general there is nothing
known about  stationary
configurations besides  three-vortex case and some special cases of $N \geq 4$.

In this study, we are interested in the  number of stationary
configurations for giving vorticities. All stationary configurations   for  two and
three point vortices are known  explicitly \cite{grobli1877specielle,synge1949motion,novikov1975dynamics,novikov1979vortex,aref1979motion,aref2003vortex}.
In \cite{o1987stationary} O'Neil  proved that for almost
every  choice of vorticities of the $N$-vortex problem, there are finite equilibria, rigidly translating configurations and collinear relative equilibria.
In\cite{palmore1982relative,o1987stationary} lower bounds on the number of relative equilibria have been established for the $N$-vortex problem with all
positive or large positive and small negative vorticities by the use of topological arguments. Aside from special cases with certain symmetries \cite{aref2005vortex},
the only general work on stationary
configurations are for the four-vortex problem: O'Neil  \cite{o2007relative}  and Hampton and Moeckel  \cite{hampton2009finiteness}  independently proved that for almost
every  choice of vorticities  of the four-vortex problem, there are finite equilibria, rigidly translating configurations and relative equilibria. In particular, little is known for  collapse configurations. More specifically, O'Neil proved 
\begin{theorem}\label{O'Neil}\emph{(\textbf{O'Neil})}
If the vorticities $\Gamma_n$ $(n\in\{1,2,3,4\})$ are nonzero then the four-vortex problem has:
\begin{itemize}
  \item at most 6 collinear relative equilibria when $\Gamma = 0$
  \item at most 14 strictly planar relative equilibria when $\Gamma = 0$
  \item at most 56  planar relative equilibria when $\Gamma \neq 0$ provided $L\neq 0$, $\Gamma_1\Gamma_3 -\Gamma_2\Gamma_4\neq 0$, $\Gamma_1\Gamma_4 -\Gamma_2\Gamma_3\neq 0$, $\Gamma_j+\Gamma_k \neq 0$ $(j,k\in\{1,2,3,4\})$, $\Gamma_1+\Gamma_2+\Gamma_l \neq 0$ $(l\in\{3,4\})$ and $\Gamma_1\Gamma_2+\Gamma_l(\Gamma_1+\Gamma_2) \neq 0$ $(l\in\{3,4\})$;
\end{itemize}
\end{theorem}
and
Hampton and Moeckel proved 
\begin{theorem}\label{Hampton-Moeckel}\emph{(\textbf{Hampton-Moeckel})}
If the vorticities $\Gamma_n$ $(n\in\{1,2,3,4\})$ are nonzero then the four-vortex problem has:
\begin{itemize}
  \item exactly 2 equilibria when the necessary condition $L = 0$ holds
  \item at most 6 rigidly translating configurations the necessary condition $\Gamma = 0$
holds
  \item at most 12 collinear relative equilibria
  \item at most 14 strictly planar relative equilibria when $\Gamma = 0$
  \item at most 74 strictly planar relative equilibria when $\Gamma \neq 0$ provided $\Gamma_j+\Gamma_k \neq 0$
and $\Gamma_j+\Gamma_k +\Gamma_l\neq 0$ for all distinct indices $j,k,l\in\{1,2,3,4\}$.
\end{itemize}
\end{theorem}
Here a configuration is called strictly
planar if it is planar but not collinear, and planar configurations include collinear and strictly
planar configurations.

It is well known that a continuum of relative equilibria can exist for the five-vortex problem \cite{roberts1999continuum}, thus the finiteness of relative equilibria and/or stationary
configurations  would be expected only for generic in general. However the main purposes of the
present study is to show that there are finitely many  stationary configurations for the four-vortex problem:
\begin{theorem}\label{Main}
If the vorticities $\Gamma_n$ $(n\in\{1,2,3,4\})$ are nonzero then the four-vortex problem has finitely many  stationary configurations.
\end{theorem}

It suffices to focus on the relative equilibria
 and collapse configurations  of the four-vortex problem. The proof of Theorem \ref{Main} is motivated by the elegant method of  Albouy and  Kaloshin for celestial mechanics \cite{Albouy2012Finiteness}.  The principle of
the method is to follow a possible continuum of central configurations in the
complex domain and to study its possible singularities there.

Both proofs of Theorem \ref{Hampton-Moeckel} and Theorem \ref{Main}  borrow   the similarity between the four-vortex problem and the Newtonian four-body problem in relative equilibria,  and in both proofs  one
follows a continuum of relative equilibria in the complex domain until it
reaches a singularity. Our analysis of the singularities is different from Hampton and Moeckel's.
 In contrast to the method of BKK theory by Hampton and Moeckel  \cite{hampton2009finiteness}, our proof
of Theorem \ref{Main} does not require any difficult computation.

We embed equations of relative equilibria
 and collapse configurations into a polynomial system (see the following (\ref{stationaryconfiguration2})),  then except two cases, a continuum of  relative equilibria
 and collapse configurations is excluded by analysis of the singularities. For the two exceptional cases, we   directly solve a  polynomial system  equivalent to the system (\ref{stationaryconfiguration2}) and prove the corresponding finiteness. The two exceptional cases are directly checked by simply solving an equivalent   polynomial system of (\ref{stationaryconfiguration2}) by employing  standard commands in Mathematica
on a desktop computer, and this computation takes less than ten seconds.

%As a by-product, one of the two exceptional cases shows that, in contrast to the four-body problem,  the four-vortex problem may not have any  stationary configuration in essence for some vorticities (see the following Theorem \ref{MainGammaLnot0} and Remark \ref{counter-example}).

%There are two cases that corresponding continuum of relative equilibria can not be  directly excluded by analysis of the singularities. We can  directly

In general the $N$-vortex problem is
simpler than the Newtonian $N$-body problem
for a given $N$, especially    when  problem of relative equilibria is concerned.  For example,  the way that the equations of relative equilibria reduce to polynomial systems  is simpler for  the $N$-vortex problem than for the Newtonian $N$-body problem \cite{hampton2006finiteness,hampton2009finiteness}. On the other hand, the Newtonian $N$-body problem is
simpler than the $N$-vortex problem
for a given $N$. For example, the Hamiltonian $H$ of the $N$-vortex problem is a transcendental function rather than   that  of the Newtonian $N$-body problem; in particular, the larger set of parameter values ($\Gamma_n<0$ is allowed) yields that there is  a continuum of relative equilibria in a certain $N$-vortex problem \cite{roberts1999continuum}.

Once the finiteness is proved, an explicit upper bound on the number of
relative equilibria
 and collapse configurations is obtained by direct application of   B\'{e}zout Theorems.   However, such a bound is not optimal for relative equilibria or collapse configurations. Nevertheless, we provide upper bounds in the following   summary  result:
\begin{corollary}\label{upperbounds}
If the vorticities $\Gamma_n$ $(n\in\{1,2,3,4\})$ are nonzero,  then the four-vortex problem has:
\begin{itemize}
  \item exactly 2 equilibria when the necessary condition $L = 0$ holds
  \item at most 6 rigidly translating configurations the necessary condition $\Gamma = 0$
holds
  \item at most 12 collinear relative equilibria, more precisely, \begin{itemize}
                                                   \item[i.] at most 12 collinear relative equilibria when $L \neq 0$
                                                   \item[ii.] at most 10 collinear relative equilibria when $L = 0$
                                                   \item[iii.] at most 6 collinear relative equilibria when $L \neq 0$ and $\Gamma = 0$
                                                 \end{itemize}
  \item at most 74 strictly planar  relative equilibria,  furthermore, at most 14 strictly  planar relative equilibria when $L \neq 0$ and $\Gamma = 0$
  \item at most 130  collapse configurations when the necessary condition $L = 0$ holds.
\end{itemize}
\end{corollary}
Note that there is no collinear collapse configuration for the general $N$-vortex problem.

The paper is structured as follows. In \textbf{Section 2}, we give some notations and definitions. In particular, following Albouy and  Kaloshin  \cite{Albouy2012Finiteness}, we introduce singular sequences of normalized central configurations.  In \textbf{Section 3}, we  discuss  some tools to classify the singular sequences. In \textbf{Section 4}, we   study all possibilities for a singular sequence and reduce the problem to the ten diagrams in Figure \ref{fig:Problematicdiagrams}. In \textbf{Section 5}, we obtain the constraints on the vorticities corresponding to each of the ten diagrams. In \textbf{Section 6}, based upon the prior work, we prove the main result on finiteness. Finally, in \textbf{Section 7}, we investigate  upper bounds on the number of
relative equilibria
 and collapse configurations.

\section{Preliminaries}
\label{Preliminaries}
\indent\par
In this section we give some notations and definitions that will be needed later.
\subsection{Stationary  configurations}
\label{Stationary configurations}
\indent\par
First it is  more convenient to consider the vortex positions $\mathbf{r}_n\in \mathbb{R}^2$ as complex
numbers $z_n\in \mathbb{C}$ for us,
 in which case the dynamics are given by $\dot{z}_n =-\textbf{i}V_n$ where
\begin{equation}\label{vectorfield}
    V_n= \sum_{1 \leq j \leq N, j \neq n} \frac{\Gamma_j z_{jn}}{r_{jn}^2}= \sum_{ j \neq n} \frac{\Gamma_j }{{\overline{z}_{jn}}},
\end{equation}
$z_{jn}=z_{n}-z_{j}$, $r_{jn}=|z_{jn}|=\sqrt{z_{jn}{\overline{z}_{jn}}}$, $\textbf{i}=\sqrt{-1}$ and the overbar denotes complex conjugation.

Let $\mathbb{C}^N= \{ z = (z_1,  \cdots, z_N):z_j \in \mathbb{C}, j = 1,  \cdots, N \}$ denote the space of configurations for $N$ point vortex. Let $\Delta=\{ z \in \mathbb{C}^N:z_j=z_k ~~\emph{for some}~~ j\neq k  \} $ be the collision set in $\mathbb{C}^N $. Then
the set $\mathbb{C}^N \backslash \Delta$ is the space of collision-free configurations.

\begin{definition}The following quantities are defined:
\begin{center}
$\begin{array}{cc}
  \text{Total vorticity} & \Gamma =\sum_{j=1}^{N}\Gamma_j  \\
  \text{Total vortex angular momentum} & L =\sum_{1\leq j<k\leq N}\Gamma_j\Gamma_k  \\
 \text{ Moment of vorticity }& M =\sum_{j=1}^{N}\Gamma_j z_j \\
 \text{ Angular impulse }& I =\sum_{j=1}^{N}\Gamma_j |z_j|^2=\sum_{j=1}^{N}\Gamma_j z_j{\overline{z}_j}
\end{array}$
\end{center}

\end{definition}
Then it is easy to see that
\begin{equation}\label{GammaIr}
\Gamma I-M \overline{M} =\sum_{1\leq j<k\leq N}\Gamma_j\Gamma_k z_{jk}{\overline{z}_{jk}}=\sum_{1\leq j<k\leq N}\Gamma_j\Gamma_k r_{jk}^2\triangleq S,
\end{equation}
and \begin{equation}\label{GammaL}
\Gamma^2- 2L >0.
\end{equation}

Following O'Neil  \cite{o1987stationary}  we will call a configuration stationary if the relative shape
remains constant, i.e., if the ratios of intervortex distances $r_{jk}/r_{lm}$ remain constant (such
solutions are often called homographic). More  precisely,
\begin{definition}
A configuration $z \in \mathbb{C}^N \backslash \Delta$ is stationary if there exists a constant $\Lambda\in {\mathbb{C}}$ such that
\begin{equation}\label{stationaryconfiguration}
V_j-V_k=\Lambda(z_j-z_k), ~~~~~~~~~~ 1\leq j, k\leq N.
\end{equation}
\end{definition}

It is shown that the only stationary
configurations of vortices are
equilibria, rigidly translating configurations, relative equilibria (uniformly rotating configurations) and collapse configurations \cite{o1987stationary}.
\begin{definition}
\begin{itemize}
  \item[i.] $z \in \mathbb{C}^N \backslash \Delta$ is an \emph{equilibrium} if $V_1=\cdots=V_N=0$.
  \item[ii.] $z \in \mathbb{C}^N \backslash \Delta$ is \emph{rigidly translating} if $V_1=\cdots=V_N=V$ for some $V\in \mathbb{C}\backslash\{0\}$. (The
vortices are said to move with common velocity $V$.)
  \item[iii.] $z \in \mathbb{C}^N \backslash \Delta$ is  a \emph{relative equilibrium} if there exist constants $\lambda\in \mathbb{R}\backslash\{0\},z_0\in \mathbb{C}$ such that $V_n=\lambda(z_n-z_0),~~~~~~~~~~ 1\leq n\leq N$.
  \item[iv.] $z \in \mathbb{C}^N \backslash \Delta$ is a \emph{collapse configuration} if there exist constants $\Lambda,z_0\in \mathbb{C}$ with $\emph{Im}(\Lambda)\neq0$ such that $V_n=\Lambda(z_n-z_0),~~~~~~~~~~ 1\leq n\leq N$.
\end{itemize}
\end{definition}

It is easy to see that  $L  =0$ is a necessary condition
for the existence of equilibria, and  $\Gamma  =0$ is a necessary condition
for the existence of configurations \cite{o1987stationary}.
\begin{proposition}\label{necessaryconditions}
Every equilibrium has vorticities satisfying $L  =0$; every rigidly
translating configuration has vorticities satisfying $\Gamma  =0$
\end{proposition}

If $z'$ differs from $z$ only by translation, rotation,
and change of scale (dilation) in the plane, then it is easy to see that  $z$ is stationary if and only if $z'$ is stationary.
\begin{definition}
A configuration $z$ is equivalent to a configuration $z'$ if for some $a,b\in \mathbb{C}$ with $b\neq 0$, $z'_n=b(z_n+a),~~~ 1\leq n\leq N$.

 $z\in \mathbb{C}^N \backslash \Delta$ is a translation-normalized configuration  if $M=0$; $z\in \mathbb{C}^N \backslash \Delta$ is a rotation-normalized configuration  if $z_{12}\in\mathbb{R}$.  Fixing the scale of a configuration, we can give the definition of dilation-normalized configuration, however, we do not specify the scale here.

 A  configuration, which is translation-normalized, rotation-normalized and dilation-normalized, is called a \textbf{normalized configuration}.
\end{definition}

Note that for a given configuration $z$, there always is certain rotation-normalized configuration and dilation-normalized configuration being  equivalent to $z$; if $\Gamma\neq0$, then there always is certain translation-normalized configuration  being  equivalent to $z$ too, at this time, there is exactly  one normalized configuration being  equivalent to $z$.
\subsection{Central configurations}

\indent\par
In this paper we consider only relative equilibria and collapse configurations, then it is easy to see that there always is certain translation-normalized configuration  being  equivalent to a given configuration.

The  equations of relative equilibria and collapse configurations can be unified into the following formular
\begin{equation}\label{stationaryconfiguration1}
V_n=\Lambda(z_n-z_0),~~~~~~~~~~ 1\leq n\leq N,
\end{equation}
$\Lambda\in \mathbb{R}\backslash\{0\}$ corresponds to relative equilibria and $\Lambda\in \mathbb{C}\backslash\mathbb{R}$ corresponds to  collapse configurations.
\begin{definition}
Relative equilibria and collapse configurations are both called \textbf{central configurations}.
\end{definition}

The equations (\ref{stationaryconfiguration1}) can be reducible to
\begin{equation}\label{stationaryconfiguration2}
\Lambda z_n= V_n,~~~~~~~~~~ 1\leq n\leq N,
\end{equation}
if we replace $z_n$ by $z_n+z_0$, i.e., the solutions of  equations (\ref{stationaryconfiguration2}) have been  removed the translation freedoms.
In fact, it is easy to see that the solutions of  equations (\ref{stationaryconfiguration2}) satisfy
\begin{equation}\label{center0}
M=0,
\end{equation}
\begin{equation}\label{LI}
\Lambda I= L.
\end{equation}

To remove the dilation freedoms, we set $|\Lambda|=1$. Following Albouy and  Kaloshin \cite{Albouy2012Finiteness}  we introduce
\begin{definition}\label{positivenormalizedcentralconfiguration}
A real normalized central configuration of the planar
$N$-vortex problem is a solution of (\ref{stationaryconfiguration2}) satisfying $z_{12}\in\mathbb{R}$ and $|\Lambda|=1$.
\end{definition}
\begin{remark}
\begin{itemize}
  \item[1.] Note that  solutions $z$ (real normalized central configurations) of (\ref{stationaryconfiguration1}) come in a pair: $-z\longmapsto z$ sends
solution on solution, that is, central configurations is determined up to a common factor $\pm 1$ by normalizing here.
Thus we count  the total central configurations up to a common factor $\pm 1$ below.

  \item[2.]
The word ``real"  refers to the reality hypothesis, in general it is  omitted in a real context. However, we  will study complex central configurations and establish in Section \ref{Finitenessresult} strong statements about their finiteness. Note that the distances $r_{jk}=\sqrt{z_{jk}{\overline{z}_{jk}}}$ are now bi-valued, but we do not need the positivity condition of the
distances $r_{jk}$ at all, so  it is unnecessary to  introduce the notion ``positive normalized central configuration" as in \cite{Albouy2012Finiteness}.
\end{itemize}
\end{remark}

%Then \begin{equation}\label{LI0}
%I=0\Longleftrightarrow L=0,
%\end{equation}
%and it also follows that collapse configurations satisfy \begin{equation}\label{LI0c}
%I= L=0.
%\end{equation}
We conclude this subsection with the following simple fact.
\begin{proposition}\label{Iis0}
Collapse configurations satisfy $\Gamma\neq 0$ and \begin{equation}\label{LI0whole}
S=I= L=0.
\end{equation}

For relative equilibria we have
$S=0 ~~~ \Longleftrightarrow ~~~  \left\{
             \begin{array}{lr}
             \Gamma\neq 0   &\\
             I= L=0 &
             \end{array}
\right.  \text{or} ~~~\left\{
             \begin{array}{lr}
             \Gamma= 0   &\\
             I\neq 0, L\neq 0 &
             \end{array}
\right.$
\end{proposition}
{\bf Proof.} %of Theorem \ref{asymptic2}:}

The proof is trivial by (\ref{GammaIr}), (\ref{GammaL})  and (\ref{LI}).

$~~~~~~~~~~~~~~~~~~~~~~~~~~~~~~~~~~~~~~~~~~~~~~~~~~~~~~~~~~~~~~~~~~~~~~~~~~~~~~~~~~~~~~~~~~~~~~~~~~~~~~~~~~~~~~~~~~~~~~~~~~~~~~~~~~~~~~~~~~~~~~~~~~~~\Box$\\

\subsection{Complex central configurations}

\indent\par

To eliminate complex conjugation in (\ref{stationaryconfiguration2}) we introduce a new set of variables $w_n$ and a ``conjugate"
relation:
\begin{equation}\label{stationaryconfiguration3}
\begin{array}{c}
  \Lambda z_n=\sum_{ j \neq n} \frac{\Gamma_j }{{w_{jn}}},~~~~~~~~~~ 1\leq n\leq N, \\
 \overline{\Lambda} w_n=\sum_{ j \neq n} \frac{\Gamma_j }{{z_{jn}}},~~~~~~~~~~ 1\leq n\leq N,
\end{array}
\end{equation}
where $z_{jn}=z_{n}-z_{j}$ and $w_{jn}=w_{n}-w_{j}$.

The rotation freedom is expressed in $z_n,w_n$ variables as the invariance of
(\ref{stationaryconfiguration3}) by the map $ R_a: (z_n, w_n) \mapsto (a z_n, a^{-1}w_n) $ for any $a\in\mathbb{C}\backslash \{0\}$   and any $n=1,2,\cdots,N$. The condition $z_{12}\in\mathbb{R}$ we proposed to remove this rotation freedom
becomes $z_{12}=w_{12}$.

To the variables $z_n,w_n\in \mathbb{C}$ we add the variables $Z_{jk},W_{jk}\in \mathbb{C}$ $(1\leq j< k\leq N)$ such that
$Z_{jk}=1/w_{jk}, W_{jk}=1/z_{jk}$. For $1\leq k< j\leq N$ we set $Z_{jk}=-Z_{kj}, W_{jk}=-W_{kj}$. Then equations (\ref{stationaryconfiguration2}) together with the condition $z_{12}\in\mathbb{R}$ and $|\Lambda|=1$ becomes
\begin{equation}\label{stationaryconfigurationmain}
\begin{array}{cc}
  \Lambda z_n=\sum_{ j \neq n} \Gamma_j Z_{jn},&1\leq n\leq N, \\
  \overline{\Lambda} w_n=\Lambda^{-1} w_n=\sum_{ j \neq n} \Gamma_j W_{jn},& 1\leq n\leq N, \\
  Z_{jk} w_{jk}=1,&1\leq j< k\leq N, \\
  W_{jk} z_{jk}=1,&1\leq j< k\leq N, \\
  z_{jk}=z_k-z_j,~~~  w_{jk}=w_k-w_j,&1\leq j, k\leq N, \\
  Z_{jk}=-Z_{kj},~~~ W_{jk}=-W_{kj},&1\leq k< j\leq N, \\
  z_{12}=w_{12}.
\end{array}
\end{equation}
This is a polynomial system in the  variables $\mathcal{Q}=(\mathcal{Z},\mathcal{W})\in(\mathbb{C}^{N}\times\mathbb{C}^{N(N-1)/2})^2$, here
\begin{center}
$\mathcal{Z}=(z_1,z_2,\cdots,z_N,Z_{12},Z_{13},\cdots,Z_{(N-1)N})$, $\mathcal{W}=(w_1,w_2,\cdots,w_N,W_{12},W_{13},\cdots,W_{(N-1)N}).$
\end{center}

It is easy to see that a positive normalized central configuration   of (\ref{stationaryconfiguration2}) is a  solution $\mathcal{Q}=(\mathcal{Z},\mathcal{W})$ of (\ref{stationaryconfigurationmain}) such that $z_n={\overline{w}}_n$   and vice versa.

Following Albouy and  Kaloshin \cite{Albouy2012Finiteness}  we introduce
\begin{definition}[Normalized central configuration]\label{normalizedcentralconfiguration}
 A normalized central configuration is a solution $\mathcal{Q}=(\mathcal{Z},\mathcal{W})$ of (\ref{stationaryconfigurationmain}). A real
normalized central configuration is a normalized central configuration such that
$z_n={\overline{w}}_n$ for any $n=1,2,\cdots,N$. %A positive normalized central configuration
%is a real normalized central configuration such that $r_{jk}=\pm\sqrt{z_{jk}{w_{jk}}}$ is positive for any $j, k,j\neq k$
\end{definition}
Definition \ref{normalizedcentralconfiguration} of a real normalized central configuration coincides with
Definition \ref{positivenormalizedcentralconfiguration}.

\begin{definition}
 We will use the name ``distance" for the $r_{jk}=\sqrt{z_{jk}{w_{jk}}}$. We will use the name $z$-separation (respectively $w$-separation) for
the $z_{jk}$'s (respectively the $w_{jk}$'s) in the complex plane.
\end{definition}

Note that solutions $\mathcal{Q}=(\mathcal{Z},\mathcal{W})$ of (\ref{stationaryconfigurationmain}) come in a pair: $(-\mathcal{Z},-\mathcal{W})\longmapsto(\mathcal{Z},\mathcal{W})$ sends
solution on solution, that is, a solution of (\ref{stationaryconfigurationmain}) is determined up to a common factor $\pm 1$.

\subsection{Elimination theory}

\indent\par

Let $m$ be a positive integer. Following \cite{mumford1995algebraic}, we define
a closed algebraic subset of the affine space $\mathbb{C}^m$ as the set of common zeroes
of a system of polynomials on $\mathbb{C}^m$.

The polynomial system (\ref{stationaryconfigurationmain}) defines a closed algebraic subset $\mathcal{A}\subset(\mathbb{C}^{N}\times\mathbb{C}^{N(N-1)/2})^2$. For the planar four-vortex problem, we will prove that  this subset is finite, then positive normalized central configurations is finite.
To distinguish the two possibilities, finitely many or infinitely many points, we
will only use the following result (see \cite{Albouy2012Finiteness}) from elimination theory.
\begin{lemma}\label{Eliminationtheory}
Let $\mathcal{X}$ be a closed algebraic subset of $\mathbb{C}^m$ and $f:\mathbb{C}^m\rightarrow \mathbb{C}$ be a
polynomial. Either the image $F(\mathcal{X})\subset\mathbb{ C}$ is a finite set, or it is the complement
of a finite set. In the second case one says that f is dominating.
\end{lemma}

\subsection{Singular sequences of normalized central configurations}

\indent\par
 We consider a solution $\mathcal{Q}=(\mathcal{Z},\mathcal{W})$ of (\ref{stationaryconfigurationmain}), this is a normalized central configuration. Let $\mathfrak{N}=N(N+1)/2$. Set
 \begin{center}
$\mathcal{Z}=(\mathcal{Z}_{1},\mathcal{Z}_{2},\cdots,\mathcal{Z}_{\mathfrak{N}})=(z_1,z_2,\cdots,z_N,Z_{12},Z_{13},\cdots,Z_{(N-1)N})$,

$\mathcal{W}=(\mathcal{W}_{1},\mathcal{W}_{2},\cdots,\mathcal{W}_{\mathfrak{N}})=(w_1,w_2,\cdots,w_N,W_{12},W_{13},\cdots,W_{(N-1)N})$.
\end{center}
 Let $\|\mathcal{Z}\|=\max_{j=1,2,\cdots,\mathfrak{N}}|\mathcal{Z}_{j}|$ be the modulus of the maximal component of
the vector $\mathcal{Z}\in \mathbb{C}^\mathfrak{N}$. Similarly, set  $\|\mathcal{W}\|=\max_{k=1,2,\cdots,\mathfrak{N}}|\mathcal{W}_{k}|$.

 Consider
a sequence $\mathcal{Q}^{(n)}$, $n=1,2,\cdots$, of normalized central configurations. Extract a sub-sequence such that the maximal component of $\mathcal{Z}^{(n)}$ is always the same, i.e., $\|\mathcal{Z}^{(n)}\|=|\mathcal{Z}^{(n)}_{j}|$ for a $j\in \{1,2,\cdots,\mathfrak{N}\}$ that does not depend
on $n$. Extract again in such a way that the vector sequence $\mathcal{Z}^{(n)}/\|\mathcal{Z}^{(n)}\|$ converges. Extract again in such a way that there is similarly an integer $k\in \{1,2,\cdots,\mathfrak{N}\}$ such that $\|\mathcal{W}^{(n)}\|=|\mathcal{W}^{(n)}_{k}|$ for all $n$. Extract a last time in
such a way that the vector sequence $\mathcal{W}^{(n)}/\|\mathcal{W}^{(n)}\|$ converges.

If the initial sequence is such that $\mathcal{Z}^{(n)}$ or $\mathcal{W}^{(n)}$ is unbounded, so is the extracted
sequence. Note that $\|\mathcal{Z}^{(n)}\|$ and $\|\mathcal{W}^{(n)}\|$ are bounded away from zero: if the first $N$
components of the vector $\mathcal{Z}^{(n)}$ or $\|\mathcal{W}^{(n)}\|$ all go to zero, then the denominator $z_{12}=w_{12}$ of the
component $Z_{12}=W_{12}$ go to zero and $\mathcal{Z}^{(n)}$ and $\mathcal{W}^{(n)}$ are unbounded. There are two possibilities for the
extracted sub-sequences above:\begin{itemize}
                                \item $\mathcal{Z}^{(n)}$ and $\mathcal{W}^{(n)}$ are bounded,
                                \item at least one of $\mathcal{Z}^{(n)}$ and $\mathcal{W}^{(n)}$ is unbounded.
                              \end{itemize}
\begin{definition}[Singular sequence]
 Consider a sequence of normalized central configurations. A
sub-sequence extracted by the above process, in the unbounded case, is called
a \emph{singular sequence}.
\end{definition}

Our method to prove the finiteness of the central configurations consists
essentially of two steps. First, we study all possibilities for a singular sequence.
We show that such an unbounded sequence is impossible for the planar four-vortex problem. Second, we use Lemma \ref{Eliminationtheory} to prove that if there are infinitely many normalized
central configurations, there exist singular sequences, and even singular sequences where some distance goes to zero or to infinity.
Consequently there are finitely many normalized central configurations for
the planar four-vortex problem.

\section{Rules of colored diagram}

\indent\par

\begin{definition}[Notation of asymptotic estimates]

\begin{center}

\begin{description}
  \item[$a\sim b$]  means $a/b\rightarrow 1$
  \item[$a\prec b$]   means $a/b\rightarrow 0$
  \item[$a\preceq b$]  means $a/b$ is bounded
  \item[$a\approx b$]  means $a\preceq b$ and $a\succeq b$
\end{description}
\end{center}

\end{definition}

\begin{definition}[Strokes and circles.]

We pick a singular sequence. We
write the indices of the bodies in a figure and use two colors for edges and
vertices.

The first color, the $z$-color, is used to mark the maximal order components
of
\begin{center}
$\mathcal{Z}=(z_1,z_2,\cdots,z_N,Z_{12},Z_{13},\cdots,Z_{(N-1)N})$.
\end{center}
 They correspond to the components of
the converging  vector sequence $\mathcal{Z}^{(n)}/\|\mathcal{Z}^{(n)}\|$ that do not tend to zero. We draw a circle around
the name of vertex $\textbf{j}$ if the term $z^{(n)}_j$ is of maximal order among all the components of $\mathcal{Z}^{(n)}$. We draw a stroke between the names $\textbf{k}$ and $\textbf{l}$ if the term
$Z^{(n)}_{kl}$ is of maximal order among all the components of $\mathcal{Z}^{(n)}$.
\end{definition}

%\subsection{Rules of colored diagram}

\indent\par

The following rules mainly concern $z$-diagram, but they apply as well to
the $w$-diagram.

If there
is a maximal order term in an equation, there should be another one. This
gives immediately the following Rule I.
\begin{description}
  \item[{Rule I}]
 There is something at each end of any $z$-stroke: another $z$-stroke
or/and a $z$-circle drawn around the name of the vertex. A $z$-circle cannot be isolated; there must be a $z$-stroke emanating from it. There is at least one
$z$-stroke in the $z$-diagram.
\end{description}

\begin{definition}[$z$-close]

 Consider a singular sequence. We say that bodies $\textbf{k}$ and $\textbf{l}$
are close in $z$-coordinate, or $z$-close, or that $z_k$ and $z_l$ are close, if $z^{(n)}_{kl}\prec \|\mathcal{Z}^{(n)}\|$.
\end{definition}

The following statement is obvious.
\begin{description}
  \item[{Rule II}] If bodies $\textbf{k}$ and $\textbf{l}$
are  $z$-close, they are both $z$-circled or both not
$z$-circled.
\end{description}

\begin{definition}[Isolated component]\label{isolatedcomponent}
An isolated component of the $z$-diagram is a subset of vertices
such that no $z$-stroke is joining a vertex of this subset to a vertex of the
complement.
\end{definition}

\begin{description}
  \item[{Rule III}]  The moment of vorticity of a set of bodies forming an isolated component of the $z$-diagram is $z$-close to the origin.
\end{description}

\begin{description}
  \item[{Rule IV}]  Consider the $z$-diagram or an isolated component of it. If there
is a $z$-circled vertex, there is another one. The $z$-circled vertices can all be
$z$-close together only if the total vorticity of these vertices is zero.
\end{description}

\begin{definition}[Maximal $z$-stroke]
Consider a $z$-stroke from vertex   $\textbf{k}$ to vertex $\textbf{l}$. We say it is
a maximal $z$-stroke if $\textbf{k}$ and $\textbf{l}$ are not $z$-close.
\end{definition}

\begin{description}
  \item[{Rule V}]  There is at least one $z$-circle at certain end of any maximal $z$-stroke. As a result,
if an isolated component of the $z$-diagram has no $z$-circled vertex,
then it has no maximal $z$-stroke.
\end{description}

 On the same diagram we also draw
$w$-strokes and $w$-circles. Graphically we use another color. The previous rules
and definitions apply to $w$-strokes and $w$-circles. What we will call simply the
diagram is the superposition of the $z$-diagram and the $w$-diagram. We will,
for example, adapt Definition \ref{isolatedcomponent} of an isolated component: a subset of bodies
forms an isolated component of the diagram if and only if it forms an isolated
component of the $z$-diagram and an isolated component of the $w$-diagram.

\begin{definition}[Edges and strokes]
There is an edge between vertex $\textbf{k}$ and vertex $\textbf{l}$ if there
is either a $z$-stroke, or a $w$-stroke, or both. There are three types of edges,
$z$-edges, $w$-edges and $zw$-edges, and only two types of strokes, represented with
two different colors.
\end{definition}
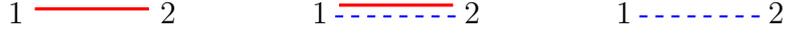
\begin{figure}[!h]
	\centering
	\begin{tikzpicture}
		\vspace*{0cm}\hspace*{0cm} % diagram 1
	\draw  (-4,  0)  node [black]{$1$}; %z-circle
	\draw  (-2,  0)  node [black]{$2$}; %z-circle
	\draw [red,very thick] (-3.75,0.05)--(-2.25,0.05); % z-edge

\vspace*{0cm}\hspace*{0cm} % diagram 1
	\draw  (0,  0)  node [black]{$1$}; %z-circle
	\draw  (2,  0)  node [black]{$2$}; %z-circle
	\draw [red,very thick] (0.25,0.1)--(1.75,0.1); % z-edge
		\draw [blue, dashed, thick] (0.2,-0.05)--(1.8,-0.05); % w-edge	

\vspace*{0cm}\hspace*{0cm} % diagram 1
	\draw  (4,  0)  node [black]{$1$}; %z-circle
	\draw  (6,  0)  node [black]{$2$}; %z-circle
		\draw [blue, dashed, thick] (4.2,-0.05)--(5.8,-0.05); % w-edge
	\end{tikzpicture}
\caption{A $z$-stroke, a $z$-stroke plus a $z$-stroke, a $w$-stroke,
forming respectively a z-edge, a $zw$-edge, a $w$-edge.  }
	\label{fig:C=0}
\end{figure}

\subsection{New normalization. Main estimates.}

\indent\par

 One does not change a central configuration by multiplying the $z$ coordinates by $a\in\mathbb{C}\backslash\{0\}$ and the $w$ coordinates
by $a^{-1}$. Our diagram is invariant by such an operation, as it considers the
$z$-coordinates and the $z$-coordinates separately.

We used the normalization $z_{12}=w_{12}$ in the previous considerations. In
the following we will normalize instead with $\|\mathcal{Z}\|=\|\mathcal{W}\|$. We start with a central configuration normalized with the condition $z_{12}=w_{12}$, then multiply the $z$-coordinates by $a > 0$, the $w$-coordinates by $a^{-1}$, in such a way that the
maximal component of $\mathcal{Z}$ and the maximal component of $\mathcal{W}$ have the same
modulus, i.e., $\|\mathcal{Z}\|=\|\mathcal{W}\|$.

A singular sequence was defined by the condition either $\|\mathcal{Z}^{(n)}\|\rightarrow \infty$ or $\|\mathcal{W}^{(n)}\|\rightarrow \infty$.
We also remarked that both $\|\mathcal{Z}^{(n)}\|$ and $\|\mathcal{W}^{(n)}\|$ were bounded away from zero. With the
new normalization, a singular sequence is simply characterized by $\|\mathcal{Z}^{(n)}\|=\|\mathcal{W}^{(n)}\|\rightarrow \infty$. From now on we
only discuss singular sequences.

Set $\|\mathcal{Z}^{(n)}\|=\|\mathcal{W}^{(n)}\|=1/\epsilon^2$, then $\epsilon\rightarrow 0$.

\begin{proposition}[Estimate]\label{Estimate1}
For any $(k,l)$, $1\leq k<l\leq N$, we have $\epsilon^2\preceq z_{kl}\preceq \epsilon^{-2}$, $\epsilon^2\preceq w_{kl}\preceq \epsilon^{-2}$ and $\epsilon^2\preceq r_{kl}\preceq \epsilon^{-2}$.

There is a $z$-stroke between $\textbf{k}$ and $\textbf{l}$ if and only if $w_{kl}\approx \epsilon^{2}$, then $ r_{kl}\preceq 1$.

 There is a maximal $z$-stroke between $\textbf{k}$ and $\textbf{l}$ if and only if $z_{kl}\approx \epsilon^{-2}, w_{kl}\approx \epsilon^{2}$, then $ r_{kl}\approx 1$.

  There is a $z$-edge between $\textbf{k}$ and $\textbf{l}$ if and only if $z_{kl}\succ \epsilon^{2},w_{kl}\approx \epsilon^{2}$, then $\epsilon^{2}\prec r_{kl}\preceq 1 $.

There is a maximal $z$-edge between $\textbf{k}$ and $\textbf{l}$ if and only if $z_{kl}\approx \epsilon^{-2},w_{kl}\approx \epsilon^{2}$, then $ r_{kl}\approx 1$.

There is a $zw$-edge between $\textbf{k}$ and $\textbf{l}$ if and only if $z_{kl},w_{kl}\approx \epsilon^{2}$, this can be  characterized as $ r_{kl}\approx \epsilon^{2}$.

\end{proposition}
{\bf Proof.} %of Theorem \ref{asymptic2}:}

The proof is trivial, so it is omitted.

$~~~~~~~~~~~~~~~~~~~~~~~~~~~~~~~~~~~~~~~~~~~~~~~~~~~~~~~~~~~~~~~~~~~~~~~~~~~~~~~~~~~~~~~~~~~~~~~~~~~~~~~~~~~~~~~~~~~~~~~~~~~~~~~~~~~~~~~~~~~~~~~~~~~~\Box$\\

\begin{remark}
By the estimates above, the strokes in a $zw$-edge are not maximal. A maximal $z$-stroke is exactly a maximal $z$-edge.
\end{remark}

\begin{description}
  \item[{Rule VI}]
  If there are two consecutive $z$-stroke, there is a third $z$-stroke closing the triangle.
\end{description}
{\bf Proof.} %of Theorem \ref{asymptic2}:}

Suppose $Z_{12}\approx \epsilon^{-2}$ and $Z_{13}\approx \epsilon^{-2}$, then $w_{12}\approx \epsilon^{2}$ and $w_{13}\approx \epsilon^{2}$.

Therefore, $w_{23}=w_{13}-w_{12}\preceq \epsilon^{2}$, but $w_{23}\succeq \epsilon^{2}$, thus $w_{23}\approx \epsilon^{2}$, or $Z_{23}\approx \epsilon^{-2}$.

$~~~~~~~~~~~~~~~~~~~~~~~~~~~~~~~~~~~~~~~~~~~~~~~~~~~~~~~~~~~~~~~~~~~~~~~~~~~~~~~~~~~~~~~~~~~~~~~~~~~~~~~~~~~~~~~~~~~~~~~~~~~~~~~~~~~~~~~~~~~~~~~~~~~~\Box$\\

\section{Systematic exclusion of four-vortex diagrams}

\subsection{Possible diagrams}
\indent\par
We call a bicolored vertex of the diagram a vertex which connects at least
a stroke of $z$-color with at least a stroke of $w$-color. The number of edges from
a bicolored vertex is at least 1 and at most $N-1$. The number of strokes from
a bicolored vertex is at least 2 and at most $2N-2$. Given a diagram, we
define $C$ as the maximal number of strokes from a bicolored vertex. We use
this number to classify all possible diagrams.

Recall that the $z$-diagram indicates the maximal terms among a finite set
of terms. It is nonempty. If there is a circle, there is an edge of the same
color emanating from it. So there is at least a $z$-stroke, and similarly, at least a
$w$-stroke.

\subsubsection{$C=0$}

\indent\par

 If there is no bicolored vertex, then
{$C$ is not defined or we can say $C=0$}, there are at most two strokes and they are ¡°parallel¡± by Rule I, VI and $C=0$. Thus
the only possible diagram is the following Figure \ref{fig:C=0}.
\begin{figure}[!h]
	\centering
	\begin{tikzpicture}
		\vspace*{0cm}\hspace*{0cm} % diagram 1
	\draw [red,very thick] (-1,  0) circle (0.25) node [black]{$1$}; %z-circle
	\draw [red,very thick] (1,  0) circle (0.25) node [black]{$2$}; %z-circle
	\draw [red,very thick] (-0.75,0.05)--(0.75,0.05); % z-edge
	\draw [blue, dashed, thick] (-1,  -1.5) circle (0.2) node [black]{$4$}; %w-circle
	\draw [blue, dashed, thick] (1,  -1.5) circle (0.2) node [black]{$3$}; %w-circle
		\draw [blue, dashed, thick] (-0.8,-1.55)--(0.8,-1.55); % w-edge	
	\end{tikzpicture}
\caption{$C=0$  }
	\label{fig:C=0}
\end{figure}
\subsubsection{$C=2$}

\indent\par
 There are two cases: a $zw$-edge exists or not.

If it is present, it should be isolated.   Thus
the only possible diagrams are that in the following Figure \ref{fig:C=21}.
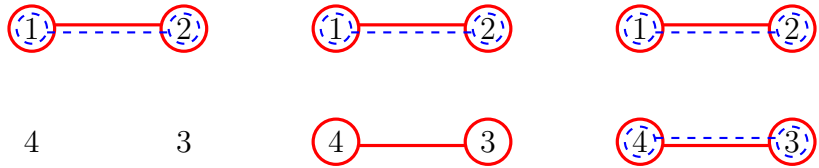
\begin{figure}[!h]
	\centering
	\begin{tikzpicture}
		\vspace*{0cm}\hspace*{0cm} % diagram 1
	\draw [red,very thick] (-5,  0) circle (0.3) node [black]{$1$}; %z-circle
	\draw [red,very thick] (-3,  0) circle (0.3) node [black]{$2$}; %z-circle
\draw [blue, dashed, thick] (-5, 0) circle (0.2); %w-circle
	\draw [blue, dashed, thick] (-3, 0) circle (0.2) ; %w-circle
	\draw [red,very thick] (-4.7,0.05)--(-3.3,0.05); % z-edge
\draw [blue, dashed, thick] (-4.8,-0.05)--(-3.2,-0.05); % w-edge
\draw  (-5,  -1.5) node [black]{$4$}; %z-circle
	\draw  (-3,  -1.5)  node [black]{$3$}; %z-circle

\vspace*{0cm}\hspace*{0cm} % diagram 1
		\draw [red,very thick] (-1,  0) circle (0.3) node [black]{$1$}; %z-circle
	\draw [red,very thick] (1,  0) circle (0.3) node [black]{$2$}; %z-circle
\draw [blue, dashed, thick] (-1, 0) circle (0.2); %w-circle
	\draw [blue, dashed, thick] (1, 0) circle (0.2) ; %w-circle
	\draw [red,very thick] (-0.7,0.05)--(0.7,0.05); % z-edge
\draw [blue, dashed, thick] (-0.8,-0.05)--(0.8,-0.05); % w-edge
\draw [red,very thick] (-1,  -1.5) circle (0.3) node [black]{$4$}; %z-circle
	\draw [red,very thick] (1,  -1.5) circle (0.3) node [black]{$3$}; %z-circle
%\draw [blue, dashed, thick] (-1, -1.5) circle (0.2); %w-circle
%	\draw [blue, dashed, thick] (1, -1.5) circle (0.2) ; %w-circle
	\draw [red,very thick] (-0.7,-1.55)--(0.7,-1.55); % z-edge
%\draw [blue, dashed, thick] (3.2,-1.45)--(4.8,-1.45); % w-edge	

	\vspace*{0cm}\hspace*{0cm} % diagram 1
	\draw [red,very thick] (3,  0) circle (0.3) node [black]{$1$}; %z-circle
	\draw [red,very thick] (5,  0) circle (0.3) node [black]{$2$}; %z-circle
\draw [blue, dashed, thick] (3, 0) circle (0.2); %w-circle
	\draw [blue, dashed, thick] (5, 0) circle (0.2) ; %w-circle
	\draw [red,very thick] (3.3,0.05)--(4.7,0.05); % z-edge
\draw [blue, dashed, thick] (3.2,-0.05)--(4.8,-0.05); % w-edge
\draw [red,very thick] (3,  -1.5) circle (0.3) node [black]{$4$}; %z-circle
	\draw [red,very thick] (5,  -1.5) circle (0.3) node [black]{$3$}; %z-circle
\draw [blue, dashed, thick] (3, -1.5) circle (0.2); %w-circle
	\draw [blue, dashed, thick] (5, -1.5) circle (0.2) ; %w-circle
	\draw [red,very thick] (3.3,-1.55)--(4.7,-1.55); % z-edge
\draw [blue, dashed, thick] (3.2,-1.45)--(4.8,-1.45); % w-edge	
	\end{tikzpicture}
\caption{$C=2$, $zw$-edge appears }
	\label{fig:C=21}
\end{figure}

If it is not present, there are adjacent $z$-edges and $w$-edges. From any
such adjacency there is no other edge  by Rule VI. By trying to continue it, we see
that the only diagram is the following Figure \ref{fig:C=22}.
\begin{figure}[!h]
	\centering
	\begin{tikzpicture}
		\vspace*{0cm}\hspace*{0cm} % diagram 1
	\draw [blue, dashed, thick] (0,  0) circle (0.2) node [black]{$1$};
	\draw [red,very thick] (0,  0) circle (0.3);
	\draw [blue, dashed, thick] (2,  0) circle (0.2) node [black]{$2$};
	\draw [red,very thick] (2,  0) circle (0.3);
	\draw [blue, dashed, thick] (0.2,0)--(1.8,0);
	\draw [blue, dashed, thick] (0.2,-1.5)--(1.8,-1.5);
	\draw [blue, dashed, thick] (0,  -1.5) circle (0.2) node [black]{$4$};
	\draw [red,very thick] (0,  -1.5) circle (0.3);
	\draw[blue, dashed, thick] (2,  -1.5) circle (0.2) node [black]{$3$};
	\draw [red,very thick] (2,  -1.5) circle (0.3);
	\draw [red,very thick] (0, -1.2)--(0,-0.3);
	\draw [red,very thick] (2, -1.2)--(2,-0.3);
	\end{tikzpicture}
\caption{$C=2$, no $zw$-edge  }
	\label{fig:C=22}
\end{figure}
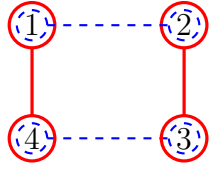

\subsubsection{$C=3$}

\indent\par
Consider a bicolored vertex (let us say, $\textbf{1}$) with three strokes.
 There are two cases: a $zw$-edge exists or not.

If it is not present, suppose two $z$-edges respectively connected  vertex $\textbf{2}$ and vertex $\textbf{3}$,  with  a $w$-edge connected  vertex $\textbf{4}$. Then vertices  $\textbf{1,2,3}$ formed a $z$-color triangle and there are no other strokes between vertices  $\textbf{1,2,3}$ by Rule VI and $zw$-edge is not present. So vertices  $\textbf{1,2,3}$ are all $w$-circled by Rule I, II and Estimate \ref{Estimate1}. Then by Rule I again,  vertex  $\textbf{2}$ and vertex  $\textbf{3}$ are both formed $w$-strokes with vertex  $\textbf{4}$, but it follows that vertex  $\textbf{2}$ and vertex  $\textbf{3}$ formed a $w$-stroke, this is a contradiction.

If it is present, suppose the $zw$-edge connected  vertex $\textbf{2}$, with  a $z$-edge connected  vertex $\textbf{3}$. Then vertices  $\textbf{1,2,3}$ formed a $z$-color triangle and there is no any other stroke between vertices  $\textbf{1,2,3}$ by Rule VI and $C=3$. So vertices  $\textbf{1,2,3}$ are all $w$-circled by Rule I, II and Estimate \ref{Estimate1}, thus vertex  $3$ and vertex  $4$ formed a  $w$-edge and vertex  $4$ is $w$-circled. Hence
the only possible diagrams are that in the following Figure \ref{fig:C=31}.
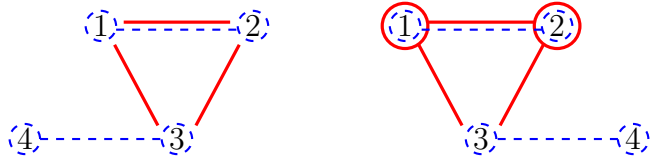
\begin{figure}[!h]
	\centering
	\begin{tikzpicture}
		\vspace*{0cm}\hspace*{0cm} % diagram 1
	\draw [blue, dashed, thick] (-3,  0) circle (0.2) node [black]{$1$};
	\draw [blue, dashed, thick] (-1,0) circle (0.2) node [black]{$2$};
	\draw [blue, dashed, thick] (-2,-1.5) circle (0.2) node [black]{$3$};
	\draw [blue, dashed, thick] (-4,-1.5) circle (0.2) node [black]{$4$};

	\draw [blue, dashed, thick] (-2.8,-0.05)--(-1.2,-0.05);
	\draw [blue, dashed, thick] (-3.8,-1.5)--(-2.2,-1.5);
		\draw [red,very thick] (-2.7,0.05)--(-1.3,0.05);
	\draw  [red,very thick]  (-1.18,-0.25)--(-1.75,-1.32);
	\draw  [red,very thick]  (-2.82, -0.25)--(-2.25,-1.32);

\vspace*{0cm}\hspace*{0cm} % diagram 1
	\draw [blue, dashed, thick] (1,  0) circle (0.2) node [black]{$1$};
	\draw [blue, dashed, thick] (3,0) circle (0.2) node [black]{$2$};
	\draw [blue, dashed, thick] (4,-1.5) circle (0.2) node [black]{$4$};
	\draw [blue, dashed, thick] (2,-1.5) circle (0.2) node [black]{$3$};
	\draw [red,very thick] (1,  0) circle (0.3) ;
	\draw [red,very thick] (3,0) circle (0.3) ;
	
	\draw [blue, dashed, thick] (2.8,-0.05)--(1.2,-0.05);
	\draw [blue, dashed, thick] (3.8,-1.5)--(2.2,-1.5);
		\draw [red,very thick] (2.7,0.05)--(1.3,0.05);
	\draw  [red,very thick]  (1.18,-0.25)--(1.75,-1.32);
	\draw  [red,very thick]  (2.82, -0.25)--(2.25,-1.32);

	\end{tikzpicture}
\caption{$C=3$, $zw$-edge appears }
	\label{fig:C=31}
\end{figure}

 \subsubsection{$C=4$}

\indent\par
Consider a bicolored vertex (let us say, $\textbf{1}$ ) with four strokes.

In the first case, vertex $\textbf{1}$  has two adjacent $zw$-edges connected  vertex $\textbf{2}$ and  vertex $\textbf{3}$ separately. A third
$zw$-edge closes the triangles by Rule VI.  As $C=4$ there is no any stroke connected  vertex $\textbf{4}$, thus vertex  $\textbf{4}$ is neither $z$-circled nor $w$-circled.  Hence
the only possible diagrams are that in the following Figure \ref{fig:C=41}.

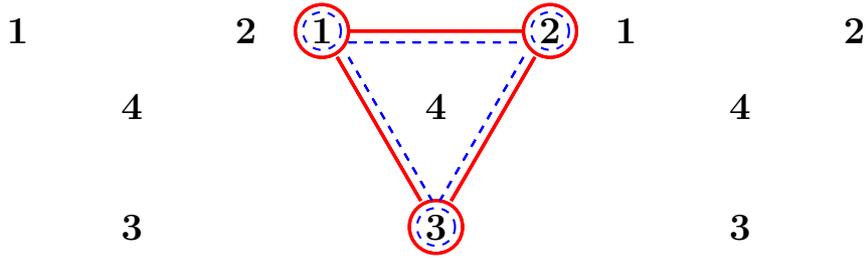
\begin{figure}[!h]
	\centering
	\begin{tikzpicture}
	\vspace*{0cm}\hspace*{-4cm}
\draw	  (-3/2,0)    node {\large\textbf{1}};
\draw		(3/2,0) node {\large\textbf{2}};
\draw		(0,-3/2*1.732) node {\large\textbf{3}};
\draw		(0,-1) node {\large\textbf{4}};

\draw [red,very thick] (-3/2+.35,0)--(3/2-.35,0);
\draw [blue, dashed,thick] (-3/2+0.35,-.15)--(3/2-.35,-0.15);

\draw [red,very thick] (-3/2+.2,-0.2*1.732)--(-.2,-3/2*1.732+.2*1.732);
\draw [blue, dashed,thick] (-3/2+0.15+.2, -0.2*1.732)--(0.15-.2,-3/2*1.732+0.2*1.732);

\draw [red,very thick] (3/2-0.2,-0.2*1.732)--(0.2,-3/2*1.732+0.2*1.732);
\draw [blue, dashed,thick] (3/2-0.15-0.2,-.2*1.732)--(-0.15+0.2,-3/2*1.732+.2*1.732);

	\vspace*{0cm}\hspace*{4cm}

\draw	[red,very thick]  (-3/2,0)  circle (0.35);
\draw  (-3/2,0)  node {\large\textbf{1}};

\draw	[red,very thick]  (3/2,0)  circle (0.35);
\draw		(3/2,0) node {\large\textbf{2}};

\draw	[red,very thick] (0,-3/2*1.732) circle (0.35);
\draw		(0,-3/2*1.732) node {\large\textbf{3}};

\draw		(0,-1) node {\large\textbf{4}};

\draw [red,very thick] (-3/2+.35,0)--(3/2-.35,0);
\draw [blue, dashed,thick] (-3/2+0.35,-.15)--(3/2-.35,-0.15);

\draw [red,very thick] (-3/2+.2,-0.2*1.732)--(-.2,-3/2*1.732+.2*1.732);
\draw [blue, dashed,thick] (-3/2+0.15+.2, -0.2*1.732)--(0.15-.2,-3/2*1.732+0.2*1.732);

\draw [red,very thick] (3/2-0.2,-0.2*1.732)--(0.2,-3/2*1.732+0.2*1.732);
\draw [blue, dashed,thick] (3/2-0.15-0.2,-.2*1.732)--(-0.15+0.2,-3/2*1.732+.2*1.732);

	\vspace*{0cm}\hspace*{4cm}
	\draw	[blue,dashed, thick]  (-3/2,0)  circle (0.25);
\draw	[red,very thick]  (-3/2,0)  circle (0.35);
\draw  (-3/2,0)  node {\large\textbf{1}};

	\draw	[blue,dashed, thick]  (3/2,0)  circle (0.25);
\draw	[red,very thick]  (3/2,0)  circle (0.35);
\draw		(3/2,0) node {\large\textbf{2}};

	\draw	[blue,dashed, thick]  (0,-3/2*1.732) circle (0.25);
\draw	[red,very thick] (0,-3/2*1.732) circle (0.35);
\draw		(0,-3/2*1.732) node {\large\textbf{3}};

\draw		(0,-1) node {\large\textbf{4}};

\draw [red,very thick] (-3/2+.35,0)--(3/2-.35,0);
\draw [blue, dashed,thick] (-3/2+0.35,-.15)--(3/2-.35,-0.15);

\draw [red,very thick] (-3/2+.2,-0.2*1.732)--(-.2,-3/2*1.732+.2*1.732);
\draw [blue, dashed,thick] (-3/2+0.15+.2, -0.2*1.732)--(0.15-.2,-3/2*1.732+0.2*1.732);

\draw [red,very thick] (3/2-0.2,-0.2*1.732)--(0.2,-3/2*1.732+0.2*1.732);
\draw [blue, dashed,thick] (3/2-0.15-0.2,-.2*1.732)--(-0.15+0.2,-3/2*1.732+.2*1.732);

	\end{tikzpicture}
	\caption{$C=4$, three $zw$-edges}
\label{fig:C=41}
\end{figure}
In the second case, vertex $\textbf{1}$  has one adjacent $zw$-edge connected  vertex $\textbf{2}$, and  two $z$-edges connected vertex $\textbf{3}$ and vertex $\textbf{4}$ separately. Then vertices $\textbf{1,2,3,4}$ form a fully $z$-stroked diagram by Rule VI, and there is no any $w$-stroke between vertex $\textbf{j}$ and vertex $\textbf{k}$ $(j=1,2,k=3,4)$ by $C=4$. By Rule I there are $w$-circle at vertex $\textbf{1}$ and vertex $\textbf{2}$, then there are $w$-circle at vertex $\textbf{3}$ and vertex $\textbf{4}$ too. Thus vertex $\textbf{3}$ and vertex $\textbf{4}$ form a $w$-stroke by Rule I. According to if there are $z$-circle at vertices, the only possible diagrams are that in the following Figure \ref{fig:C=42}.

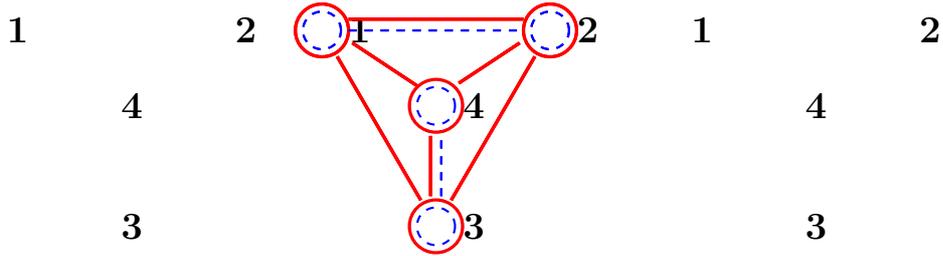
\begin{figure}[!h]
	\centering
	\begin{tikzpicture}
	\vspace*{0cm}\hspace*{-4cm}
	\draw	[blue,dashed,thick]  (-3/2,0)  circle (0.35);
	\draw	  (-3/2,0)    node {\large\textbf{1}};
	
		\draw	[blue,dashed,thick]  (3/2,0)  circle (0.35);
	\draw		(3/2,0) node {\large\textbf{2}};
	
		\draw	[blue,dashed,thick]  (0,-3/2*1.732)   circle (0.35);
	\draw		(0,-3/2*1.732) node {\large\textbf{3}};
	
		\draw	[blue,dashed,thick]  (0,-1)  circle (0.35);
	\draw		(0,-1) node {\large\textbf{4}};

	\draw [red,very thick] (-3/2+.35,0.15)--(3/2-.35,0.15);
\draw [blue, dashed,thick] (-3/2+0.35,0)--(3/2-.35,0);
	
	\draw [red,very thick] (-3/2+.2,-0.2*1.732)--(-.2,-3/2*1.732+.2*1.732);
%	\draw [red,very thick] (-3/2-0.15+.25, -0.25*1.732)--(-0.15-.25,-3/2*1.732+0.25*1.732);
	
	\draw [red,very thick] (3/2-0.2,-0.2*1.732)--(0.2,-3/2*1.732+0.2*1.732);
%	\draw [red,very thick] (3/2+0.15-0.25,-0.25*1.732)--(0.15+0.25,-3/2*1.732+.25*1.732);

		\draw [red,very thick]  (-3/2+0.25+0.15,-0.25*2/3)--(0-0.4+0.15,-1+0.4*2/3);  %k=(1, -2/3)
	%		\draw [red,very thick]  (-3/2+0.4-0.1,-0.4*2/3)--(0-0.3-0.1,-1+0.3*2/3);  %k=(1, -2/3)
		
		\draw [red,very thick]  (3/2-0.25-.15,-.25*2/3)--(0.45 -.15,-1+.45*2/3);  %k=(1, 2/3)
%			\draw [blue,dashed,thick]   (3/2-0.4+.1,-.4*2/3)--(0.3 +.1,-1+.3*2/3);  %k=(1, 2/3)

		\draw [red,very thick] (-.07,-3/2*1.732+.4)--(-.07,-1-.4);
			\draw [blue,dashed,thick]  (.07,-3/2*1.732+0.4)--(.07,-1-0.4);

	\vspace*{0cm}\hspace*{4.5cm}
	
	\draw	[blue,dashed, thick]  (-3/2,0)  circle (0.25);
\draw	[red,very thick]  (-3/2,0)  circle (0.35);
\draw  (-3/2,0)  node {\large\textbf{1}};

\draw	[blue,dashed, thick]  (3/2,0)  circle (0.25);
\draw	[red,very thick]  (3/2,0)  circle (0.35);
\draw		(3/2,0) node {\large\textbf{2}};
	
		\draw	[blue,dashed,thick]  (0,-3/2*1.732)   circle (0.35);
\draw		(0,-3/2*1.732) node {\large\textbf{3}};

\draw	[blue,dashed,thick]  (0,-1)  circle (0.35);
\draw		(0,-1) node {\large\textbf{4}};

\draw [red,very thick] (-3/2+.35,0.15)--(3/2-.35,0.15);
\draw [blue, dashed,thick] (-3/2+0.35,0)--(3/2-.35,0);

\draw [red,very thick] (-3/2+.2,-0.2*1.732)--(-.2,-3/2*1.732+.2*1.732);
%	\draw [red,very thick] (-3/2-0.15+.25, -0.25*1.732)--(-0.15-.25,-3/2*1.732+0.25*1.732);

\draw [red,very thick] (3/2-0.2,-0.2*1.732)--(0.2,-3/2*1.732+0.2*1.732);
%	\draw [red,very thick] (3/2+0.15-0.25,-0.25*1.732)--(0.15+0.25,-3/2*1.732+.25*1.732);

\draw [red,very thick]  (-3/2+0.25+0.15,-0.25*2/3)--(0-0.4+0.15,-1+0.4*2/3);  %k=(1, -2/3)
%		\draw [red,very thick]  (-3/2+0.4-0.1,-0.4*2/3)--(0-0.3-0.1,-1+0.3*2/3);  %k=(1, -2/3)

\draw [red,very thick]  (3/2-0.25-.15,-.25*2/3)--(0.45 -.15,-1+.45*2/3);  %k=(1, 2/3)
%			\draw [blue,dashed,thick]   (3/2-0.4+.1,-.4*2/3)--(0.3 +.1,-1+.3*2/3);  %k=(1, 2/3)

\draw [red,very thick] (-.07,-3/2*1.732+.4)--(-.07,-1-.4);
\draw [blue,dashed,thick]  (.07,-3/2*1.732+0.4)--(.07,-1-0.4);

	\vspace*{0cm}\hspace*{4.5cm}
	\draw	[blue,dashed, thick]  (-3/2,0)  circle (0.25);
	\draw	[red,very thick]  (-3/2,0)  circle (0.35);
	\draw  (-3/2,0)  node {\large\textbf{1}};
	
	\draw	[blue,dashed, thick]  (3/2,0)  circle (0.25);
	\draw	[red,very thick]  (3/2,0)  circle (0.35);
	\draw		(3/2,0) node {\large\textbf{2}};
	
	\draw	[blue,dashed, thick]  (0,-3/2*1.732) circle (0.25);
	\draw	[red,very thick] (0,-3/2*1.732) circle (0.35);
	\draw		(0,-3/2*1.732) node {\large\textbf{3}};
	
		\draw	[blue,dashed, thick]  (0,-1) circle (0.25);
		\draw	[red,very thick](0,-1) circle (0.35);
	\draw		(0,-1) node {\large\textbf{4}};

\draw [red,very thick] (-3/2+.35,0.15)--(3/2-.35,0.15);
\draw [blue, dashed,thick] (-3/2+0.35,0)--(3/2-.35,0);

\draw [red,very thick] (-3/2+.2,-0.2*1.732)--(-.2,-3/2*1.732+.2*1.732);
%	\draw [red,very thick] (-3/2-0.15+.25, -0.25*1.732)--(-0.15-.25,-3/2*1.732+0.25*1.732);

\draw [red,very thick] (3/2-0.2,-0.2*1.732)--(0.2,-3/2*1.732+0.2*1.732);
%	\draw [red,very thick] (3/2+0.15-0.25,-0.25*1.732)--(0.15+0.25,-3/2*1.732+.25*1.732);

\draw [red,very thick]  (-3/2+0.25+0.15,-0.25*2/3)--(0-0.4+0.15,-1+0.4*2/3);  %k=(1, -2/3)
%		\draw [red,very thick]  (-3/2+0.4-0.1,-0.4*2/3)--(0-0.3-0.1,-1+0.3*2/3);  %k=(1, -2/3)

\draw [red,very thick]  (3/2-0.25-.15,-.25*2/3)--(0.45 -.15,-1+.45*2/3);  %k=(1, 2/3)
%			\draw [blue,dashed,thick]   (3/2-0.4+.1,-.4*2/3)--(0.3 +.1,-1+.3*2/3);  %k=(1, 2/3)

\draw [red,very thick] (-.07,-3/2*1.732+.4)--(-.07,-1-.4);
\draw [blue,dashed,thick]  (.07,-3/2*1.732+0.4)--(.07,-1-0.4);

	\end{tikzpicture}
	\caption{$C=4$, two $zw$-edges}
\label{fig:C=42}
\end{figure}

In the  last case, vertex $\textbf{1}$  has one adjacent $zw$-edge connected  vertex $\textbf{2}$,   one $z$-edges connected vertex $\textbf{3}$ and one $w$-edges connected vertex $\textbf{4}$ separately. Then vertices  $\textbf{1,2,3}$ formed a $z$-color triangle and $\textbf{1,2,4}$ formed a $w$-color triangle by Rule VI. It is easy to see that
there is no any other stroke between vertices  by Rule VI and $C=4$, furthermore, there is no any circle at vertices, otherwise, there will be new stroke between vertices  by Rule IV, I and  Estimate \ref{Estimate1}. Thus the only possible diagram is  the following Figure \ref{fig:C=43}.

\begin{figure}[!h]
	\centering
	\begin{tikzpicture}

	\vspace*{0cm}\hspace*{0cm}

	\draw  (-3/2,0)  node {\large\textbf{1}};

	\draw		(3/2,0) node {\large\textbf{2}};

	\draw		(0,-3/2*1.732) node {\large\textbf{3}};

	\draw		(0,-1) node {\large\textbf{4}};
	
		\draw [red,very thick] (-3/2+.35,0.15)--(3/2-.35,0.15);
	\draw [blue, dashed,thick] (-3/2+0.35,0)--(3/2-.35,0);
	
	\draw [red,very thick] (-3/2+.2,-0.2*1.732)--(-.2,-3/2*1.732+.2*1.732);
%	\draw [red,very thick] (-3/2-0.15+.25, -0.25*1.732)--(-0.15-.25,-3/2*1.732+0.25*1.732);
	
	\draw [red,very thick] (3/2-0.2,-0.2*1.732)--(0.2,-3/2*1.732+0.2*1.732);
%	\draw [red,very thick] (3/2+0.15-0.25,-0.25*1.732)--(0.15+0.25,-3/2*1.732+.25*1.732);

	\draw [blue,dashed,thick]  (-3/2+0.3+0.15,-0.3*2/3)--(0-0.35+0.15,-1+0.35*2/3);  %k=(1, -2/3)
%	\draw [red,very thick]  (-3/2+0.4-0.1,-0.4*2/3)--(0-0.3-0.1,-1+0.3*2/3);  %k=(1, -2/3)
	
	\draw [blue,dashed,thick]  (3/2-0.25-.15,-.25*2/3)--(0.4 -.15,-1+.4*2/3);  %k=(1, 2/3)
%	\draw [blue,dashed,thick]   (3/2-0.4+.1,-.4*2/3)--(0.3 +.1,-1+.3*2/3);  %k=(1, 2/3)

%	\draw [red,very thick] (-.07,-3/2*1.732+.4)--(-.07,-1-.4);
%	\draw [blue,dashed,thick]  (.07,-3/2*1.732+0.4)--(.07,-1-0.4);

	\end{tikzpicture}
	\caption{$C=4$, one $zw$-edge}
\label{fig:C=43}
\end{figure}
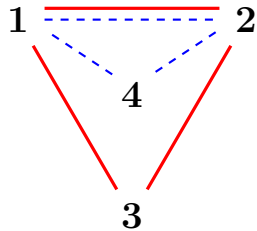

\subsubsection{$C=5$}

\indent\par
Consider a bicolored vertex (let us say, $\textbf{1}$ ) with five strokes.

Suppose  vertex $\textbf{1}$  has one adjacent $z$-edge connected  vertex $\textbf{4}$, and two adjacent $zw$-edges connected  vertex $\textbf{2}$ and  vertex $\textbf{3}$ separately. Then vertices  $\textbf{1,2,3}$ formed a $w$-color triangle and vertices $\textbf{1,2,3,4}$ form a fully $z$-stroked diagram by Rule VI.  As $C=5$ there is no any $w$-stroke connected  vertex $\textbf{4}$. Thus there is no any $w$-circle at vertices by Rule II and  Estimate \ref{Estimate1}. According to if there are $z$-circle at vertices, the only possible diagrams are that in the following Figure \ref{fig:C=51}.

\begin{figure}[!h]
	\centering
	\begin{tikzpicture}
	\vspace*{0cm}\hspace*{-4cm}
	%\draw	[blue,dashed,thick]  (-3/2,0)  circle (0.35);
	\draw	  (-3/2,0)    node {\large\textbf{1}};
	
%	\draw	[blue,dashed,thick]  (3/2,0)  circle (0.35);
	\draw		(3/2,0) node {\large\textbf{2}};
	
%	\draw	[blue,dashed,thick]  (0,-3/2*1.732)   circle (0.35);
	\draw		(0,-3/2*1.732) node {\large\textbf{3}};
	
%	\draw	[blue,dashed,thick]  (0,-1)  circle (0.35);
	\draw		(0,-1) node {\large\textbf{4}};

	\draw [red,very thick] (-3/2+.35,0.15)--(3/2-.35,0.15);
\draw [blue, dashed,thick] (-3/2+0.35,0)--(3/2-.35,0);

\draw [blue, dashed,thick] (-3/2+.2,-0.2*1.732)--(-.2,-3/2*1.732+.2*1.732);
\draw [red,very thick] (-3/2-0.15+.25, -0.25*1.732)--(-0.15-.25,-3/2*1.732+0.25*1.732);

\draw [blue, dashed,thick] (3/2-0.2,-0.2*1.732)--(0.2,-3/2*1.732+0.2*1.732);
\draw [red,very thick] (3/2+0.15-0.25,-0.25*1.732)--(0.15+0.25,-3/2*1.732+.25*1.732);

\draw [red,very thick]  (-3/2+0.3+0.15,-0.3*2/3)--(0-0.35+0.15,-1+0.35*2/3);  %k=(1, -2/3)
%	\draw [red,very thick]  (-3/2+0.4-0.1,-0.4*2/3)--(0-0.3-0.1,-1+0.3*2/3);  %k=(1, -2/3)

\draw [red,very thick]  (3/2-0.25-.15,-.25*2/3)--(0.4 -.15,-1+.4*2/3);  %k=(1, 2/3)
%	\draw [blue,dashed,thick]   (3/2-0.4+.1,-.4*2/3)--(0.3 +.1,-1+.3*2/3);  %k=(1, 2/3)

\draw [red,very thick] (0,-3/2*1.732+.4)--(0,-1-.4);
%	\draw [blue,dashed,thick]  (.07,-3/2*1.732+0.4)--(.07,-1-0.4);

	\vspace*{0cm}\hspace*{4.5cm}
	
%	\draw	[blue,dashed, thick]  (-3/2,0)  circle (0.25);
	\draw	[red,very thick]  (-3/2,0)  circle (0.35);
	\draw  (-3/2,0)  node {\large\textbf{1}};
	
%	\draw	[blue,dashed, thick]  (3/2,0)  circle (0.25);
	\draw	[red,very thick]  (3/2,0)  circle (0.35);
	\draw		(3/2,0) node {\large\textbf{2}};
	
	\draw	[red,very thick]  (0,-3/2*1.732)   circle (0.35);
	\draw		(0,-3/2*1.732) node {\large\textbf{3}};
	
%	\draw	[blue,dashed,thick]  (0,-1)  circle (0.35);
	\draw		(0,-1) node {\large\textbf{4}};
	
	\draw [red,very thick] (-3/2+.35,0.15)--(3/2-.35,0.15);
\draw [blue, dashed,thick] (-3/2+0.35,0)--(3/2-.35,0);

\draw [blue, dashed,thick] (-3/2+.2,-0.2*1.732)--(-.2,-3/2*1.732+.2*1.732);
\draw [red,very thick] (-3/2-0.15+.25, -0.25*1.732)--(-0.15-.25,-3/2*1.732+0.25*1.732);

\draw [blue, dashed,thick] (3/2-0.2,-0.2*1.732)--(0.2,-3/2*1.732+0.2*1.732);
\draw [red,very thick] (3/2+0.15-0.25,-0.25*1.732)--(0.15+0.25,-3/2*1.732+.25*1.732);

\draw [red,very thick]  (-3/2+0.3+0.15,-0.3*2/3)--(0-0.35+0.15,-1+0.35*2/3);  %k=(1, -2/3)
%	\draw [red,very thick]  (-3/2+0.4-0.1,-0.4*2/3)--(0-0.3-0.1,-1+0.3*2/3);  %k=(1, -2/3)

\draw [red,very thick]  (3/2-0.25-.15,-.25*2/3)--(0.4 -.15,-1+.4*2/3);  %k=(1, 2/3)
%	\draw [blue,dashed,thick]   (3/2-0.4+.1,-.4*2/3)--(0.3 +.1,-1+.3*2/3);  %k=(1, 2/3)

\draw [red,very thick] (0,-3/2*1.732+.4)--(0,-1-.4);
%	\draw [blue,dashed,thick]  (.07,-3/2*1.732+0.4)--(.07,-1-0.4);

	\vspace*{0cm}\hspace*{4.5cm}
%	\draw	[blue,dashed, thick]  (-3/2,0)  circle (0.25);
	\draw	[red,very thick]  (-3/2,0)  circle (0.35);
	\draw  (-3/2,0)  node {\large\textbf{1}};
	
%	\draw	[blue,dashed, thick]  (3/2,0)  circle (0.25);
	\draw	[red,very thick]  (3/2,0)  circle (0.35);
	\draw		(3/2,0) node {\large\textbf{2}};
	
%	\draw	[blue,dashed, thick]  (0,-3/2*1.732) circle (0.25);
	\draw	[red,very thick] (0,-3/2*1.732) circle (0.35);
	\draw		(0,-3/2*1.732) node {\large\textbf{3}};
	
%	\draw	[blue,dashed, thick]  (0,-1) circle (0.25);
	\draw	[red,very thick](0,-1) circle (0.35);
	\draw		(0,-1) node {\large\textbf{4}};
	
	\draw [red,very thick] (-3/2+.35,0.15)--(3/2-.35,0.15);
	\draw [blue, dashed,thick] (-3/2+0.35,0)--(3/2-.35,0);
	
	\draw [blue, dashed,thick] (-3/2+.2,-0.2*1.732)--(-.2,-3/2*1.732+.2*1.732);
	\draw [red,very thick] (-3/2-0.15+.25, -0.25*1.732)--(-0.15-.25,-3/2*1.732+0.25*1.732);
	
	\draw [blue, dashed,thick] (3/2-0.2,-0.2*1.732)--(0.2,-3/2*1.732+0.2*1.732);
	\draw [red,very thick] (3/2+0.15-0.25,-0.25*1.732)--(0.15+0.25,-3/2*1.732+.25*1.732);

	\draw [red,very thick]  (-3/2+0.3+0.15,-0.3*2/3)--(0-0.4+0.15,-1+0.4*2/3);  %k=(1, -2/3)
%	\draw [red,very thick]  (-3/2+0.4-0.1,-0.4*2/3)--(0-0.3-0.1,-1+0.3*2/3);  %k=(1, -2/3)
	
	\draw [red,very thick]  (3/2-0.25-.15,-.25*2/3)--(0.4 -.15,-1+.4*2/3);  %k=(1, 2/3)
%	\draw [blue,dashed,thick]   (3/2-0.4+.1,-.4*2/3)--(0.3 +.1,-1+.3*2/3);  %k=(1, 2/3)

	\draw [red,very thick] (0,-3/2*1.732+.4)--(0,-1-.4);
%	\draw [blue,dashed,thick]  (.07,-3/2*1.732+0.4)--(.07,-1-0.4);

	\end{tikzpicture}
	\caption{$C=5$}
\label{fig:C=51}
\end{figure}

\subsubsection{$C=6$}

\indent\par
Consider a bicolored vertex  with six strokes. Then   vertices $\textbf{1,2,3,4}$ form a fully $zw$-edged diagram by Rule VI. According to if there are $z$-circle or $w$-circle at vertices, the only possible diagrams are that in the following Figure \ref{fig:C=61}.

\begin{figure}[!h]
	\centering
	\begin{tikzpicture}
	\vspace*{0cm}\hspace*{-4cm}
%	\draw	[blue,dashed,thick]  (-3/2,0)  circle (0.35);
	\draw	  (-3/2,0)    node {\large\textbf{1}};
	
%	\draw	[blue,dashed,thick]  (3/2,0)  circle (0.35);
	\draw		(3/2,0) node {\large\textbf{2}};
	
%	\draw	[blue,dashed,thick]  (0,-3/2*1.732)   circle (0.35);
	\draw		(0,-3/2*1.732) node {\large\textbf{3}};
	
%	\draw	[blue,dashed,thick]  (0,-1)  circle (0.35);
	\draw		(0,-1) node {\large\textbf{4}};

	\draw [red,very thick] (-3/2+.35,0.15)--(3/2-.35,0.15);
	\draw [blue, dashed,thick] (-3/2+0.35,0)--(3/2-.35,0);
	
	\draw [blue, dashed,thick] (-3/2+.2,-0.2*1.732)--(-.2,-3/2*1.732+.2*1.732);
	\draw [red,very thick] (-3/2-0.15+.25, -0.25*1.732)--(-0.15-.25,-3/2*1.732+0.25*1.732);
	
	\draw [blue, dashed,thick] (3/2-0.2,-0.2*1.732)--(0.2,-3/2*1.732+0.2*1.732);
	\draw [red,very thick] (3/2+0.15-0.25,-0.25*1.732)--(0.15+0.25,-3/2*1.732+.25*1.732);

	\draw [blue,dashed,thick]  (-3/2+0.3+0.15,-0.3*2/3)--(0-0.35+0.15,-1+0.35*2/3);  %k=(1, -2/3)
	\draw [red,very thick]  (-3/2+0.4-0.1,-0.4*2/3)--(0-0.3-0.1,-1+0.3*2/3);  %k=(1, -2/3)
	
	\draw [red,very thick]  (3/2-0.25-.15,-.25*2/3)--(0.4 -.15,-1+.4*2/3);  %k=(1, 2/3)
	\draw [blue,dashed,thick]   (3/2-0.4+.1,-.4*2/3)--(0.3 +.1,-1+.3*2/3);  %k=(1, 2/3)

	\draw [red,very thick] (-.07,-3/2*1.732+.4)--(-.07,-1-.4);
	\draw [blue,dashed,thick]  (.07,-3/2*1.732+0.4)--(.07,-1-0.4);

	\vspace*{0cm}\hspace*{4.5cm}
	
%	\draw	[blue,dashed, thick]  (-3/2,0)  circle (0.25);
	\draw	[red,very thick]  (-3/2,0)  circle (0.35);
	\draw  (-3/2,0)  node {\large\textbf{1}};
	
%	\draw	[blue,dashed, thick]  (3/2,0)  circle (0.25);
	\draw	[red,very thick]  (3/2,0)  circle (0.35);
	\draw		(3/2,0) node {\large\textbf{2}};
	
	\draw	[red,very thick]  (0,-3/2*1.732)   circle (0.35);
	\draw		(0,-3/2*1.732) node {\large\textbf{3}};
	
	\draw	[red,very thick]  (0,-1)  circle (0.35);
	\draw		(0,-1) node {\large\textbf{4}};
	
	\draw [red,very thick] (-3/2+.35,0.15)--(3/2-.35,0.15);
	\draw [blue, dashed,thick] (-3/2+0.35,0)--(3/2-.35,0);
	
	\draw [blue, dashed,thick] (-3/2+.2,-0.2*1.732)--(-.2,-3/2*1.732+.2*1.732);
	\draw [red,very thick] (-3/2-0.15+.25, -0.25*1.732)--(-0.15-.25,-3/2*1.732+0.25*1.732);
	
	\draw [blue, dashed,thick] (3/2-0.2,-0.2*1.732)--(0.2,-3/2*1.732+0.2*1.732);
	\draw [red,very thick] (3/2+0.15-0.25,-0.25*1.732)--(0.15+0.25,-3/2*1.732+.25*1.732);

	\draw [blue,dashed,thick]  (-3/2+0.3+0.15,-0.3*2/3)--(0-0.35+0.15,-1+0.35*2/3);  %k=(1, -2/3)
	\draw [red,very thick]  (-3/2+0.4-0.1,-0.4*2/3)--(0-0.3-0.1,-1+0.3*2/3);  %k=(1, -2/3)
	
	\draw [red,very thick]  (3/2-0.25-.15,-.25*2/3)--(0.4 -.15,-1+.4*2/3);  %k=(1, 2/3)
	\draw [blue,dashed,thick]   (3/2-0.4+.1,-.4*2/3)--(0.3 +.1,-1+.3*2/3);  %k=(1, 2/3)

	\draw [red,very thick] (-.07,-3/2*1.732+.4)--(-.07,-1-.4);
	\draw [blue,dashed,thick]  (.07,-3/2*1.732+0.4)--(.07,-1-0.4);

	\vspace*{0cm}\hspace*{4.5cm}
	\draw	[blue,dashed, thick]  (-3/2,0)  circle (0.25);
	\draw	[red,very thick]  (-3/2,0)  circle (0.35);
	\draw  (-3/2,0)  node {\large\textbf{1}};
	
	\draw	[blue,dashed, thick]  (3/2,0)  circle (0.25);
	\draw	[red,very thick]  (3/2,0)  circle (0.35);
	\draw		(3/2,0) node {\large\textbf{2}};
	
	\draw	[blue,dashed, thick]  (0,-3/2*1.732) circle (0.25);
	\draw	[red,very thick] (0,-3/2*1.732) circle (0.35);
	\draw		(0,-3/2*1.732) node {\large\textbf{3}};
	
	\draw	[blue,dashed, thick]  (0,-1) circle (0.25);
	\draw	[red,very thick](0,-1) circle (0.35);
	\draw		(0,-1) node {\large\textbf{4}};
	
	\draw [red,very thick] (-3/2+.35,0.15)--(3/2-.35,0.15);
	\draw [blue, dashed,thick] (-3/2+0.35,0)--(3/2-.35,0);
	
	\draw [blue, dashed,thick] (-3/2+.2,-0.2*1.732)--(-.2,-3/2*1.732+.2*1.732);
	\draw [red,very thick] (-3/2-0.15+.25, -0.25*1.732)--(-0.15-.25,-3/2*1.732+0.25*1.732);
	
	\draw [blue, dashed,thick] (3/2-0.2,-0.2*1.732)--(0.2,-3/2*1.732+0.2*1.732);
	\draw [red,very thick] (3/2+0.15-0.25,-0.25*1.732)--(0.15+0.25,-3/2*1.732+.25*1.732);

	\draw [blue,dashed,thick]  (-3/2+0.3+0.15,-0.3*2/3)--(0-0.35+0.15,-1+0.35*2/3);  %k=(1, -2/3)
	\draw [red,very thick]  (-3/2+0.4-0.1,-0.4*2/3)--(0-0.3-0.1,-1+0.3*2/3);  %k=(1, -2/3)
	
	\draw [red,very thick]  (3/2-0.25-.15,-.25*2/3)--(0.4 -.15,-1+.4*2/3);  %k=(1, 2/3)
	\draw [blue,dashed,thick]   (3/2-0.4+.1,-.4*2/3)--(0.3 +.1,-1+.3*2/3);  %k=(1, 2/3)

	\draw [red,very thick] (-.07,-3/2*1.732+.4)--(-.07,-1-.4);
	\draw [blue,dashed,thick]  (.07,-3/2*1.732+0.4)--(.07,-1-0.4);

	\end{tikzpicture}
	\caption{$C=6$}
\label{fig:C=61}
\end{figure}

\subsection{Exclusion of  diagrams}
\indent\par
 We derive a list of problematic diagrams which we cannot exclude without further hypotheses
on the vorticities, are incorporated into the list of diagrams in Figure \ref{fig:Problematicdiagrams}.

\begin{proposition}\label{relationstwovertices}
Suppose a diagram has two $z$-circled vertices (say $\textbf{1}$ and $\textbf{2}$) which are also $z$-close,   if none of all the other vertices is $z$-close with them,  then $\Gamma_1+\Gamma_2\neq 0$ and $\overline{\Lambda}z_{12}w_{12}\sim \Gamma_1+\Gamma_2$. In particular, vertices $\textbf{1}$ and $\textbf{2}$ cannot form a $z$-stroke.
\end{proposition}
{\bf Proof.} %of Theorem \ref{asymptic2}:}
According to  $\overline{\Lambda} w_1=\sum_{ j \neq 1} \Gamma_j W_{j1}$ and $\overline{\Lambda} w_2=\sum_{ j \neq 2} \Gamma_j W_{j2}$, it follows that
\begin{equation}\label{formulartwovertices}
    \overline{\Lambda} w_{12}=(\Gamma_1+\Gamma_2)W_{12}+\sum_{ j > 2} \Gamma_j (W_{j2}-W_{j1}).
\end{equation}

It is easy to see that $W_{j2}-W_{j1}=\frac{1}{z_{j2}}-\frac{1}{z_{j1}}\prec \epsilon^2$ for any $j > 2$, therefore,
\begin{center}
$\overline{\Lambda} w_{12}\sim(\Gamma_1+\Gamma_2)/z_{12}$.
\end{center}

Note that $\epsilon^2\preceq z_{kl}\prec \epsilon^{-2}$, $\epsilon^2\preceq w_{kl}\preceq \epsilon^{-2}$, the proof is trivial now.

$~~~~~~~~~~~~~~~~~~~~~~~~~~~~~~~~~~~~~~~~~~~~~~~~~~~~~~~~~~~~~~~~~~~~~~~~~~~~~~~~~~~~~~~~~~~~~~~~~~~~~~~~~~~~~~~~~~~~~~~~~~~~~~~~~~~~~~~~~~~~~~~~~~~~\Box$\\

\begin{proposition}\label{relationsonvorticities1}
Suppose a diagram has an isolated $z$-stroke, then vertices of it  are both $z$-circled; if the two vertices are $z$-close (for example, provided the two vertices are connected by $w$-stroke), then the total vorticity of them is zero.
\end{proposition}
{\bf Proof.} %of Theorem \ref{asymptic2}:}

The proof is trivial by Rule I, IV  and Estimate \ref{Estimate1}.

$~~~~~~~~~~~~~~~~~~~~~~~~~~~~~~~~~~~~~~~~~~~~~~~~~~~~~~~~~~~~~~~~~~~~~~~~~~~~~~~~~~~~~~~~~~~~~~~~~~~~~~~~~~~~~~~~~~~~~~~~~~~~~~~~~~~~~~~~~~~~~~~~~~~~\Box$\\

\begin{proposition}\label{relationsonvorticities2}
Suppose a diagram has an isolated $z$-color triangle, and none of vertices (say $\textbf{1,2,3}$) of it  are $z$-circled, then $\frac{1}{\Gamma_1}+\frac{1}{\Gamma_2}+\frac{1}{\Gamma_3}=0$ or $\Gamma_1\Gamma_2+\Gamma_2 \Gamma_3+ \Gamma_3 \Gamma_1=0$.
\end{proposition}
{\bf Proof.} %of Theorem \ref{asymptic2}:}

It follows from  $\Lambda z_n=\sum_{ j \neq n} \Gamma_j Z_{jn}$ that
\begin{center}
$\Gamma_2 Z_{12}+\Gamma_3 Z_{13}\prec \epsilon^{-2}$, ~~~~~~~$\Gamma_1 Z_{21}+\Gamma_3 Z_{23}\prec \epsilon^{-2}$, ~~~~~~~$ \Gamma_1 Z_{31}+\Gamma_2 Z_{32}\prec \epsilon^{-2}$.
\end{center}
Thus
\begin{equation}
    \frac{Z_{12}}{\Gamma_3}\sim  \frac{Z_{23}}{\Gamma_1}\sim  \frac{Z_{31}}{\Gamma_2}, ~~~~~~~~~~~~~~~~~\text{or} ~~~~~~~~~~~~~~~~~w_{12}\Gamma_3\sim w_{23}\Gamma_1\sim  w_{31}\Gamma_2.
\end{equation}
According to the fact that $w_{12}+w_{23}+w_{31}=0$, it follows that $\frac{1}{\Gamma_1}+\frac{1}{\Gamma_2}+\frac{1}{\Gamma_3}=0$.

$~~~~~~~~~~~~~~~~~~~~~~~~~~~~~~~~~~~~~~~~~~~~~~~~~~~~~~~~~~~~~~~~~~~~~~~~~~~~~~~~~~~~~~~~~~~~~~~~~~~~~~~~~~~~~~~~~~~~~~~~~~~~~~~~~~~~~~~~~~~~~~~~~~~~\Box$\\

\begin{proposition}\label{relationsonvorticities3}
Suppose a fully $z$-stroked sub-diagram with four vertices exists in isolation  in a diagram, and none of vertices (say $\textbf{1,2,3,4}$) of it  are $z$-circled, then
\begin{equation}\label{L1234}
  L_{1234}= \Gamma_1\Gamma_2+\Gamma_2 \Gamma_3+ \Gamma_3 \Gamma_1+\Gamma_4 (\Gamma_1+\Gamma_2+\Gamma_3)=0.
\end{equation}

\end{proposition}
{\bf Proof.} %of Theorem \ref{asymptic2}:}

According to  $\Lambda z_n=\sum_{ j \neq n} \Gamma_j Z_{jn}$, it follows that
\begin{center}
$\Gamma_2 Z_{12}+\Gamma_3 Z_{13}+\Gamma_4 Z_{14}\prec \epsilon^{-2}$, ~~~~~~~$\Gamma_1 Z_{21}+\Gamma_3 Z_{23}+\Gamma_4 Z_{24}\prec \epsilon^{-2}$, ~~~~~~~$ \Gamma_1 Z_{31}+\Gamma_2 Z_{32}+\Gamma_4 Z_{34}\prec \epsilon^{-2}$,  ~~~~~~~$\Gamma_1 Z_{41}+\Gamma_2 Z_{42}+\Gamma_3 Z_{43}\prec \epsilon^{-2}$.
\end{center}
Thus
\begin{equation}
\begin{array}{c}
   Z_{13}\sim \frac{-\Gamma_2 Z_{12}-\Gamma_4 Z_{14}}{\Gamma_3},~~~~~~~~~~~~~~~~~ Z_{24}\sim \frac{\Gamma_1 Z_{12}-\Gamma_3 Z_{23}}{\Gamma_4}, \\
  Z_{34}\sim\frac{-\Gamma_1\Gamma_2 Z_{12}-\Gamma_1\Gamma_4 Z_{14}+\Gamma_2\Gamma_3 Z_{23}}{\Gamma_3\Gamma_4}.
\end{array}\nonumber
\end{equation}
According to the fact that $w_{12}+w_{23}+w_{31}=0$, it follows that
\begin{equation}\label{f123}
    \Gamma_2 Z_{12}^2+(\Gamma_2 +\Gamma_3)Z_{23} Z_{12} \sim-\Gamma_4 Z_{14} ( Z_{12}+Z_{23}).\nonumber
\end{equation}
Then $Z_{12}+Z_{23}\approx \epsilon^{-2}$ and
\begin{equation}\label{z14f123}
    Z_{14} \sim- \frac{\Gamma_2 Z_{12}^2+(\Gamma_2 +\Gamma_3)Z_{23} Z_{12}}{\Gamma_4( Z_{12}+Z_{23})} .\nonumber
\end{equation}
Set $Z_{12}\sim\epsilon^{-2}$ and $Z_{23}\sim a\epsilon^{-2}$, where $a\in\mathbb{C}\backslash\{0,-1\}$, then
\begin{equation}
\begin{array}{c}
   Z_{13}\sim \frac{a}{a+1}\epsilon^{-2},~~~~~~~~~~~~~~~~~ Z_{24}\sim \frac{\Gamma_1-a \Gamma_3}{\Gamma_4}\epsilon^{-2}, \\
  Z_{34}\sim\frac{a ((a+1) \Gamma_2+\Gamma_1)}{(a+1) \Gamma_4}\epsilon^{-2},~~~~~~~~~~~~~~~~~    Z_{14}\sim-\frac{a \Gamma_2+a \Gamma_3+\Gamma_2}{a \Gamma_4+\Gamma_4}\epsilon^{-2}.
\end{array}\nonumber
\end{equation}

According to the facts that $w_{12}+w_{24}+w_{41}=0$, $w_{13}+w_{34}+w_{41}=0$ and $w_{23}+w_{34}+w_{42}=0$, it follows that
\begin{equation}\label{f124}
   -a^2 \Gamma_3 (\Gamma_3+\Gamma_4)+(a+1) \Gamma_2 (\Gamma_4-a \Gamma_3)+\Gamma_1 ((a+1) \Gamma_2+a (\Gamma_3+\Gamma_4)+\Gamma_4)=0,
\end{equation}
\begin{equation}\label{f134}
   (a+1)^2 \Gamma_2^2+(a+1) \Gamma_2 (a (\Gamma_3+\Gamma_4)+\Gamma_4)+a \Gamma_3 \Gamma_4+\Gamma_1 ((a+1) \Gamma_2+a (\Gamma_3+\Gamma_4))=0,
\end{equation}
\begin{equation}\label{f234}
  -\Gamma_1 ((a+1) \Gamma_2-a \Gamma_3+\Gamma_4)+a (a+1) (\Gamma_3 \Gamma_4+\Gamma_2 (\Gamma_3+\Gamma_4))-\Gamma_1^2=0.
\end{equation}

A straightforward computation of  Groebner basis for (\ref{f124}), (\ref{f134}) and (\ref{f234}) shows that
\begin{equation}\label{g123}
    (\Gamma_2 \Gamma_3+\Gamma_1 (\Gamma_2+\Gamma_3)) (\Gamma_3 \Gamma_4+\Gamma_2 (\Gamma_3+\Gamma_4)+\Gamma_1 (\Gamma_2+\Gamma_3+\Gamma_4))=0.\nonumber
\end{equation}

By the symmetry of vertices  $\textbf{1,2,3,4}$, the following relations also hold.
\begin{equation}\label{g124}
   (\Gamma_2 \Gamma_4+\Gamma_1 (\Gamma_2+\Gamma_4)) (\Gamma_2 (\Gamma_3+\Gamma_4)+\Gamma_1 (\Gamma_2+\Gamma_3+\Gamma_4)+\Gamma_3 \Gamma_4)=0.\nonumber
\end{equation}
\begin{equation}\label{g134}
    (\Gamma_3 \Gamma_4+\Gamma_1 (\Gamma_3+\Gamma_4)) (\Gamma_1 (\Gamma_2+\Gamma_3+\Gamma_4)+(\Gamma_2+\Gamma_4) \Gamma_3+\Gamma_2 \Gamma_4)=0.\nonumber
\end{equation}
\begin{equation}\label{g234}
   (\Gamma_3 \Gamma_4+\Gamma_2 (\Gamma_3+\Gamma_4)) ((\Gamma_1+\Gamma_3+\Gamma_4) \Gamma_2+(\Gamma_1+\Gamma_4) \Gamma_3+\Gamma_1 \Gamma_4)=0.\nonumber
\end{equation}

It is easy to see that there must be $L_{1234}=0$.

$~~~~~~~~~~~~~~~~~~~~~~~~~~~~~~~~~~~~~~~~~~~~~~~~~~~~~~~~~~~~~~~~~~~~~~~~~~~~~~~~~~~~~~~~~~~~~~~~~~~~~~~~~~~~~~~~~~~~~~~~~~~~~~~~~~~~~~~~~~~~~~~~~~~~\Box$\\

\begin{remark}
In general, we claim that
\begin{proposition}
Suppose a fully $z$-stroked sub-diagram with $n$ vertices exists in isolation  in a diagram, and none of vertices (say $\textbf{1}$, $\textbf{2}$, $ \cdots $ $\textbf{n}$) of it  are $z$-circled, then
\begin{equation}
  L_{1\cdots n}=\sum_{1\leq j <k\leq n} \Gamma_j\Gamma_k=0.\nonumber
\end{equation}
\end{proposition}

However it is not easy to prove this result for general $n$.
\end{remark}

\subsubsection{Problematic diagrams}

\indent\par
First, we give  a simple result for the following discussions.

\begin{proposition}\label{simplelemma}
Suppose  $\Gamma_1,\Gamma_2,\Gamma_3,\Gamma_4$ are all nonzero,  then
\begin{center}
$ \left\{
             \begin{array}{l}
             {\Gamma_1}+{\Gamma_2}+{\Gamma_3}=0   \\
             \frac{1}{\Gamma_1}+\frac{1}{\Gamma_2}+\frac{1}{\Gamma_3}=0
             \end{array}
\right.   ~~~ \Longleftrightarrow ~~~ \left\{
             \begin{array}{l}
             {\Gamma_1}+{\Gamma_2}+{\Gamma_3}=0   \\
             L_{1234}=0
             \end{array}
\right.   ~~~ \Longleftrightarrow ~~~\left\{
             \begin{array}{l}
             L_{1234}=0   \\
             \frac{1}{\Gamma_1}+\frac{1}{\Gamma_2}+\frac{1}{\Gamma_3}=0
             \end{array}
\right. $
\end{center}
And in this case, $\Gamma_1,\Gamma_2,\Gamma_3,\Gamma_4$ cannot all be real.
\end{proposition}
{\bf Proof.} %of Theorem \ref{asymptic2}:}

The proof is simple, and we only point out that the inequality
\begin{equation}
    ({\Gamma_1}+{\Gamma_2}+{\Gamma_3})^2-2(\Gamma_1\Gamma_2+\Gamma_2 \Gamma_3+ \Gamma_3 \Gamma_1)>0\nonumber
\end{equation}
for real $\Gamma_1,\Gamma_2,\Gamma_3$.

$~~~~~~~~~~~~~~~~~~~~~~~~~~~~~~~~~~~~~~~~~~~~~~~~~~~~~~~~~~~~~~~~~~~~~~~~~~~~~~~~~~~~~~~~~~~~~~~~~~~~~~~~~~~~~~~~~~~~~~~~~~~~~~~~~~~~~~~~~~~~~~~~~~~~\Box$\\

As a result of Proposition \ref{relationstwovertices} and Proposition \ref{relationsonvorticities1}, it  is easy to see that the first two diagrams in Figure \ref{fig:C=21}, the second diagram in Figure \ref{fig:C=31} and the second diagram in Figure \ref{fig:C=42} are  impossible.

The first diagram in Figure \ref{fig:C=31} is  impossible by Proposition \ref{relationsonvorticities1} and Proposition \ref{relationsonvorticities2}.

The second diagram in Figure \ref{fig:C=41} and the second diagram in Figure \ref{fig:C=51} are  impossible by  Rule IV and Proposition \ref{relationsonvorticities2}.

The first diagram in Figure \ref{fig:C=42} is  impossible by Proposition \ref{relationsonvorticities1} and Proposition \ref{relationsonvorticities3}.

The first diagram in Figure \ref{fig:C=51} is  impossible by Proposition \ref{relationsonvorticities2} and Proposition \ref{relationsonvorticities3}.

The second diagram in Figure \ref{fig:C=61} is  impossible by  Rule IV and Proposition \ref{relationsonvorticities3}.

The conclusion of this section is that any singular sequence should converge
to one of the ten diagrams in Figure \ref{fig:Problematicdiagrams}.

\section{Problematic diagrams}

\indent\par

\begin{figure}%[!h]
	\centering
	
		\begin{subfigure}[b]{0.2\textwidth}
\centering
		\resizebox{\linewidth}{!}{
		\begin{tikzpicture}%[scale=0.7]
			\hspace{0cm}
	\draw  (-3/2,  0) node {\large\textbf{1}};%z-circle
\draw (3/2,  0) node{\large\textbf{2}};%z-circle

\draw  (-3/2,  -2)  node {\large\textbf{4}};%w-circle
\draw  (3/2,  -2) node {\large\textbf{3}}; %w-circle

	\draw [red,very thick] (-3/2,  0) circle (0.35);
	\draw [red,very thick] (3/2,  0) circle (0.35);

	\draw [blue,dashed, thick] (-3/2,  -2) circle (0.35);
	\draw [blue,dashed, thick]  (3/2,  -2) circle (0.35);

		\draw [red,very thick] (-1.2,0)--(1.2,0); % z-edge
	\draw [blue,dashed, thick] (-1.2,-2)--(1.2,-2); % z-edge
	\end{tikzpicture}
}
	\caption{Diagram I}
	\end{subfigure}
		\begin{subfigure}[b]{0.2\textwidth}
\centering
		\resizebox{\linewidth}{!}{
		\begin{tikzpicture}%[scale=0.7]
		%\hspace{-1cm}
		%	\draw	[blue,dashed,thick]  (-3/2,0)  circle (0.35);
		\draw	  (-3/2,0)    node {\large\textbf{1}};
		
		%	\draw	[blue,dashed,thick]  (3/2,0)  circle (0.35);
		\draw		(3/2,0) node {\large\textbf{2}};
		
		%	\draw	[blue,dashed,thick]  (0,-3/2*1.732)   circle (0.35);
		\draw		(0,-3/2*1.732) node {\large\textbf{3}};
		
		%	\draw	[blue,dashed,thick]  (0,-1)  circle (0.35);
		\draw		(0,-1) node {\large\textbf{4}};

		\draw [red,very thick] (-3/2+.35,0.15)--(3/2-.35,0.15);
		\draw [blue, dashed,thick] (-3/2+0.35,0)--(3/2-.35,0);
		
		\draw [blue, dashed,thick] (-3/2+.2,-0.2*1.732)--(-.2,-3/2*1.732+.2*1.732);
		\draw [red,very thick] (-3/2-0.15+.25, -0.25*1.732)--(-0.15-.25,-3/2*1.732+0.25*1.732);
		
		\draw [blue, dashed,thick] (3/2-0.2,-0.2*1.732)--(0.2,-3/2*1.732+0.2*1.732);
		\draw [red,very thick] (3/2+0.15-0.25,-0.25*1.732)--(0.15+0.25,-3/2*1.732+.25*1.732);

		\draw [blue,dashed,thick]  (-3/2+0.3+0.15,-0.3*2/3)--(0-0.35+0.15,-1+0.35*2/3);  %k=(1, -2/3)
		\draw [red,very thick]  (-3/2+0.4-0.1,-0.4*2/3)--(0-0.3-0.1,-1+0.3*2/3);  %k=(1, -2/3)
		
		\draw [red,very thick]  (3/2-0.25-.15,-.25*2/3)--(0.4 -.15,-1+.4*2/3);  %k=(1, 2/3)
		\draw [blue,dashed,thick]   (3/2-0.4+.1,-.4*2/3)--(0.3 +.1,-1+.3*2/3);  %k=(1, 2/3)

		\draw [red,very thick] (-.07,-3/2*1.732+.4)--(-.07,-1-.4);
		\draw [blue,dashed,thick]  (.07,-3/2*1.732+0.4)--(.07,-1-0.4);
			\end{tikzpicture}
}
		\caption{Diagram  II}
	\end{subfigure}

\begin{subfigure}[b]{0.2\textwidth}
		\centering
		\resizebox{\linewidth}{!}{
	\begin{tikzpicture}
	
		\draw  (-3/2,  0) node {\large\textbf{1}};%z-circle
	\draw (3/2,  0) node{\large\textbf{2}};%z-circle
	
	\draw  (-3/2,  -2)  node {\large\textbf{4}};%w-circle
	\draw  (3/2,  -2) node {\large\textbf{3}}; %w-circle

		\draw [blue,dashed, thick] (-3/2,  0) circle (0.25);
	\draw [red,very thick] (-3/2,  0) circle (0.35);
	
		\draw [blue,dashed, thick] (3/2,  0) circle (0.25);
\draw [red,very thick] (3/2,  0) circle (0.35);

\draw [blue,dashed, thick] (-3/2,  -2) circle (0.25);
\draw [red,very thick](-3/2,  -2) circle (0.35);

\draw [blue,dashed, thick] (3/2,  -2) circle (0.25);
\draw[red,very thick]  (3/2,  -2) circle (0.35);

\draw [red,very thick] (-1.2,0)--(1.2,0); % z-edge
\draw [red,very thick] (-1.2,-2)--(1.2,-2);

\draw [blue,dashed, thick] (-3/2,-.3)--(-3/2,-1.7);
\draw [blue,dashed, thick] (3/2,-.3)--(3/2,-1.7);

	\end{tikzpicture}
}
	\caption{Diagram III}
\end{subfigure}
	\begin{subfigure}[b]{0.2\textwidth}
		\centering
		\resizebox{\linewidth}{!}{
	\begin{tikzpicture}
\draw  (-3/2,  0) node {\large\textbf{1}};%z-circle
	\draw (3/2,  0) node{\large\textbf{2}};%z-circle
	
	\draw  (-3/2,  -2)  node {\large\textbf{4}};%w-circle
	\draw  (3/2,  -2) node {\large\textbf{3}}; %w-circle

	\draw [blue,dashed, thick] (-3/2,  0) circle (0.25);
	\draw [red,very thick] (-3/2,  0) circle (0.35);
	
	\draw [blue,dashed, thick] (3/2,  0) circle (0.25);
	\draw [red,very thick] (3/2,  0) circle (0.35);

	\draw [blue,dashed, thick] (-3/2,  -2) circle (0.25);
	\draw [red,very thick](-3/2,  -2) circle (0.35);
	
	\draw [blue,dashed, thick] (3/2,  -2) circle (0.25);
	\draw[red,very thick]  (3/2,  -2) circle (0.35);

	\draw [red,very thick] (-1.2,0)--(1.2,0); % z-edge
	\draw [red,very thick] (-1.2,-2)--(1.2,-2);
	
	\draw [blue,dashed, thick] (-1.2,0-0.15)--(1.2,0-0.15);
	\draw [blue,dashed, thick] (-1.2,-2+0.15)--(1.2,-2+0.15);
	\end{tikzpicture}
}
	\caption{Diagram $III'$}
\end{subfigure}
	\begin{subfigure}[b]{0.2\textwidth}
		\centering
		\resizebox{\linewidth}{!}{
	\begin{tikzpicture}
	\draw	[blue,dashed, thick]  (-3/2,0)  circle (0.25);
	\draw	[red,very thick]  (-3/2,0)  circle (0.35);
	\draw  (-3/2,0)  node {\large\textbf{1}};
	
	\draw	[blue,dashed, thick]  (3/2,0)  circle (0.25);
	\draw	[red,very thick]  (3/2,0)  circle (0.35);
	\draw		(3/2,0) node {\large\textbf{2}};
	
	\draw	[blue,dashed, thick]  (0,-3/2*1.732) circle (0.25);
	\draw	[red,very thick] (0,-3/2*1.732) circle (0.35);
	\draw		(0,-3/2*1.732) node {\large\textbf{3}};
	
	\draw	[blue,dashed, thick]  (0,-1) circle (0.25);
	\draw	[red,very thick](0,-1) circle (0.35);
	\draw		(0,-1) node {\large\textbf{4}};

	\draw [red,very thick] (-3/2+.35,0.15)--(3/2-.35,0.15);
	\draw [blue, dashed,thick] (-3/2+0.35,0)--(3/2-.35,0);
	
	\draw [red,very thick] (-3/2+.2,-0.2*1.732)--(-.2,-3/2*1.732+.2*1.732);
	%	\draw [red,very thick] (-3/2-0.15+.25, -0.25*1.732)--(-0.15-.25,-3/2*1.732+0.25*1.732);
	
	\draw [red,very thick] (3/2-0.2,-0.2*1.732)--(0.2,-3/2*1.732+0.2*1.732);
	%	\draw [red,very thick] (3/2+0.15-0.25,-0.25*1.732)--(0.15+0.25,-3/2*1.732+.25*1.732);

	\draw [red,very thick]  (-3/2+0.25+0.15,-0.25*2/3)--(0-0.4+0.15,-1+0.4*2/3);  %k=(1, -2/3)
	%		\draw [red,very thick]  (-3/2+0.4-0.1,-0.4*2/3)--(0-0.3-0.1,-1+0.3*2/3);  %k=(1, -2/3)
	
	\draw [red,very thick]  (3/2-0.25-.15,-.25*2/3)--(0.45 -.15,-1+.45*2/3);  %k=(1, 2/3)
	%			\draw [blue,dashed,thick]   (3/2-0.4+.1,-.4*2/3)--(0.3 +.1,-1+.3*2/3);  %k=(1, 2/3)

	\draw [red,very thick] (-.07,-3/2*1.732+.4)--(-.07,-1-.4);
	\draw [blue,dashed,thick]  (.07,-3/2*1.732+0.4)--(.07,-1-0.4);
	\end{tikzpicture}
}
	\caption{Diagram IV}
\end{subfigure}
	\begin{subfigure}[b]{0.2\textwidth}
		\centering
		\resizebox{\linewidth}{!}{
	\begin{tikzpicture}
	\draw	[blue,dashed, thick]  (-3/2,0)  circle (0.25);
	\draw	[red,very thick]  (-3/2,0)  circle (0.35);
	\draw  (-3/2,0)  node {\large\textbf{1}};
	
	\draw	[blue,dashed, thick]  (3/2,0)  circle (0.25);
	\draw	[red,very thick]  (3/2,0)  circle (0.35);
	\draw		(3/2,0) node {\large\textbf{2}};
	
	\draw	[blue,dashed, thick]  (0,-3/2*1.732) circle (0.25);
	\draw	[red,very thick] (0,-3/2*1.732) circle (0.35);
	\draw		(0,-3/2*1.732) node {\large\textbf{3}};
	
	\draw	[blue,dashed, thick]  (0,-1) circle (0.25);
	\draw	[red,very thick](0,-1) circle (0.35);
	\draw		(0,-1) node {\large\textbf{4}};
	
	\draw [red,very thick] (-3/2+.35,0.15)--(3/2-.35,0.15);
	\draw [blue, dashed,thick] (-3/2+0.35,0)--(3/2-.35,0);
	
	\draw [blue, dashed,thick] (-3/2+.2,-0.2*1.732)--(-.2,-3/2*1.732+.2*1.732);
	\draw [red,very thick] (-3/2-0.15+.25, -0.25*1.732)--(-0.15-.25,-3/2*1.732+0.25*1.732);
	
	\draw [blue, dashed,thick] (3/2-0.2,-0.2*1.732)--(0.2,-3/2*1.732+0.2*1.732);
	\draw [red,very thick] (3/2+0.15-0.25,-0.25*1.732)--(0.15+0.25,-3/2*1.732+.25*1.732);

	\draw [blue,dashed,thick]  (-3/2+0.3+0.15,-0.3*2/3)--(0-0.35+0.15,-1+0.35*2/3);  %k=(1, -2/3)
	\draw [red,very thick]  (-3/2+0.4-0.1,-0.4*2/3)--(0-0.3-0.1,-1+0.3*2/3);  %k=(1, -2/3)
	
	\draw [red,very thick]  (3/2-0.25-.15,-.25*2/3)--(0.4 -.15,-1+.4*2/3);  %k=(1, 2/3)
	\draw [blue,dashed,thick]   (3/2-0.4+.1,-.4*2/3)--(0.3 +.1,-1+.3*2/3);  %k=(1, 2/3)

	\draw [red,very thick] (-.07,-3/2*1.732+.4)--(-.07,-1-.4);
	\draw [blue,dashed,thick]  (.07,-3/2*1.732+0.4)--(.07,-1-0.4);
	\end{tikzpicture}
}
	\caption{Diagram V}
\end{subfigure}

	\begin{subfigure}[b]{0.2\textwidth}
		\centering
		\resizebox{\linewidth}{!}{
	\begin{tikzpicture}
	\draw	  (-3/2,0)    node {\large\textbf{1}};
	\draw		(3/2,0) node {\large\textbf{2}};
	\draw		(0,-3/2*1.732) node {\large\textbf{3}};
	\draw		(0,-1) node {\large\textbf{4}};

	\draw [red,very thick] (-3/2+.35,0)--(3/2-.35,0);
	\draw [blue, dashed,thick] (-3/2+0.35,-.15)--(3/2-.35,-0.15);
	
	\draw [red,very thick] (-3/2+.2,-0.2*1.732)--(-.2,-3/2*1.732+.2*1.732);
	\draw [blue, dashed,thick] (-3/2+0.15+.2, -0.2*1.732)--(0.15-.2,-3/2*1.732+0.2*1.732);
	
	\draw [red,very thick] (3/2-0.2,-0.2*1.732)--(0.2,-3/2*1.732+0.2*1.732);
	\draw [blue, dashed,thick] (3/2-0.15-0.2,-.2*1.732)--(-0.15+0.2,-3/2*1.732+.2*1.732);
	
	\end{tikzpicture}
}
	\caption{Diagram VI}
\end{subfigure}    %important note, here, if empty line,  means break to TIKZ
	\begin{subfigure}[b]{0.2\textwidth}
		\centering
		\resizebox{\linewidth}{!}{
	\begin{tikzpicture}
	\draw	[blue,dashed, thick]  (-3/2,0)  circle (0.25);
	\draw	[red,very thick]  (-3/2,0)  circle (0.35);
	\draw  (-3/2,0)  node {\large\textbf{1}};
	
	\draw	[blue,dashed, thick]  (3/2,0)  circle (0.25);
	\draw	[red,very thick]  (3/2,0)  circle (0.35);
	\draw		(3/2,0) node {\large\textbf{2}};
	
	\draw	[blue,dashed, thick]  (0,-3/2*1.732) circle (0.25);
	\draw	[red,very thick] (0,-3/2*1.732) circle (0.35);
	\draw		(0,-3/2*1.732) node {\large\textbf{3}};
	
	\draw		(0,-1) node {\large\textbf{4}};

	\draw [red,very thick] (-3/2+.35,0)--(3/2-.35,0);
	\draw [blue, dashed,thick] (-3/2+0.35,-.15)--(3/2-.35,-0.15);
	
	\draw [red,very thick] (-3/2+.2,-0.2*1.732)--(-.2,-3/2*1.732+.2*1.732);
	\draw [blue, dashed,thick] (-3/2+0.15+.2, -0.2*1.732)--(0.15-.2,-3/2*1.732+0.2*1.732);
	
	\draw [red,very thick] (3/2-0.2,-0.2*1.732)--(0.2,-3/2*1.732+0.2*1.732);
	\draw [blue, dashed,thick] (3/2-0.15-0.2,-.2*1.732)--(-0.15+0.2,-3/2*1.732+.2*1.732);
	\end{tikzpicture}
}
	\caption{Diagram VII}
\end{subfigure}
	\begin{subfigure}[b]{0.2\textwidth}
		\centering
		\resizebox{\linewidth}{!}{
	\begin{tikzpicture}
		\draw  (-3/2,0)  node {\large\textbf{1}};

	\draw		(3/2,0) node {\large\textbf{2}};

	\draw		(0,-3/2*1.732) node {\large\textbf{3}};

	\draw		(0,-1) node {\large\textbf{4}};
	
	\draw [red,very thick] (-3/2+.35,0.15)--(3/2-.35,0.15);
	\draw [blue, dashed,thick] (-3/2+0.35,0)--(3/2-.35,0);
	
	\draw [red,very thick] (-3/2+.2,-0.2*1.732)--(-.2,-3/2*1.732+.2*1.732);
	%	\draw [red,very thick] (-3/2-0.15+.25, -0.25*1.732)--(-0.15-.25,-3/2*1.732+0.25*1.732);
	
	\draw [red,very thick] (3/2-0.2,-0.2*1.732)--(0.2,-3/2*1.732+0.2*1.732);
	%	\draw [red,very thick] (3/2+0.15-0.25,-0.25*1.732)--(0.15+0.25,-3/2*1.732+.25*1.732);

	\draw [blue,dashed,thick]  (-3/2+0.3+0.15,-0.3*2/3)--(0-0.35+0.15,-1+0.35*2/3);  %k=(1, -2/3)
	%	\draw [red,very thick]  (-3/2+0.4-0.1,-0.4*2/3)--(0-0.3-0.1,-1+0.3*2/3);  %k=(1, -2/3)
	
	\draw [blue,dashed,thick]  (3/2-0.25-.15,-.25*2/3)--(0.4 -.15,-1+.4*2/3);  %k=(1, 2/3)
	%	\draw [blue,dashed,thick]   (3/2-0.4+.1,-.4*2/3)--(0.3 +.1,-1+.3*2/3);  %k=(1, 2/3)
	\end{tikzpicture}
}
	\caption{Diagram VIII}
\end{subfigure}
	\begin{subfigure}[b]{0.2\textwidth}

		\centering
		\resizebox{\linewidth}{!}{
	\begin{tikzpicture}
	
	%	\draw	[blue,dashed, thick]  (-3/2,0)  circle (0.25);
	\draw	[red,very thick]  (-3/2,0)  circle (0.35);
	\draw  (-3/2,0)  node {\large\textbf{1}};
	
	%	\draw	[blue,dashed, thick]  (3/2,0)  circle (0.25);
	\draw	[red,very thick]  (3/2,0)  circle (0.35);
	\draw		(3/2,0) node {\large\textbf{2}};
	
	\draw	[red,very thick]  (0,-3/2*1.732)   circle (0.35);
	\draw		(0,-3/2*1.732) node {\large\textbf{3}};
	
	\draw	[red,very thick]  (0,-1)  circle (0.35);
	\draw		(0,-1) node {\large\textbf{4}};
	
	\draw [red,very thick] (-3/2+.35,0.15)--(3/2-.35,0.15);
	\draw [blue, dashed,thick] (-3/2+0.35,0)--(3/2-.35,0);
	
	\draw [blue, dashed,thick] (-3/2+.2,-0.2*1.732)--(-.2,-3/2*1.732+.2*1.732);
	\draw [red,very thick] (-3/2-0.15+.25, -0.25*1.732)--(-0.15-.25,-3/2*1.732+0.25*1.732);
	
	\draw [blue, dashed,thick] (3/2-0.2,-0.2*1.732)--(0.2,-3/2*1.732+0.2*1.732);
	\draw [red,very thick] (3/2+0.15-0.25,-0.25*1.732)--(0.15+0.25,-3/2*1.732+.25*1.732);

	\draw [red,very thick]  (-3/2+0.3+0.15,-0.3*2/3)--(0-0.35+0.15,-1+0.35*2/3);  %k=(1, -2/3)
	%	\draw [red,very thick]  (-3/2+0.4-0.1,-0.4*2/3)--(0-0.3-0.1,-1+0.3*2/3);  %k=(1, -2/3)
	
	\draw [red,very thick]  (3/2-0.25-.15,-.25*2/3)--(0.4 -.15,-1+.4*2/3);  %k=(1, 2/3)
	%	\draw [blue,dashed,thick]   (3/2-0.4+.1,-.4*2/3)--(0.3 +.1,-1+.3*2/3);  %k=(1, 2/3)

	\draw [red,very thick] (0,-3/2*1.732+.4)--(0,-1-.4);
	%	\draw [blue,dashed,thick]  (.07,-3/2*1.732+0.4)--(.07,-1-0.4);
	\end{tikzpicture}
}
	\caption{Diagram IX}
\end{subfigure}

\caption{Problematic diagrams}
\label{fig:Problematicdiagrams}	
\end{figure}
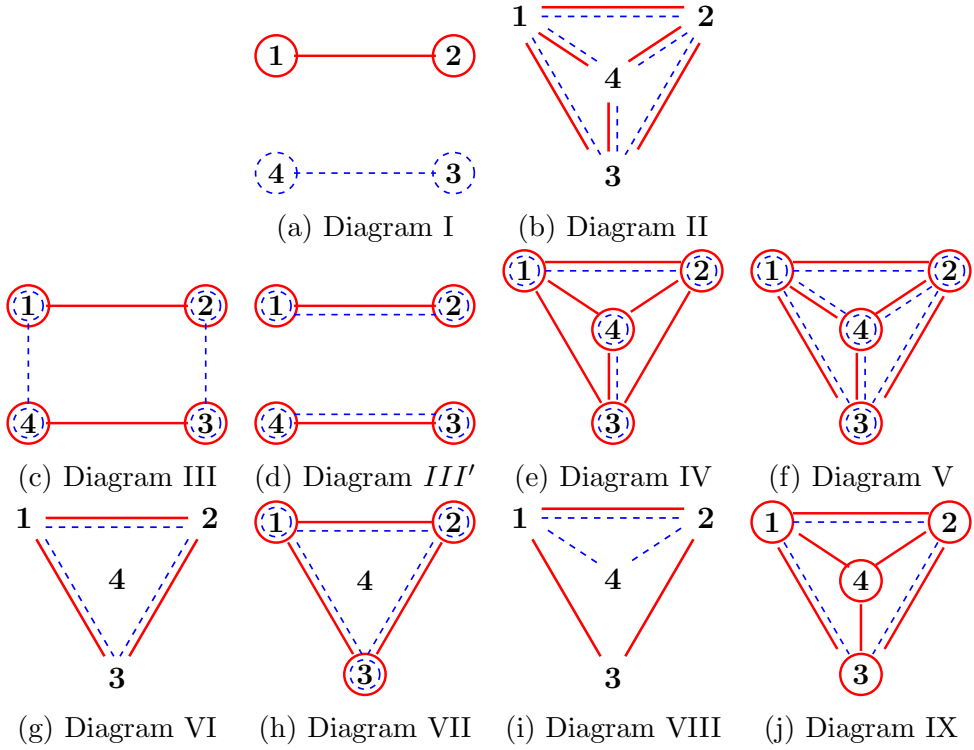

We could not eliminate the
diagrams in Figure \ref{fig:Problematicdiagrams}. Some singular sequence could still exist and approach
any of these diagrams.  In this section   we obtain the constraints on the
vorticities corresponding to each of the ten diagrams from Figure  \ref{fig:Problematicdiagrams}.

\subsection{Diagram I}
\indent\par
First,   if $\Gamma_1+\Gamma_2= 0$, then two $z$-circled vertices  $\textbf{1}$ and $\textbf{2}$  are  $z$-close, however this  contradicts Proposition \ref{relationstwovertices}. Thus $\Gamma_1+\Gamma_2\neq 0$. Similarly, we have also $\Gamma_3+\Gamma_4\neq 0$.

Without loss of generality, assume $z_1\sim -\Gamma_2\epsilon^{-2}$ and $z_2\sim \Gamma_1\epsilon^{-2}$.

According to  (\ref{formulartwovertices}), it follows that
\begin{equation}\label{formulartwoverticesd1}
    \overline{\Lambda} w_{12}=(\Gamma_1+\Gamma_2)W_{12}+ \Gamma_3 (W_{32}-W_{31})+\Gamma_4 (W_{42}-W_{41}).
\end{equation}

It is easy to see that
\begin{center}
$W_{12}=\frac{1}{z_{12}}\sim \frac{\epsilon^2}{\Gamma_1+\Gamma_2}$
\end{center}
 and
\begin{center}
$W_{j2}-W_{j1}=\frac{1}{z_{j2}}-\frac{1}{z_{j1}}\sim \frac{1}{z_{2}}-\frac{1}{z_{1}}\sim (\frac{1}{\Gamma_1}+\frac{1}{\Gamma_2})\epsilon^2$
\end{center}
for any $j \in \{3,4\}$.

By $\Lambda z_n=\sum_{ j \neq n} \Gamma_j Z_{jn}$
it follows that
\begin{equation}
{\Lambda} z_{2}\sim \Gamma_1 Z_{12}, ~~~~~~~~~~~~~~~~~\text{or} ~~~~~~~~~~~~~~~~~w_{12} \sim \frac{\epsilon^2}{\Lambda}.\nonumber
\end{equation}

Therefore,
\begin{equation}\label{formulartwoverticesd1key0}
{\overline{\Lambda}}/{\Lambda}=1+ (\Gamma_3 +\Gamma_4) (\frac{1}{\Gamma_1}+\frac{1}{\Gamma_2}).\nonumber
\end{equation}
First, it follows that $\Lambda\in \mathbb{C}\backslash \mathbb{R}$, otherwise, we have $(\Gamma_1 +\Gamma_2)(\Gamma_3 +\Gamma_4)=0$, this gives a contradiction.

As a result, we have $\Lambda=\pm \textbf{i}$ and
\begin{equation}\label{formulartwoverticesd1key1}
 (\Gamma_1 +\Gamma_2)(\Gamma_3 +\Gamma_4) +2\Gamma_1 \Gamma_2=0.
\end{equation}

On the other hand, it follows from $\Lambda\in \mathbb{C}\backslash \mathbb{R}$  that
\begin{equation}\label{angularmomentumis0}
  L =\sum_{1\leq j<k\leq 4}\Gamma_j\Gamma_k=0
\end{equation}
and  the total vorticity $\Gamma =\Gamma_1 +\Gamma_2+\Gamma_3 +\Gamma_4\neq 0$.

 By (\ref{formulartwoverticesd1key1}) and  (\ref{angularmomentumis0}), it follows that
 \begin{equation}\label{formulartwoverticesd1key2}
 \Gamma_1 \Gamma_2-\Gamma_3 \Gamma_4=0.\nonumber
\end{equation}

 To summarize, the following relations on the four vorticities should be satisfied if the Diagram I
is approached by a singular sequence:
 \begin{equation}\label{formulartwoverticesd1keyzz}
 \begin{array}{c}
   \Lambda=\pm \textbf{i}, \\
   \Gamma_1 \Gamma_2=\Gamma_3 \Gamma_4, \\
      L=0,  \\
       \Gamma\neq0,  \\
       \Gamma_1+\Gamma_2\neq 0,  \Gamma_3+\Gamma_4\neq 0.
 \end{array}
\end{equation}
Moreover, for Diagram I it is noteworthy that\begin{equation}\label{formulartwoverticesd1keyr}
 r_{12},   r_{34}\approx 1, ~~~~~~~  ~~~~~~~  r_{13}, r_{23},r_{14},r_{24}\approx \epsilon^{-2}.
\end{equation}
The proof is straightforward and we omit it.

\subsection{Diagram II}
\indent\par
By  Proposition \ref{relationsonvorticities3} it follows  that
\begin{equation}\label{formulartwoverticesd2key}
  L =\sum_{1\leq j<k\leq 4}\Gamma_j\Gamma_k=0,
\end{equation}
thus  the total vorticity $\Gamma  \neq 0$.

\subsection{Diagram III}
\indent\par

For Diagram III we can assume that
\begin{equation*}
     \begin{array}{cc}
                                                                                            z_1\sim z_4\sim -\Gamma_2 a \epsilon^{-2}, & z_2\sim z_3\sim \Gamma_1 a \epsilon^{-2}, \\
                                                                                           w_1\sim w_2\sim -\Gamma_4 b \epsilon^{-2}, & w_3\sim w_4\sim \Gamma_1 b \epsilon^{-2}.
                                                                                         \end{array}
\end{equation*}
It follows that
\begin{equation}\label{formulartwoverticesd7key0}
    \Gamma_1 \Gamma_3 =\Gamma_2 \Gamma_4.
\end{equation}

{\bfseries{Case A: $\Gamma\neq 0$ }.}

We claim that
\begin{center}
$(\Gamma_1+\Gamma_2) (\Gamma_2+\Gamma_3)  (\Gamma_3+\Gamma_4)(\Gamma_1+\Gamma_4)\neq 0$.
\end{center}
Otherwise, without loss of generality, assume $\Gamma_1+\Gamma_2=0$, by (\ref{formulartwoverticesd7key0}) it follows that $\Gamma_3+\Gamma_4=0$, this contradicts $\Gamma\neq 0$.

In this case it is easy to see that
\begin{equation}\label{DiagramIII2}
 \begin{array}{c}
   r_{12}^2\sim \frac{\Gamma_1+\Gamma_2}{\Lambda} ,~~~  r_{23}^2\sim \frac{\Gamma_2+\Gamma_3}{\overline{\Lambda}},~~~ r_{34}^2\sim \frac{\Gamma_3+\Gamma_4}{\Lambda},~~~ r_{14}^2\sim \frac{\Gamma_1+\Gamma_4}{\overline{\Lambda}} ; \\
    r_{13}^2\sim (\Gamma_1+\Gamma_2) (\Gamma_1+\Gamma_4)ab\epsilon^{-4},~~~~~~~~~~~~
    r_{24}^2\sim -(\Gamma_1+\Gamma_2) (\Gamma_1+\Gamma_4)ab\epsilon^{-4}.
 \end{array}
\end{equation}\\

{\bfseries{Case B: $\Gamma= 0$ }.}

We claim that
 \begin{center}
  $(\Gamma_1+\Gamma_2) (\Gamma_2+\Gamma_3)  (\Gamma_3+\Gamma_4)(\Gamma_1+\Gamma_4)=0$.
\end{center}

Indeed,  by (\ref{formulartwoverticesd7key0}) and $\Gamma= 0$  it follows that
\begin{center}
$(\Gamma_1+\Gamma_2)(\Gamma_1+\Gamma_4)=0$.
\end{center}
The claim above is obvious now.

Furthermore, we claim that
\begin{equation}\label{DiagramIII1}
    \Gamma_1+\Gamma_2=\Gamma_2+\Gamma_3=\Gamma_3+\Gamma_4=\Gamma_1+\Gamma_4=0.
\end{equation}

Without loss of generality, assume $\Gamma_1+\Gamma_2=0$, by (\ref{formulartwoverticesd7key0}) it follows that $\Gamma_3+\Gamma_4=0$.

If $\Gamma_2+\Gamma_3\neq0$, by (\ref{formulartwoverticesd7key0}) it follows that $\Gamma_1+\Gamma_4\neq0$.
Then
\begin{equation*}
    w_{32}\sim w_{31}\sim w_{42}\sim w_{41}\sim-( \Gamma_1 +\Gamma_4)b \epsilon^{-2}.
\end{equation*}

According to  $\Lambda z_1=\sum_{ j \neq 1} \Gamma_j Z_{j1}$ and $ \Lambda z_2=\sum_{ j \neq 2} \Gamma_j Z_{j2}$, it follows that
%\begin{equation}\label{d71}
%\Lambda z_{1}\sim  \Gamma_2Z_{21}, ~~~~~~~  ~~~~~~~\Lambda z_{2}\sim  \Gamma_1Z_{12}
%\end{equation}
%and
\begin{equation}%\label{d72}
\Lambda z_{12}=(\Gamma_1+\Gamma_2)Z_{12}+ \Gamma_3 (Z_{32}-Z_{31})+ \Gamma_4 (Z_{42}-Z_{41}).\nonumber
\end{equation}
Thus \begin{equation}%\label{d72}
\epsilon^{2} \prec \Lambda z_{12}= \Gamma_3 (\frac{1}{w_{32}}-\frac{1}{w_{31}})+ \Gamma_4 (\frac{1}{w_{42}}-\frac{1}{w_{41}})\prec \epsilon^{2},\nonumber
\end{equation}
this is a contradiction.

It follows  that
\begin{equation}\label{formulartwoverticesd7key1}
    \Gamma_1=\Gamma_3, ~~~~~~~  ~~~~~~~\Gamma_2=\Gamma_4, ~~~~~~~  ~~~~~~~\Gamma=0.\nonumber
\end{equation}
Then   $L  \neq 0$ and $\Lambda=\pm 1$.

%To summarize, the following relations on the four vorticities should be satisfied if the Diagram III
%is approached by a singular sequence:
% \begin{equation}\label{formulartwoverticesd6keyzz}
% \begin{array}{c}
%   \Lambda=\pm 1, \\
%   \Gamma_3=\Gamma_1, \Gamma_2=-\Gamma_1, \Gamma_4=-\Gamma_1,\\
%      \Gamma=0,
%       L\neq0.
% \end{array}
%\end{equation}

Moreover, in this case it is noteworthy that\begin{equation}\label{formulartwoverticesd7key2}
    r_{12}, r_{23}, r_{34},  r_{41}\prec 1.
\end{equation}
The rest $r_{24}$ and $r_{31}$ maybe any case in $\prec 1$, $\approx 1$ or $\succ 1$, but by $S=\sum_{1\leq j<k\leq 4}\Gamma_j\Gamma_k r_{jk}^2=0$ it follows that $r_{24}$ and $r_{31}$ are both $\succ 1$ or not.

\subsection{Diagram $III'$}
\indent\par
By  Proposition \ref{relationsonvorticities1} it follows  that
\begin{equation}\label{formulartwoverticesd8key1}
   \Gamma_1+\Gamma_2=0, ~~~~~~~  ~~~~~~~\Gamma_3+\Gamma_4=0.\nonumber
\end{equation}
Then $\Gamma=0$ $\Longrightarrow$  $L  \neq 0$ and $\Lambda=\pm 1$.

By  Proposition \ref{relationstwovertices} it follows  that
\begin{equation*}
     \begin{array}{c}
                                                                                            z_1\sim z_2\sim z_3\sim z_4\sim   a \epsilon^{-2}, \\
                                                                                          w_1\sim w_2\sim w_3\sim w_4\sim   b \epsilon^{-2}.
                                                                                         \end{array}
\end{equation*}

It is easy to see that\begin{equation*}
     \begin{array}{c}
                                                                                            z_{12}\approx z_{34}\approx w_{12}\approx z_{34} \approx   \epsilon^{2}, \\
                                                                                       \epsilon^{2}  \prec z_{32}\sim z_{31}\sim z_{42}\sim z_{41}\prec \epsilon^{-2},\\
                                                                                       \epsilon^{2}  \prec w_{32}\sim w_{31}\sim w_{42}\sim w_{41}\prec \epsilon^{-2}.
                                                                                         \end{array}
\end{equation*}

By\begin{equation}%\label{d72}
\Lambda z_{12}=(\Gamma_1+\Gamma_2)Z_{12}+ \Gamma_3 (Z_{32}-Z_{31})+ \Gamma_4 (Z_{42}-Z_{41}),\nonumber
\end{equation}
it follows  that
\begin{equation*}
     \frac{\Lambda z_{12}}{w_{21}}= \frac{\Gamma_3}{w_{13}w_{23}}+\frac{\Gamma_4}{w_{14}w_{24}}\prec \frac{1}{w_{13}^2},
\end{equation*}
thus $w_{13}\prec 1$. Similarly, $z_{13}\prec 1$. Hence
\begin{equation*}
  \epsilon^{4}   \prec r_{32}^2\sim r_{31}^2\sim r_{42}^2\sim r_{41}^2\prec 1.
\end{equation*}

To summarize, the following relations on the four vorticities should be satisfied if the Diagram $III'$
is approached by a singular sequence:
 \begin{equation}\label{formulartwoverticesdIII'keyzz}
 \begin{array}{c}
   \Lambda=\pm 1, \\
   \Gamma_1+\Gamma_2=0, \Gamma_3+\Gamma_4=0,\\
      \Gamma=0,
       L\neq0,\\ 
       r_{12}^2\approx r_{34}^2\approx
       \epsilon^{4}   \prec r_{32}^2\sim r_{31}^2\sim r_{42}^2\sim r_{41}^2\prec 1.
 \end{array}
\end{equation}

\subsection{Diagram IV}
\indent\par
By  Proposition \ref{relationsonvorticities1} it follows  that
\begin{equation}\label{formulartwoverticesd8key1}
   \Gamma_1+\Gamma_2=0, ~~~~~~~  ~~~~~~~\Gamma_3+\Gamma_4=0.\nonumber
\end{equation}
Then $\Gamma=0$ $\Longrightarrow$  $L  \neq 0$ and $\Lambda=\pm 1$.

Moreover, for Diagram IV it is noteworthy that\begin{equation}\label{formulartwoverticesd8key2}
    r_{14}, r_{24}, r_{34},  r_{12}, r_{23},r_{31}\preceq 1.
\end{equation}
Indeed, let us just notice that $w_{jk}\approx \epsilon^{2}, z_{jk}\preceq \epsilon^{-2}$ for any $(j,k)$, $1\leq j<k\leq 4$.

To summarize, the following relations on the four vorticities should be satisfied if the Diagram IV
is approached by a singular sequence:
 \begin{equation}\label{formulartwoverticesd8keyzz}
 \begin{array}{c}
   \Lambda=\pm 1, \\
   \Gamma_1+\Gamma_2=0, \Gamma_3+\Gamma_4=0,\\
      \Gamma=0,
       L\neq0.
 \end{array}
\end{equation}

\subsection{Diagram V}
\indent\par
By  Rule IV and Estimate \ref{Estimate1} it follows  that
\begin{equation}\label{formulartwoverticesd9key}
    \Gamma=0.
\end{equation}
Then   $L  \neq 0$ and $\Lambda=\pm 1$.

Moreover, for Diagram V \begin{equation}\label{formulartwoverticesd9key2}
    r_{14}, r_{24}, r_{34},  r_{12}, r_{23},r_{31}\approx \epsilon^2.
\end{equation}
\subsection{Diagram VI}
\indent\par
By  Proposition \ref{relationsonvorticities2} it follows  that
\begin{equation}\label{formulartwoverticesd3key}
    \frac{1}{\Gamma_1}+\frac{1}{\Gamma_2}+\frac{1}{\Gamma_3}=0.
\end{equation}
Then it is easy to see that the total vortex angular momentum $L  \neq 0$, as a result, $\Lambda=\pm 1$. %$\Lambda\in  \mathbb{R}$ or 

We claim that  the total vorticity $\Gamma  \neq 0$. If not, by (\ref{center0}) \begin{equation}
M=\sum_{j=1}^{4}\Gamma_j z_j=0,\nonumber
\end{equation}
it follows that\begin{equation}\label{d31}
\Gamma_1 z_{14}+\Gamma_2 z_{24}+\Gamma_3 z_{34}=\Gamma z_{4}=0.
\end{equation}
For Diagram VI it is easy to see that
\begin{equation}
 z_{14}, z_{24},z_{34}\succ \epsilon^2;        ~~~~~~~  ~~~~~~~  z_{12}, z_{23},z_{31}\approx \epsilon^2. \nonumber
\end{equation}
Thus\begin{equation}
 z_{14}\sim z_{24}\sim z_{34}\succ \epsilon^2. \nonumber
\end{equation}
Then by (\ref{d31}) it follows that
\begin{center}
$\Gamma_1 +\Gamma_2 +\Gamma_3=0$,
\end{center}
this contradicts (\ref{formulartwoverticesd3key}).

We also claim that \begin{equation}
 r_{14}\sim r_{24}\sim r_{34}\preceq 1. \nonumber
\end{equation}
Indeed, by $z_{14}\sim z_{24}\sim z_{34}$ and $w_{14}\sim w_{24}\sim w_{34}$, $r_{14}\sim r_{24}\sim r_{34}$ holds. If $r_{34}\succ 1$, then by $S=\sum_{1\leq j<k\leq 4}\Gamma_j\Gamma_k r_{jk}^2$  it follows that
 \begin{equation}
(\Gamma_1 +\Gamma_2 +\Gamma_3)\Gamma_4=0.\nonumber
\end{equation}
this contradicts (\ref{formulartwoverticesd3key}).

 To summarize, the following relations on the four vorticities should be satisfied if the Diagram VI
is approached by a singular sequence:
 \begin{equation}\label{formulartwoverticesd3keyzz}
 \begin{array}{c}
   \Lambda=\pm 1, \\
   \frac{1}{\Gamma_1}+\frac{1}{\Gamma_2}+\frac{1}{\Gamma_3}=0, \\
      L\neq0,
       \Gamma\neq0.
 \end{array}
\end{equation}

Moreover, for Diagram VI \begin{equation}\label{formulartwoverticesd3key2}
     r_{12}, r_{23},r_{31}\approx \epsilon^2,  ~~~~~~~    \epsilon^2 \prec r_{14}, r_{24}, r_{34}\preceq 1 ~~~\text{and }~~~ r_{12}r_{34},r_{13}r_{24},r_{14}r_{23}\prec 1.
\end{equation}

\subsection{Diagram VII}
\indent\par
By  Rule IV and Estimate \ref{Estimate1} it follows  that
\begin{equation}\label{formulartwoverticesd4key}
    {\Gamma_1}+{\Gamma_2}+{\Gamma_3}=0.
\end{equation}
Then it is easy to see that the total vortex angular momentum $L  \neq 0$ and the total vorticity $\Gamma  \neq 0$. As a result, $\Lambda\in  \mathbb{R}$ or $\Lambda=\pm 1$.

Moreover, for Diagram VII \begin{equation}\label{formulartwoverticesd4key2}
     r_{12}, r_{23},r_{31}\approx \epsilon^2,  ~~~~~~~      r_{14}, r_{24}, r_{34}\approx \epsilon^{-2} ~~~\text{and }~~~ r_{12}r_{34},r_{13}r_{24},r_{14}r_{23}\approx 1.
\end{equation}
\subsection{Diagram VIII}
\indent\par
By  Proposition \ref{relationsonvorticities2} it follows  that
\begin{equation}\label{formulartwoverticesd5key}
    \frac{1}{\Gamma_1}+\frac{1}{\Gamma_2}+\frac{1}{\Gamma_3}=\frac{1}{\Gamma_1}+\frac{1}{\Gamma_2}+\frac{1}{\Gamma_4}=0.\nonumber
\end{equation}
Then it is easy to see that the total vortex angular momentum $L  \neq 0$. As a result, $\Lambda\in  \mathbb{R}$ or $\Lambda=\pm 1$.

We claim that  the total vorticity $\Gamma  \neq 0$. If not, we have \begin{equation}
\begin{array}{c}
 \Gamma_1+ \Gamma_2=-2\Gamma_3, \\
  \Gamma_1 \Gamma_2=-\Gamma_3(\Gamma_1+ \Gamma_2 )=2\Gamma_3^2.
\end{array}
\nonumber
\end{equation}
But this contradicts $(\Gamma_1+ \Gamma_2)^2\geq 4\Gamma_1 \Gamma_2$.

Moreover, for Diagram VIII it is noteworthy that\begin{equation}\label{formulartwoverticesd5key1}
    r_{34}\approx 1
\end{equation}
and all the other $r_{jk}\prec 1$.
Indeed, it is easy to see that all   $r_{jk}\prec 1$ hold except $r_{34}$. And $r_{34}\approx 1$ holds because
$S=\Gamma I=\frac{\Gamma L}{\Lambda}\neq 0$. Thus $ r_{12}r_{34},r_{13}r_{24},r_{14}r_{23}\prec 1$

 To summarize, the following relations on the four vorticities should be satisfied if the Diagram VIII
is approached by a singular sequence:
 \begin{equation}\label{formulartwoverticesd5keyzz}
 \begin{array}{c}
   \Lambda=\pm 1, \\
   \frac{1}{\Gamma_1}+\frac{1}{\Gamma_2}+\frac{1}{\Gamma_3}=0, \\
   \Gamma_3=\Gamma_4,\\
      L\neq0,
       \Gamma\neq0.
 \end{array}
\end{equation}

\subsection{Diagram IX}
\indent\par
By  Proposition \ref{relationsonvorticities2} it follows  that
\begin{equation}\label{formulartwoverticesd6key1}
    \frac{1}{\Gamma_1}+\frac{1}{\Gamma_2}+\frac{1}{\Gamma_3}=0.\nonumber
\end{equation}
Then   $L  \neq 0$ and $\Lambda=\pm 1$.

We claim that  the total vorticity $\Gamma  \neq 0$. If not, we have \begin{equation}\label{d61}
\Gamma_1 z_{14}+\Gamma_2 z_{24}+\Gamma_3 z_{34}=\Gamma z_{4}=0.
\end{equation}
For Diagram IX it is easy to see that
\begin{equation}
 z_{14}, z_{24},z_{34}\succ \epsilon^2;        ~~~~~~~  ~~~~~~~  z_{12}, z_{23},z_{31}\approx \epsilon^2. \nonumber
\end{equation}
Thus\begin{equation}\label{d62}
 z_{14}\sim z_{24}\sim z_{34}\succ \epsilon^2.
\end{equation}
Then by (\ref{d61}) it follows that
\begin{center}
$\Gamma_1 +\Gamma_2 +\Gamma_3=0$,
\end{center}
this contradicts (\ref{formulartwoverticesd6key1}).

Moreover, by (\ref{d61}) and (\ref{d62}) it follows that
\begin{equation}
 z_{14}\sim z_{24}\sim z_{34}\approx \epsilon^{-2}. \nonumber
\end{equation}
Note that $w_{14}, w_{24}, w_{34}\approx \epsilon^{2}$, thus for Diagram IX we have \begin{equation}\label{formulartwoverticesd6key2}
  r_{14}, r_{24}, r_{34}\approx 1,  ~~~~~~~   r_{12}, r_{23},r_{31}\approx \epsilon^2 ~~~\text{and }~~~ r_{12}r_{34},r_{13}r_{24},r_{14}r_{23}\prec 1.
\end{equation}

To summarize, the following relations on the four vorticities should be satisfied if the Diagram IX
is approached by a singular sequence:
 \begin{equation}\label{formulartwoverticesd6keyzz}
 \begin{array}{c}
   \Lambda=\pm 1, \\
   \frac{1}{\Gamma_1}+\frac{1}{\Gamma_2}+\frac{1}{\Gamma_3}=0, \\
      L\neq0,
       \Gamma\neq0.
 \end{array}
\end{equation}

\section{Finiteness results}\label{Finitenessresult}
\indent\par

The main Theorem \ref{Main} in this paper is an obvious inference of the following Theorem \ref{MainLis0}, \ref{MainLis0forcollapse}, \ref{MainGammais0}, \ref{MainGammaLnot0} and \ref{MainGammaLnot01}.   And we remark that the following results of  finiteness are all on  normalized central configurations in the complex domain, more than  real   configurations.

\subsection{Central configurations with $L=0$}
\indent\par

First we consider the finiteness of central configurations with vanishing total vortex angular momentum, i.e., $L=0$. In this case only Diagram I, Diagram II  and Diagram III in Case $A$  are possible.

\begin{theorem}\label{MainLis0}
If the vorticities $\Gamma_n$ $(n\in\{1,2,3,4\})$ are nonzero such that $L=0$, then the four-vortex problem has finitely many
relative equilibria; and except perhaps  for two pairs of equal vorticities with ratios $(\sqrt{3}-2)^{\pm 1}$  (for example, $\Gamma_3 =\Gamma_4=(\sqrt{3}-2)^{\pm 1}\Gamma_1=(\sqrt{3}-2)^{\pm 1}\Gamma_2$) with $\Lambda=\pm \textbf{\emph{i}}$,  the four-vortex problem has finitely many collapse configurations.
\end{theorem}
{\bf Proof.}

Elementary geometry shows that giving five of the  $r_{jk}^2$'s $(1\leq j<k\leq 4)$, the six squares of the mutual distances, determines finitely many geometrical configurations up to rotation. If there are infinitely many solutions of (\ref{stationaryconfigurationmain}), at least two of the $r_{jk}^2$'s should take
infinitely many values. We suppose $r_{12}^2$ does, and we take it as the polynomial
function in Lemma \ref{Eliminationtheory}, thus $r_{12}^2$ is dominating. There is a sequence of normalized central configurations
such that $r^{(n)}_{12}\longrightarrow 0$, i.e., $z^{(n)}_{12}w^{(n)}_{12}\longrightarrow 0$. Whatever the renormalization is, $\mathcal{Z}^{(n)}$
or $\mathcal{W}^{(n)}$ is unbounded on this sequence. We extract a singular sequence. It corresponds to one of the diagrams in Figure \ref{fig:Problematicdiagrams}, in fact, only Diagram II is possible. Therefore, all of the  $r_{jk}^2$'s   should  go to zero. Thus all of them  take infinitely many values. As a result, the  $r_{jk}^2$'s are all dominating. Similarly,  $r_{12}^2r_{34}^2$, $r_{13}^2r_{24}^2$ and $r_{14}^2r_{23}^2$ are all dominating.

Diagram II and Diagram III may  correspond to  relative equilibria or collapse configurations.

{\bfseries{Case 1: When corresponding to  relative equilibria. }}

Note that in this case  only  Diagram II and Diagram III in Figure \ref{fig:Problematicdiagrams} are possible for $L=0$.

By considering $r_{12}^2\longrightarrow \infty$, we are in Diagram III, then
\begin{equation}\label{IIIA1}
    \Gamma_1 \Gamma_2=\Gamma_3 \Gamma_4.
\end{equation}
Similarly, push $r_{13}^2$ and $r_{14}^2$ to infinity, we have
\begin{equation}\label{IIIA2}
   \begin{array}{c}
     \Gamma_1 \Gamma_3=\Gamma_2 \Gamma_4, \\
      \Gamma_1 \Gamma_4=\Gamma_2 \Gamma_3.
   \end{array}
\end{equation}
Note that $\Gamma \neq0$, by (\ref{IIIA1}) and (\ref{IIIA2}) it follows that
\begin{equation*}
    \Gamma_1= \Gamma_2=\Gamma_3 =\Gamma_4.
\end{equation*}
However, this contradicts $L=0$.\\

{\bfseries{Case 2: When corresponding to  collapse configurations. }}

In this case Diagram I,   Diagram II and Diagram III in Figure \ref{fig:Problematicdiagrams} are all possible for $L=0$.

If Diagram I does not occur,  a similar discussion as above shows the Theorem. Thus, without lose of generality, we assume that there always exists a  singular sequence of collapse configurations approaching  Diagram I  below. Then $\Lambda =\pm \mathbf{i}$.

Consider $r_{12}^2r_{34}^2$. Push it to infinity. We are in Diagram I or Diagram III. It is easy to see that the constraint on the vorticities in Diagram I is
$\Gamma_1 \Gamma_3=\Gamma_2 \Gamma_4$ or $\Gamma_1 \Gamma_4=\Gamma_2 \Gamma_3$; the constraint on the vorticities in Diagram III is
$\Gamma_1 \Gamma_2=\Gamma_3 \Gamma_4$. So we have
\begin{equation}\label{1234}
   (\Gamma_1 \Gamma_2-\Gamma_3 \Gamma_4) (\Gamma_1 \Gamma_3-\Gamma_2 \Gamma_4)(\Gamma_1 \Gamma_4-\Gamma_2 \Gamma_3)=0.\nonumber
\end{equation}

We claim that \emph{at least two of $\Gamma_1 \Gamma_2-\Gamma_3 \Gamma_4$, $\Gamma_1 \Gamma_3-\Gamma_2 \Gamma_4$ and $\Gamma_1 \Gamma_4-\Gamma_2 \Gamma_3$ are zero}.

If the claim is  true, without lose of generality,  assume
\begin{equation}\label{13,14}
   \Gamma_1 \Gamma_3-\Gamma_2 \Gamma_4=0,~~~~~\Gamma_1 \Gamma_4-\Gamma_2 \Gamma_3=0.
\end{equation}

In addition to these equations, we also have \begin{equation}\label{Lis0a}
    L=\Gamma_1 \Gamma_2+\Gamma_2 \Gamma_3+\Gamma_3\Gamma_1+(\Gamma_1 +\Gamma_2+\Gamma_3) \Gamma_4=0.
\end{equation}

A  straightforward computation shows that the solutions of the equations (\ref{13,14}) and (\ref{Lis0a}) are
\begin{equation*}
    \Gamma_1 =\Gamma_2,~~~~~~~~~~\Gamma_3 =\Gamma_4=(-\sqrt{3}-2)\Gamma_1 ~\text{or} ~\Gamma_3 =\Gamma_4=(\sqrt{3}-2)\Gamma_1.
\end{equation*}

\begin{figure}%[!h]
	\centering
	
		\begin{subfigure}[b]{0.2\textwidth}
\centering
		\resizebox{\linewidth}{!}{
		\begin{tikzpicture}%[scale=0.7]
			\hspace{-3cm}
	\draw  (-3/2,  0) node {\large\textbf{1}};%z-circle
\draw (3/2,  0) node{\large\textbf{2}};%z-circle

\draw  (-3/2,  -2)  node {\large\textbf{4}};%w-circle
\draw  (3/2,  -2) node {\large\textbf{3}}; %w-circle

	\draw [red,very thick] (-3/2,  0) circle (0.35);
	\draw [red,very thick] (3/2,  0) circle (0.35);

	\draw [blue,dashed, thick] (-3/2,  -2) circle (0.35);
	\draw [blue,dashed, thick]  (3/2,  -2) circle (0.35);

		\draw [red,very thick] (-1.2,0)--(1.2,0); % z-edge
	\draw [blue,dashed, thick] (-1.2,-2)--(1.2,-2); % z-edge
	
			\hspace{6cm}
	\draw  (-3/2,  0) node {\large\textbf{4}};%z-circle
\draw (3/2,  0) node{\large\textbf{3}};%z-circle

\draw  (-3/2,  -2)  node {\large\textbf{1}};%w-circle
\draw  (3/2,  -2) node {\large\textbf{2}}; %w-circle

	\draw [red,very thick] (-3/2,  0) circle (0.35);
	\draw [red,very thick] (3/2,  0) circle (0.35);

	\draw [blue,dashed, thick] (-3/2,  -2) circle (0.35);
	\draw [blue,dashed, thick]  (3/2,  -2) circle (0.35);

		\draw [red,very thick] (-1.2,0)--(1.2,0); % z-edge
	\draw [blue,dashed, thick] (-1.2,-2)--(1.2,-2); % z-edge
	\end{tikzpicture}
}
	\caption{Diagram I}
	\end{subfigure}

		\begin{subfigure}[b]{0.2\textwidth}
\centering
		\resizebox{\linewidth}{!}{
		\begin{tikzpicture}%[scale=0.7]
		%\hspace{-1cm}
		%	\draw	[blue,dashed,thick]  (-3/2,0)  circle (0.35);
		\draw	  (-3/2,0)    node {\large\textbf{1}};
		
		%	\draw	[blue,dashed,thick]  (3/2,0)  circle (0.35);
		\draw		(3/2,0) node {\large\textbf{2}};
		
		%	\draw	[blue,dashed,thick]  (0,-3/2*1.732)   circle (0.35);
		\draw		(0,-3/2*1.732) node {\large\textbf{3}};
		
		%	\draw	[blue,dashed,thick]  (0,-1)  circle (0.35);
		\draw		(0,-1) node {\large\textbf{4}};

		\draw [red,very thick] (-3/2+.35,0.15)--(3/2-.35,0.15);
		\draw [blue, dashed,thick] (-3/2+0.35,0)--(3/2-.35,0);
		
		\draw [blue, dashed,thick] (-3/2+.2,-0.2*1.732)--(-.2,-3/2*1.732+.2*1.732);
		\draw [red,very thick] (-3/2-0.15+.25, -0.25*1.732)--(-0.15-.25,-3/2*1.732+0.25*1.732);
		
		\draw [blue, dashed,thick] (3/2-0.2,-0.2*1.732)--(0.2,-3/2*1.732+0.2*1.732);
		\draw [red,very thick] (3/2+0.15-0.25,-0.25*1.732)--(0.15+0.25,-3/2*1.732+.25*1.732);

		\draw [blue,dashed,thick]  (-3/2+0.3+0.15,-0.3*2/3)--(0-0.35+0.15,-1+0.35*2/3);  %k=(1, -2/3)
		\draw [red,very thick]  (-3/2+0.4-0.1,-0.4*2/3)--(0-0.3-0.1,-1+0.3*2/3);  %k=(1, -2/3)
		
		\draw [red,very thick]  (3/2-0.25-.15,-.25*2/3)--(0.4 -.15,-1+.4*2/3);  %k=(1, 2/3)
		\draw [blue,dashed,thick]   (3/2-0.4+.1,-.4*2/3)--(0.3 +.1,-1+.3*2/3);  %k=(1, 2/3)

		\draw [red,very thick] (-.07,-3/2*1.732+.4)--(-.07,-1-.4);
		\draw [blue,dashed,thick]  (.07,-3/2*1.732+0.4)--(.07,-1-0.4);
			\end{tikzpicture}
}
		\caption{Diagram  II}
	\end{subfigure}

\begin{subfigure}[b]{0.2\textwidth}
		\centering
		\resizebox{\linewidth}{!}{
	\begin{tikzpicture}
	\hspace{-3cm}
		\draw  (-3/2,  0) node {\large\textbf{1}};%z-circle
	\draw (3/2,  0) node{\large\textbf{4}};%z-circle
	
	\draw  (-3/2,  -2)  node {\large\textbf{3}};%w-circle
	\draw  (3/2,  -2) node {\large\textbf{2}}; %w-circle

		\draw [blue,dashed, thick] (-3/2,  0) circle (0.25);
	\draw [red,very thick] (-3/2,  0) circle (0.35);
	
		\draw [blue,dashed, thick] (3/2,  0) circle (0.25);
\draw [red,very thick] (3/2,  0) circle (0.35);

\draw [blue,dashed, thick] (-3/2,  -2) circle (0.25);
\draw [red,very thick](-3/2,  -2) circle (0.35);

\draw [blue,dashed, thick] (3/2,  -2) circle (0.25);
\draw[red,very thick]  (3/2,  -2) circle (0.35);

\draw [red,very thick] (-1.2,0)--(1.2,0); % z-edge
\draw [red,very thick] (-1.2,-2)--(1.2,-2);

\draw [blue,dashed, thick] (-3/2,-.3)--(-3/2,-1.7);
\draw [blue,dashed, thick] (3/2,-.3)--(3/2,-1.7);

	\hspace{6cm}
	
		\draw  (-3/2,  0) node {\large\textbf{1}};%z-circle
	\draw (3/2,  0) node{\large\textbf{3}};%z-circle
	
	\draw  (-3/2,  -2)  node {\large\textbf{4}};%w-circle
	\draw  (3/2,  -2) node {\large\textbf{2}}; %w-circle

		\draw [blue,dashed, thick] (-3/2,  0) circle (0.25);
	\draw [red,very thick] (-3/2,  0) circle (0.35);
	
		\draw [blue,dashed, thick] (3/2,  0) circle (0.25);
\draw [red,very thick] (3/2,  0) circle (0.35);

\draw [blue,dashed, thick] (-3/2,  -2) circle (0.25);
\draw [red,very thick](-3/2,  -2) circle (0.35);

\draw [blue,dashed, thick] (3/2,  -2) circle (0.25);
\draw[red,very thick]  (3/2,  -2) circle (0.35);

\draw [red,very thick] (-1.2,0)--(1.2,0); % z-edge
\draw [red,very thick] (-1.2,-2)--(1.2,-2);

\draw [blue,dashed, thick] (-3/2,-.3)--(-3/2,-1.7);
\draw [blue,dashed, thick] (3/2,-.3)--(3/2,-1.7);

	\end{tikzpicture}
}\caption{Diagram III}
\end{subfigure}

\caption{Complete problematic diagrams for vorticities satisfied (\ref{1234})}
\label{fig:Problematicdiagramscollapseconfigurations}	
\end{figure}
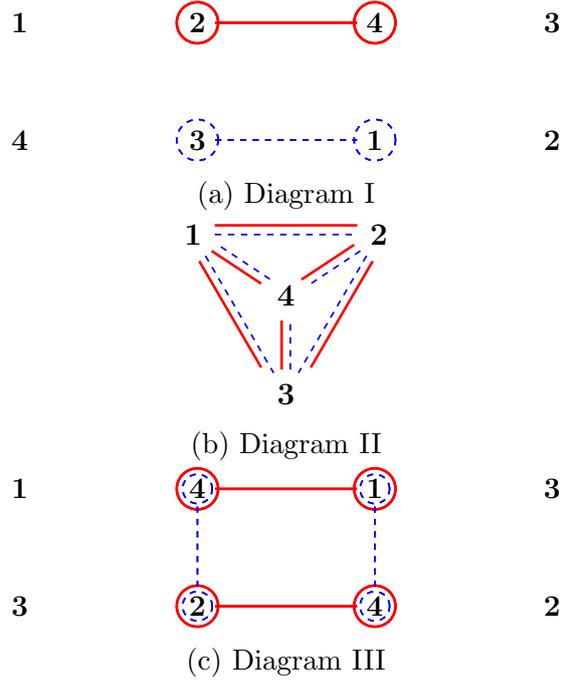

{\bfseries{By reduction to absurdity to prove the claim: }}
If the claim is not true, without lose of generality,  assume
\begin{equation}\label{1234}
   \Gamma_1 \Gamma_2-\Gamma_3 \Gamma_4=0,~~~~~(\Gamma_1 \Gamma_3-\Gamma_2 \Gamma_4)(\Gamma_1 \Gamma_4-\Gamma_2 \Gamma_3)\neq0.
\end{equation}
Then the complete
diagrams may be approached by  some singular sequence are the  Figure \ref{fig:Problematicdiagramscollapseconfigurations}.

We take $r_{12}^2+r_{34}^2$ as the polynomial
function in Lemma \ref{Eliminationtheory}, and claim that it is not dominating. Otherwise, by considering $r_{12}^2+r_{34}^2\longrightarrow \infty$, it is only possible that we are in Diagram III of Figure \ref{fig:Problematicdiagramscollapseconfigurations}. However, by 
\begin{equation*}
    S-\Gamma_1\Gamma_3r_{13}^2-\Gamma_1\Gamma_4r_{14}^2-\Gamma_2\Gamma_3r_{23}^2-\Gamma_2\Gamma_4r_{24}^2=\Gamma_1\Gamma_2(r_{12}^2+r_{34}^2)
\end{equation*}
and the results in (\ref{DiagramIII2}) we know that this is a contradiction.

Thus we consider the system  (\ref{stationaryconfigurationmain}) by restricting to some  level set $r_{12}^2+r_{34}^2\equiv const$ in the following. It is easy to see that it suffices to consider the  level set $r_{12}^2+r_{34}^2\equiv 0$.
That is, if system  (\ref{stationaryconfigurationmain}) possesses  infinitely many solutions, then some level set $r_{12}^2+r_{34}^2\equiv  0$ also includes infinitely many solutions of system  (\ref{stationaryconfigurationmain}).

By considering $r_{12}^2\longrightarrow \infty$, we are in Diagram III of Figure \ref{fig:Problematicdiagramscollapseconfigurations}. Recall that
\begin{equation*}
\begin{array}{c}
  S=\sum_{1\leq j<k\leq 4}\Gamma_j\Gamma_k r_{jk}^2=0, \\
  r_{14}^2\sim \frac{\Gamma_1+\Gamma_4}{\pm \mathbf{i}} ,~~~  r_{42}^2\sim \frac{\Gamma_4+\Gamma_2}{\mp\mathbf{i}},~~~ r_{23}^2\sim \frac{\Gamma_2+\Gamma_3}{\pm\mathbf{i}},~~~ r_{31}^2\sim \frac{\Gamma_1+\Gamma_3}{\mp\mathbf{i}}.
\end{array}
\end{equation*}
It follows that
\begin{equation*}
    \Gamma_1\Gamma_3(\Gamma_1+\Gamma_3)- \Gamma_2\Gamma_3(\Gamma_2+\Gamma_3)-\Gamma_1\Gamma_4(\Gamma_1+\Gamma_4)+ \Gamma_2\Gamma_4(\Gamma_2+\Gamma_4)=0
\end{equation*}
or
\begin{equation}\label{1j23j4}
    (\Gamma_1-\Gamma_2) (\Gamma_3-\Gamma_4) \Gamma=0.
\end{equation}

However, there is no solution  for the equations (\ref{1j23j4}), (\ref{1234}) and (\ref{Lis0a}). This prove the claim.

$~~~~~~~~~~~~~~~~~~~~~~~~~~~~~~~~~~~~~~~~~~~~~~~~~~~~~~~~~~~~~~~~~~~~~~~~~~~~~~~~~~~~~~~~~~~~~~~~~~~~~~~~~~~~~~~~~~~~~~~~~~~~~~~~~~~~~~~~~~~~~~~~~~~~\Box$\\

\begin{theorem}\label{MainLis0forcollapse}
Suppose $\Gamma_3 =\Gamma_4=(\sqrt{3}-2)^{\pm 1}\Gamma_1=(\sqrt{3}-2)^{\pm 1}\Gamma_2$, then the four-vortex problem with $\Lambda=\pm \textbf{\emph{i}}$ dose not have any collapse configurations.

\end{theorem}
{\bf Proof.}
We consider the system (\ref{stationaryconfigurationmainequ3})  with  $\Lambda=\pm \textbf{\emph{i}}$.

 Without loss of generality, suppose  $\Gamma_1=\Gamma_2=1, \Gamma_3 =\Gamma_4=\kappa$, where $\kappa=(\sqrt{3}-2)^{\pm 1}$ are the roots of the polynomial $f=\kappa^2+4\kappa+1$. Then a simple computation by Mathematica shows that  the system (\ref{stationaryconfigurationmainequ3}) combined the equation $f=0$ does not have any solution.

$~~~~~~~~~~~~~~~~~~~~~~~~~~~~~~~~~~~~~~~~~~~~~~~~~~~~~~~~~~~~~~~~~~~~~~~~~~~~~~~~~~~~~~~~~~~~~~~~~~~~~~~~~~~~~~~~~~~~~~~~~~~~~~~~~~~~~~~~~~~~~~~~~~~~\Box$\\

\subsection{Central configurations with $\Gamma=0$}
\indent\par

Next we consider the finiteness of central configurations with vanishing total vorticity, i.e., $\Gamma=0$. In this case only Diagram III in Case B, Diagram $III'$, Diagram IV and Diagram V are possible.
These Diagrams can only correspond to  relative equilibria, so the following result of finiteness has been proved essentially by O'Neil  \cite{o2007relative}  and Hampton and Moeckel  \cite{hampton2009finiteness}  independently.

\begin{theorem}\label{MainGammais0}
If the vorticities $\Gamma_n$ $(n\in\{1,2,3,4\})$ are nonzero such that $\Gamma=0$, then the four-vortex problem has finitely many  central configurations.
\end{theorem}
{\bf Proof.}

A similar argument as in the proof of Theorem \ref{MainLis0} shows that
 at least two of the $r_{jk}^2$'s should take
infinitely many values. We suppose $r_{13}^2$ does, then $r_{13}^2$ is dominating. There is a sequence of normalized central configurations
such that $r^{(n)}_{13}\longrightarrow \infty$, i.e., $z^{(n)}_{13}w^{(n)}_{13}\longrightarrow \infty$. Then we can extract a singular sequence. It  corresponds to one of the diagrams in Figure \ref{fig:Problematicdiagrams}, in fact, only Diagram III is possible. In this case, it follows that  $r_{13}^2\longrightarrow \infty$, $r_{24}^2\longrightarrow \infty$ and all of the rest of  $r_{jk}^2$'s   should  go to zero. Thus all of them  take infinitely many values. As a result, the  $r_{jk}^2$'s are all dominating. Similarly,  $r_{12}^2r_{34}^2$, $r_{13}^2r_{24}^2$ and $r_{14}^2r_{23}^2$ are all dominating.

Consider $r_{13}^2r_{24}^2$. Push it to infinity. We are in Diagram III. The constraints on the vorticities in Diagram III are
\begin{equation}\label{13and24}
    \Gamma_1= \Gamma_3,~~~~~~~~\Gamma_2 =\Gamma_4,~~~~~~~~\Gamma =0.\nonumber
\end{equation}

Similarly, push $r_{12}^2r_{34}^2$ to infinity, we have \begin{equation}\label{12and34}
  \Gamma_1= \Gamma_2,~~~~~~~~\Gamma_3 =\Gamma_4,~~~~~~~~\Gamma=0;\nonumber
\end{equation} push $r_{14}^2r_{23}^2$ to infinity, we have \begin{equation}\label{14and24}
   \Gamma_1 =\Gamma_4,~~~~~~~~\Gamma_2= \Gamma_3,~~~~~~~~ \Gamma=0.\nonumber
\end{equation}

However it is easy to see that  the equations above have only zero solution: $\Gamma_1= \Gamma_2=\Gamma_3 =\Gamma_4=0$.  This is a contradiction.

$~~~~~~~~~~~~~~~~~~~~~~~~~~~~~~~~~~~~~~~~~~~~~~~~~~~~~~~~~~~~~~~~~~~~~~~~~~~~~~~~~~~~~~~~~~~~~~~~~~~~~~~~~~~~~~~~~~~~~~~~~~~~~~~~~~~~~~~~~~~~~~~~~~~~\Box$\\

\subsection{Central configurations with $\Gamma\neq0$ and $L\neq0$. }

\indent\par

At last we consider the finiteness of central configurations with nonvanishing total vorticity and total vortex angular momentum, i.e., $\Gamma\neq0$ and $L\neq0$. In this case only  Diagram III in Case A, Diagram VI, Diagram VII, Diagram VIII and Diagram IX are possible.

First, we establish the following result:\begin{lemma}\label{nonadjacentdistancescj}
If some product $r_{jk}^2r_{lm}^2$ of two nonadjacent
distances'squares is not dominating on the closed algebraic subset $\mathcal A$. Then system (\ref{stationaryconfigurationmain}) possesses  finitely many solutions.
\end{lemma}
\noindent\emph{Proof of Lemma \ref{nonadjacentdistancescj}}:

If system  (\ref{stationaryconfigurationmain}) possesses  infinitely many solutions, without lose of generality, assume that $r_{12}^2r_{34}^2$ is not dominating on the closed algebraic subset $\mathcal A$. Then some level set $r_{12}^2r_{34}^2\equiv const\neq 0$ also includes infinitely many solutions of system  (\ref{stationaryconfigurationmain}).

\begin{figure}%[!h]
	\centering

\begin{subfigure}[b]{0.2\textwidth}
		\centering
		\resizebox{\linewidth}{!}{
	\begin{tikzpicture}
	
		\draw  (-3/2,  0) node {\large\textbf{1}};%z-circle
	\draw (3/2,  0) node{\large\textbf{4}};%z-circle
	
	\draw  (-3/2,  -2)  node {\large\textbf{2}};%w-circle
	\draw  (3/2,  -2) node {\large\textbf{3}}; %w-circle

		\draw [blue,dashed, thick] (-3/2,  0) circle (0.25);
	\draw [red,very thick] (-3/2,  0) circle (0.35);
	
		\draw [blue,dashed, thick] (3/2,  0) circle (0.25);
\draw [red,very thick] (3/2,  0) circle (0.35);

\draw [blue,dashed, thick] (-3/2,  -2) circle (0.25);
\draw [red,very thick](-3/2,  -2) circle (0.35);

\draw [blue,dashed, thick] (3/2,  -2) circle (0.25);
\draw[red,very thick]  (3/2,  -2) circle (0.35);

\draw [red,very thick] (-1.2,0)--(1.2,0); % z-edge
\draw [red,very thick] (-1.2,-2)--(1.2,-2);

\draw [blue,dashed, thick] (-3/2,-.3)--(-3/2,-1.7);
\draw [blue,dashed, thick] (3/2,-.3)--(3/2,-1.7);

	\end{tikzpicture}
}
\caption{Diagram III1}	
\end{subfigure}
	\begin{subfigure}[b]{0.2\textwidth}
		\centering
		\resizebox{\linewidth}{!}{
	\begin{tikzpicture}
	
		\draw  (-3/2,  0) node {\large\textbf{1}};%z-circle
	\draw (3/2,  0) node{\large\textbf{2}};%z-circle
	
	\draw  (-3/2,  -2)  node {\large\textbf{4}};%w-circle
	\draw  (3/2,  -2) node {\large\textbf{3}}; %w-circle

		\draw [blue,dashed, thick] (-3/2,  0) circle (0.25);
	\draw [red,very thick] (-3/2,  0) circle (0.35);
	
		\draw [blue,dashed, thick] (3/2,  0) circle (0.25);
\draw [red,very thick] (3/2,  0) circle (0.35);

\draw [blue,dashed, thick] (-3/2,  -2) circle (0.25);
\draw [red,very thick](-3/2,  -2) circle (0.35);

\draw [blue,dashed, thick] (3/2,  -2) circle (0.25);
\draw[red,very thick]  (3/2,  -2) circle (0.35);

\draw [red,very thick] (-1.2,0)--(1.2,0); % z-edge
\draw [red,very thick] (-1.2,-2)--(1.2,-2);

\draw [blue,dashed, thick] (-3/2,-.3)--(-3/2,-1.7);
\draw [blue,dashed, thick] (3/2,-.3)--(3/2,-1.7);

	\end{tikzpicture}
}\caption{Diagram III2}
\end{subfigure}

 %important note, here, if empty line,  means break to TIKZ
	\begin{subfigure}[b]{0.2\textwidth}
		\centering
		\resizebox{\linewidth}{!}{
	\begin{tikzpicture}
	\draw	[blue,dashed, thick]  (-3/2,0)  circle (0.25);
	\draw	[red,very thick]  (-3/2,0)  circle (0.35);
	\draw  (-3/2,0)  node {\large\textbf{4}};
	
	\draw	[blue,dashed, thick]  (3/2,0)  circle (0.25);
	\draw	[red,very thick]  (3/2,0)  circle (0.35);
	\draw		(3/2,0) node {\large\textbf{2}};
	
	\draw	[blue,dashed, thick]  (0,-3/2*1.732) circle (0.25);
	\draw	[red,very thick] (0,-3/2*1.732) circle (0.35);
	\draw		(0,-3/2*1.732) node {\large\textbf{3}};
	
	\draw		(0,-1) node {\large\textbf{1}};

	\draw [red,very thick] (-3/2+.35,0)--(3/2-.35,0);
	\draw [blue, dashed,thick] (-3/2+0.35,-.15)--(3/2-.35,-0.15);
	
	\draw [red,very thick] (-3/2+.2,-0.2*1.732)--(-.2,-3/2*1.732+.2*1.732);
	\draw [blue, dashed,thick] (-3/2+0.15+.2, -0.2*1.732)--(0.15-.2,-3/2*1.732+0.2*1.732);
	
	\draw [red,very thick] (3/2-0.2,-0.2*1.732)--(0.2,-3/2*1.732+0.2*1.732);
	\draw [blue, dashed,thick] (3/2-0.15-0.2,-.2*1.732)--(-0.15+0.2,-3/2*1.732+.2*1.732);
	\end{tikzpicture}
}
	\caption{Diagram VII1}
\end{subfigure}\begin{subfigure}[b]{0.2\textwidth}
		\centering
		\resizebox{\linewidth}{!}{
	\begin{tikzpicture}
	\draw	[blue,dashed, thick]  (-3/2,0)  circle (0.25);
	\draw	[red,very thick]  (-3/2,0)  circle (0.35);
	\draw  (-3/2,0)  node {\large\textbf{1}};
	
	\draw	[blue,dashed, thick]  (3/2,0)  circle (0.25);
	\draw	[red,very thick]  (3/2,0)  circle (0.35);
	\draw		(3/2,0) node {\large\textbf{4}};
	
	\draw	[blue,dashed, thick]  (0,-3/2*1.732) circle (0.25);
	\draw	[red,very thick] (0,-3/2*1.732) circle (0.35);
	\draw		(0,-3/2*1.732) node {\large\textbf{3}};
	
	\draw		(0,-1) node {\large\textbf{2}};

	\draw [red,very thick] (-3/2+.35,0)--(3/2-.35,0);
	\draw [blue, dashed,thick] (-3/2+0.35,-.15)--(3/2-.35,-0.15);
	
	\draw [red,very thick] (-3/2+.2,-0.2*1.732)--(-.2,-3/2*1.732+.2*1.732);
	\draw [blue, dashed,thick] (-3/2+0.15+.2, -0.2*1.732)--(0.15-.2,-3/2*1.732+0.2*1.732);
	
	\draw [red,very thick] (3/2-0.2,-0.2*1.732)--(0.2,-3/2*1.732+0.2*1.732);
	\draw [blue, dashed,thick] (3/2-0.15-0.2,-.2*1.732)--(-0.15+0.2,-3/2*1.732+.2*1.732);
	\end{tikzpicture}
}
	\caption{Diagram VII2}
\end{subfigure}\begin{subfigure}[b]{0.2\textwidth}
		\centering
		\resizebox{\linewidth}{!}{
	\begin{tikzpicture}
	\draw	[blue,dashed, thick]  (-3/2,0)  circle (0.25);
	\draw	[red,very thick]  (-3/2,0)  circle (0.35);
	\draw  (-3/2,0)  node {\large\textbf{1}};
	
	\draw	[blue,dashed, thick]  (3/2,0)  circle (0.25);
	\draw	[red,very thick]  (3/2,0)  circle (0.35);
	\draw		(3/2,0) node {\large\textbf{2}};
	
	\draw	[blue,dashed, thick]  (0,-3/2*1.732) circle (0.25);
	\draw	[red,very thick] (0,-3/2*1.732) circle (0.35);
	\draw		(0,-3/2*1.732) node {\large\textbf{4}};
	
	\draw		(0,-1) node {\large\textbf{3}};

	\draw [red,very thick] (-3/2+.35,0)--(3/2-.35,0);
	\draw [blue, dashed,thick] (-3/2+0.35,-.15)--(3/2-.35,-0.15);
	
	\draw [red,very thick] (-3/2+.2,-0.2*1.732)--(-.2,-3/2*1.732+.2*1.732);
	\draw [blue, dashed,thick] (-3/2+0.15+.2, -0.2*1.732)--(0.15-.2,-3/2*1.732+0.2*1.732);
	
	\draw [red,very thick] (3/2-0.2,-0.2*1.732)--(0.2,-3/2*1.732+0.2*1.732);
	\draw [blue, dashed,thick] (3/2-0.15-0.2,-.2*1.732)--(-0.15+0.2,-3/2*1.732+.2*1.732);
	\end{tikzpicture}
}
	\caption{Diagram VII3}
\end{subfigure}\begin{subfigure}[b]{0.2\textwidth}
		\centering
		\resizebox{\linewidth}{!}{
	\begin{tikzpicture}
	\draw	[blue,dashed, thick]  (-3/2,0)  circle (0.25);
	\draw	[red,very thick]  (-3/2,0)  circle (0.35);
	\draw  (-3/2,0)  node {\large\textbf{1}};
	
	\draw	[blue,dashed, thick]  (3/2,0)  circle (0.25);
	\draw	[red,very thick]  (3/2,0)  circle (0.35);
	\draw		(3/2,0) node {\large\textbf{2}};
	
	\draw	[blue,dashed, thick]  (0,-3/2*1.732) circle (0.25);
	\draw	[red,very thick] (0,-3/2*1.732) circle (0.35);
	\draw		(0,-3/2*1.732) node {\large\textbf{3}};
	
	\draw		(0,-1) node {\large\textbf{4}};

	\draw [red,very thick] (-3/2+.35,0)--(3/2-.35,0);
	\draw [blue, dashed,thick] (-3/2+0.35,-.15)--(3/2-.35,-0.15);
	
	\draw [red,very thick] (-3/2+.2,-0.2*1.732)--(-.2,-3/2*1.732+.2*1.732);
	\draw [blue, dashed,thick] (-3/2+0.15+.2, -0.2*1.732)--(0.15-.2,-3/2*1.732+0.2*1.732);
	
	\draw [red,very thick] (3/2-0.2,-0.2*1.732)--(0.2,-3/2*1.732+0.2*1.732);
	\draw [blue, dashed,thick] (3/2-0.15-0.2,-.2*1.732)--(-0.15+0.2,-3/2*1.732+.2*1.732);
	\end{tikzpicture}
}
	\caption{Diagram VII4}
\end{subfigure}

\caption{Complete problematic diagrams for $r_{12}^2r_{34}^2\equiv const$}
\label{fig:Problematicdiagramsnonadjacentdistancescj}
	
\end{figure}
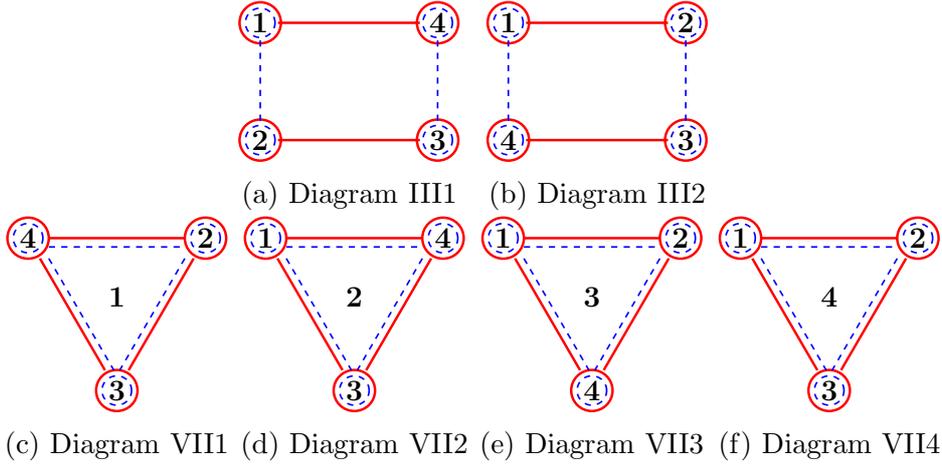

In this level set
 the complete
diagrams may be approached by  some singular sequence are the  Figure \ref{fig:Problematicdiagramsnonadjacentdistancescj}.

Similar as previous discussion, at least two of $r_{kl}^2$'s  must take infinitely many values and thus are dominating.
Suppose that $r_{kl}^2$ is dominating for some  $1\le k<l\le 4$.  There must exist a  singular sequence of central configurations with $r_{kl}^2\to 0$,  which happens only in the Diagram VII. In each case, note that all of $r_{kl}^2$'s are  dominating.

Considering  $r_{12}^2 \to \infty$, then $r_{34}^2 \to 0$, we are in Diagram VII1 or Diagram VII2. Without lose of generality,  assume that we are in Diagram VII1, thus
%\begin{equation*}
  %  ( {\Gamma_2}+{\Gamma_3}+\Gamma_4) ( {\Gamma_1}+{\Gamma_3}+{\Gamma_4})=0
%\end{equation*}
\begin{equation*}
     {\Gamma_2}+{\Gamma_3}+\Gamma_4=0.
\end{equation*}
Similarly, considering  $r_{12}^2 \to 0$, then $r_{34}^2 \to \infty$, we are in Diagram VII3 or Diagram VII4, assume that we are in Diagram VII3,then
\begin{equation*}
     {\Gamma_1}+{\Gamma_2}+{\Gamma_4}=0.
\end{equation*}

By further considering  $r_{13}^2 \to 0$,   we are in Diagram VII2 or Diagram VII4. Note that we can not be both in Diagram VII2 or Diagram VII4 now. Without lose of generality,  assume that we are in Diagram VII2,thus
\begin{equation*}
   {\Gamma_1}+{\Gamma_3}+\Gamma_4=0.
\end{equation*}

As a result,
\begin{equation*}
    \Gamma_1=\Gamma_2=\Gamma_3=-\frac{1}{2}\Gamma_4.
\end{equation*}

These vorticities are not compatible with Diagram III. So only Diagram VII1, Diagram VII2 and Diagram VII3  are possible for us. However, by considering  $r_{14}^2r_{24}^2r_{34}^2$ we know $r_{14}^2r_{24}^2r_{34}^2\longrightarrow 0$ and $r_{14}^2r_{24}^2r_{34}^2$ is dominating. But we will be out of   Diagram VII1, Diagram VII2 and Diagram VII3  if $r_{14}^2r_{24}^2r_{34}^2\longrightarrow \infty$, this is a contradiction.

$~~~~~~~~~~~~~~~~~~~~~~~~~~~~~~~~~~~~~~~~~~~~~~~~~~~~~~~~~~~~~~~~~~~~~~~~~~~~~~~~~~~~~~~~~~~~~~~~~~~~~~~~~~~~~~~~~~~~~~~~~~~~~~~~~~~~~~~~~~~~~~~~~~~~\Box$\\

So we consider all of $r_{jk}^2r_{lm}^2$'s are dominating from now on.\\

\begin{theorem}\label{MainGammaLnot0}
If the vorticities $\Gamma_n$ $(n\in\{1,2,3,4\})$ are nonzero such that $\Gamma\neq0$ and $L\neq0$, then the four-vortex problem has finitely many  central configurations except perhaps if three of  the vorticities are all negatively a half the rest vorticity, (for example, $-\frac{1}{2}\Gamma_1= \Gamma_2=\Gamma_3 =\Gamma_4$).

\end{theorem}
{\bf Proof.}

{\bfseries{Case 1: If Diagram III does not occur. }}

Since at least two of the $r_{jk}^2$'s should take
infinitely many values. We suppose $r_{14}^2$ does, then $r_{14}^2$ is dominating. There is a sequence of normalized central configurations
such that $r^{(n)}_{14}\longrightarrow \infty$. Then we can extract a singular sequence   corresponding to one of the diagrams in Figure \ref{fig:Problematicdiagrams}. In Diagram VI, Diagram VIII and Diagram IX in Figure  \ref{fig:Problematicdiagrams}, no distance is going to
infinity, only Diagram VII is possible. In this case, without loss of generality, we consider
\begin{equation}
     r_{12}, r_{23},r_{31}\approx \epsilon^2,  ~~~~~~~      r_{14}, r_{24}, r_{34}\approx \epsilon^{-2}.\nonumber
\end{equation}
 As a result, the  $r_{jk}^2$'s are all dominating and
\begin{equation}\label{123}
    \Gamma_1+\Gamma_2+ \Gamma_3=0.\nonumber
\end{equation}

Consider $r_{12}^2$. Push it to infinity. We are in Diagram VII again. The constraint on the vorticities this time is
$\Gamma_4+\Gamma_2+ \Gamma_3=0$ or $\Gamma_1+\Gamma_4+ \Gamma_3=0$. In other words
\begin{equation}\label{12}
    (\Gamma_1- \Gamma_4)(\Gamma_2 -\Gamma_4)=0.\nonumber
\end{equation}

Similarly, push $r_{23}^2$ to infinity, we have \begin{equation}\label{23}
 (\Gamma_3- \Gamma_4)(\Gamma_2 -\Gamma_4)=0;\nonumber
\end{equation} push $r_{13}^2$ to infinity, we have \begin{equation}\label{13}
   (\Gamma_1- \Gamma_4)(\Gamma_3 -\Gamma_4)=0.\nonumber
\end{equation}

It is easy to see that  the equations above have only the untrivial  solutions of the following type:
\begin{center}
$-\frac{1}{2}\Gamma_1= \Gamma_2=\Gamma_3 =\Gamma_4$.
\end{center}

{\bfseries{Case 2: If Diagram III  occurs. }}

Without lose of generality, we assume that there always exists a  singular sequence of central configurations approaching  Diagram III such that $r_{13}\to \infty$ below. Then
  \begin{equation}\label{DIII13}
  \Gamma_1 \Gamma_3-\Gamma_2 \Gamma_4=0,
\end{equation}
and $r_{13}^2$, $r_{24}^2$ and $r_{13}^2r_{24}^2$  are all dominating.

Push $r_{13}^2\longrightarrow 0$, then we are in Diagram VI, Diagram VII, Diagram VIII and Diagram IX.  Thus other distances,
say $r_{12}$ and $r_{23}$, should also go to zero. They also take infinitely many values. So $r_{12}^2$ and $r_{23}^2$ dominating. It is easy to see that in these diagrams  the product of any two nonadjacent
distances tends to zero except in Diagram VII. However by Lemma \ref{nonadjacentdistancescj} we can consider that $r_{12}^2r_{34}^2$ and $r_{14}^2r_{23}^2$   are both dominating.

Considering $r_{12}^2r_{34}^2\longrightarrow\infty$ we are in Diagram III and  the constraint on the vorticities are
\begin{equation}\label{DIII12}
  \Gamma_1 \Gamma_2-\Gamma_3 \Gamma_4=0;
\end{equation}
similarly, considering $r_{14}^2r_{23}^2\longrightarrow\infty$ we are in Diagram III and  the constraint on the vorticities are
\begin{equation}\label{DIII14}
  \Gamma_1 \Gamma_4-\Gamma_2 \Gamma_3=0.
\end{equation}
Note that $\Gamma \neq 0$ it follows that
\begin{equation*}
   \Gamma_1= \Gamma_2=\Gamma_3 =\Gamma_4.
\end{equation*}
However, these vorticitie are not compatible with Diagram VI, Diagram VII, Diagram VIII and Diagram IX.  This is a contradiction.

$~~~~~~~~~~~~~~~~~~~~~~~~~~~~~~~~~~~~~~~~~~~~~~~~~~~~~~~~~~~~~~~~~~~~~~~~~~~~~~~~~~~~~~~~~~~~~~~~~~~~~~~~~~~~~~~~~~~~~~~~~~~~~~~~~~~~~~~~~~~~~~~~~~~~\Box$\\

\begin{theorem}\label{MainGammaLnot01}
If the vorticities $\Gamma_n$ $(n\in\{1,2,3,4\})$ are nonzero such that $-\frac{1}{2}\Gamma_1= \Gamma_2=\Gamma_3 =\Gamma_4$, then the four-vortex problem has exactly 44  central configurations and six of them are real central configurations.

\end{theorem}
{\bf Proof.}

We consider the system (\ref{stationaryconfigurationmainequ3})  with  $\Lambda=\pm1$.

 Without loss of generality, suppose  $\Gamma_1=-2,\Gamma_2=\Gamma_3 =\Gamma_4=1$. Then a simple computation by Mathematica shows that  the system (\ref{stationaryconfigurationmainequ3})   has exactly 88   solutions and 12 of  them satisfy conjugate conditions $Z_{jk}=\overline{W}_{jk}$.

$~~~~~~~~~~~~~~~~~~~~~~~~~~~~~~~~~~~~~~~~~~~~~~~~~~~~~~~~~~~~~~~~~~~~~~~~~~~~~~~~~~~~~~~~~~~~~~~~~~~~~~~~~~~~~~~~~~~~~~~~~~~~~~~~~~~~~~~~~~~~~~~~~~~~\Box$\\

%\begin{remark}\label{counter-example}
%As a by-product, by this result and Proposition \ref{necessaryconditions}, it follows  that, in contrast to the four-body problem,  the four-vortex problem may not have any  stationary configuration in essence.
%\end{remark}

\section{Upper bounds}
\indent\par
\subsection{An equivalent form of (\ref{stationaryconfigurationmain}) and a coarse upper bound by B\'{e}zout Theorem}
\indent\par
If $\Gamma\neq 0$, the system (\ref{stationaryconfigurationmain}) is equivalent to
\begin{equation}\label{stationaryconfigurationmainequ1}
\begin{array}{cc}
  \Gamma z_n=\sum_{ j \neq n} \Gamma_j z_{jn},~~~ \Gamma w_n=\sum_{ j \neq n} \Gamma_j w_{jn},& 1\leq n \leq N, \\
  \Lambda z_{nN}=(\Gamma_n+\Gamma_N)Z_{nN}+\sum_{j<N, j \neq n} \Gamma_j (Z_{jN}-Z_{jn}),&1\leq n< N, \\
  \Lambda^{-1} w_{nN}=(\Gamma_n+\Gamma_N)W_{nN}+\sum_{j<N, j \neq n} \Gamma_j (W_{jN}-W_{jn}),&1\leq n< N, \\
  z_{jk}=z_{jN}-z_{kN},~~~  w_{jk}=w_{jN}-w_{kN},&1\leq j, k< N, \\
  Z_{jk} w_{jk}=1,~~~W_{jk} z_{jk}=1,&1\leq j< k\leq N, \\
 z_{jk}=-z_{kj},~~~ w_{jk}=-w_{kj},&1\leq k< j\leq N, \\
  Z_{jk}=-Z_{kj},~~~ W_{jk}=-W_{kj},&1\leq k< j\leq N, \\
  z_{12}=w_{12},~~~Z_{12}=W_{12};
\end{array}
\end{equation}
or the following closed system in the variables $z_{jk},w_{jk},Z_{jk},W_{jk}$ $(1\leq j< k\leq N)$:
\begin{equation}\label{stationaryconfigurationmainequ2}
\begin{array}{cc}
  \Lambda z_{nN}=(\Gamma_n+\Gamma_N)Z_{nN}+\sum_{j<N, j \neq n} \Gamma_j (Z_{jN}-Z_{jn}),&1\leq n< N, \\
  \Lambda^{-1} w_{nN}=(\Gamma_n+\Gamma_N)W_{nN}+\sum_{j<N, j \neq n} \Gamma_j (W_{jN}-W_{jn}),&1\leq n< N, \\
  z_{jk}=z_{jN}-z_{kN},~~~  w_{jk}=w_{jN}-w_{kN},&1\leq j, k< N, \\
  Z_{jk} w_{jk}=1,~~~W_{jk} z_{jk}=1,&1\leq j< k\leq N, \\
 z_{jk}=-z_{kj},~~~ w_{jk}=-w_{kj},&1\leq k< j\leq N, \\
  Z_{jk}=-Z_{kj},~~~ W_{jk}=-W_{kj},&1\leq k< j\leq N, \\
  z_{12}=w_{12},~~~Z_{12}=W_{12}.
\end{array}
\end{equation}

The system (\ref{stationaryconfigurationmainequ2}) above can essentially be  regarded as a closed system in the variables $Z_{jk},W_{jk}$ $(1\leq j< k\leq N)$, for example, when $N=4$,  the system (\ref{stationaryconfigurationmainequ2}) is equivalent to
\begin{equation}\label{stationaryconfigurationmainequ3}
\begin{array}{r}
 Z_{12} \left((\Gamma_1+\Gamma_2) W_{12}+\Gamma_3 W_{13}-\Gamma_3 W_{23}+\Gamma_4 W_{14}-\Gamma_4 W_{24}\right)={\Lambda^{-1} } ,\\
 Z_{13} \left(\Gamma_2 W_{12}+\Gamma_2 W_{23}+(\Gamma_1+\Gamma_3) W_{13}+\Gamma_4 W_{14}-\Gamma_4 W_{34}\right)={\Lambda^{-1} },\\
    Z_{14} \left(\Gamma_2 W_{12}+\Gamma_2 W_{24}+\Gamma_3 W_{13}+\Gamma_3 W_{34}+(\Gamma_1+\Gamma_4) W_{14}\right)={\Lambda^{-1} },\\
    Z_{23} \left(\Gamma_1 (-W_{12})+\Gamma_1 W_{13}+\Gamma_2 W_{23}+\Gamma_3 W_{23}+\Gamma_4 W_{24}-\Gamma_4 W_{34}\right)={\Lambda^{-1} },\\
    Z_{24} \left(\Gamma_1 (-W_{12})+\Gamma_1 W_{14}+\Gamma_2 W_{24}+\Gamma_3 W_{23}+\Gamma_3 W_{34}+\Gamma_4 W_{24}\right)={\Lambda^{-1} },\\
   Z_{34} \left(\Gamma_1 (-W_{13})+\Gamma_1 W_{14}-\Gamma_2 W_{23}+\Gamma_2 W_{24}+\Gamma_3 W_{34}+\Gamma_4 W_{34}\right)={\Lambda^{-1} },\\
{W_{12} \left((\Gamma_1+\Gamma_2) Z_{12}+\Gamma_3 Z_{13}-\Gamma_3 Z_{23}+\Gamma_4 Z_{14}-\Gamma_4 Z_{24}\right)}={\Lambda },\\
{W_{13} \left(\Gamma_2 Z_{12}+\Gamma_2 Z_{23}+(\Gamma_1+\Gamma_3) Z_{13}+\Gamma_4 Z_{14}-\Gamma_4 Z_{34}\right)}={\Lambda },\\
{W_{14} \left(\Gamma_2 Z_{12}+\Gamma_2 Z_{24}+\Gamma_3 Z_{13}+\Gamma_3 Z_{34}+(\Gamma_1+\Gamma_4) Z_{14}\right)}={\Lambda },\\
{W_{23} \left(\Gamma_1 (-Z_{12})+\Gamma_1 Z_{13}+\Gamma_2 Z_{23}+\Gamma_3 Z_{23}+\Gamma_4 Z_{24}-\Gamma_4 Z_{34}\right)}={\Lambda },\\
{W_{24} \left(\Gamma_1 (-Z_{12})+\Gamma_1 Z_{14}+\Gamma_2 Z_{24}+\Gamma_3 Z_{23}+\Gamma_3 Z_{34}+\Gamma_4 Z_{24}\right)}={\Lambda },\\
{W_{34} \left(\Gamma_1 (-Z_{13})+\Gamma_1 Z_{14}-\Gamma_2 Z_{23}+\Gamma_2 Z_{24}+\Gamma_3 Z_{34}+\Gamma_4 Z_{34}\right)}={\Lambda },\\
 \left((\Gamma_1+\Gamma_2) Z_{12}+\Gamma_3 Z_{13}-\Gamma_3 Z_{23}+\Gamma_4 Z_{14}-\Gamma_4 Z_{24}\right)\\
 ={\Lambda^2 }\left((\Gamma_1+\Gamma_2) W_{12}+\Gamma_3 W_{13}-\Gamma_3 W_{23}+\Gamma_4 W_{14}-\Gamma_4 W_{24}\right),\\
   Z_{12}=W_{12}.
\end{array}
\end{equation}

We embed the system (\ref{stationaryconfigurationmainequ3}) above into a polynomial system in the projective space $\mathbb{P}^{12}_{\mathbb{C}}$:
\begin{equation}\label{stationaryconfigurationmainequ4}
\begin{array}{r}
 Z_{12} \left((\Gamma_1+\Gamma_2) W_{12}+\Gamma_3 W_{13}-\Gamma_3 W_{23}+\Gamma_4 W_{14}-\Gamma_4 W_{24}\right)={\Lambda^{-1} }T^2 ,\\
 Z_{13} \left(\Gamma_2 W_{12}+\Gamma_2 W_{23}+(\Gamma_1+\Gamma_3) W_{13}+\Gamma_4 W_{14}-\Gamma_4 W_{34}\right)={\Lambda^{-1} }T^2,\\
    Z_{14} \left(\Gamma_2 W_{12}+\Gamma_2 W_{24}+\Gamma_3 W_{13}+\Gamma_3 W_{34}+(\Gamma_1+\Gamma_4) W_{14}\right)={\Lambda^{-1} }T^2,\\
    Z_{23} \left(\Gamma_1 (-W_{12})+\Gamma_1 W_{13}+\Gamma_2 W_{23}+\Gamma_3 W_{23}+\Gamma_4 W_{24}-\Gamma_4 W_{34}\right)={\Lambda^{-1} }T^2,\\
    Z_{24} \left(\Gamma_1 (-W_{12})+\Gamma_1 W_{14}+\Gamma_2 W_{24}+\Gamma_3 W_{23}+\Gamma_3 W_{34}+\Gamma_4 W_{24}\right)={\Lambda^{-1} }T^2,\\
   Z_{34} \left(\Gamma_1 (-W_{13})+\Gamma_1 W_{14}-\Gamma_2 W_{23}+\Gamma_2 W_{24}+\Gamma_3 W_{34}+\Gamma_4 W_{34}\right)={\Lambda^{-1} }T^2,\\
{W_{12} \left((\Gamma_1+\Gamma_2) Z_{12}+\Gamma_3 Z_{13}-\Gamma_3 Z_{23}+\Gamma_4 Z_{14}-\Gamma_4 Z_{24}\right)}={\Lambda }T^2,\\
{W_{13} \left(\Gamma_2 Z_{12}+\Gamma_2 Z_{23}+(\Gamma_1+\Gamma_3) Z_{13}+\Gamma_4 Z_{14}-\Gamma_4 Z_{34}\right)}={\Lambda }T^2,\\
{W_{14} \left(\Gamma_2 Z_{12}+\Gamma_2 Z_{24}+\Gamma_3 Z_{13}+\Gamma_3 Z_{34}+(\Gamma_1+\Gamma_4) Z_{14}\right)}={\Lambda }T^2,\\
{W_{23} \left(\Gamma_1 (-Z_{12})+\Gamma_1 Z_{13}+\Gamma_2 Z_{23}+\Gamma_3 Z_{23}+\Gamma_4 Z_{24}-\Gamma_4 Z_{34}\right)}={\Lambda }T^2,\\
{W_{24} \left(\Gamma_1 (-Z_{12})+\Gamma_1 Z_{14}+\Gamma_2 Z_{24}+\Gamma_3 Z_{23}+\Gamma_3 Z_{34}+\Gamma_4 Z_{24}\right)}={\Lambda }T^2,\\
{W_{34} \left(\Gamma_1 (-Z_{13})+\Gamma_1 Z_{14}-\Gamma_2 Z_{23}+\Gamma_2 Z_{24}+\Gamma_3 Z_{34}+\Gamma_4 Z_{34}\right)}={\Lambda }T^2,\\
 \left((\Gamma_1+\Gamma_2) Z_{12}+\Gamma_3 Z_{13}-\Gamma_3 Z_{23}+\Gamma_4 Z_{14}-\Gamma_4 Z_{24}\right)\\
 ={\Lambda^2 }\left((\Gamma_1+\Gamma_2) W_{12}+\Gamma_3 W_{13}-\Gamma_3 W_{23}+\Gamma_4 W_{14}-\Gamma_4 W_{24}\right),\\
   Z_{12}=W_{12}.
\end{array}
\end{equation}
Here the system (\ref{stationaryconfigurationmainequ3}) is just an affine piece of the system (\ref{stationaryconfigurationmainequ4}) for $T\neq 0$.

After deleting  the first or the seventh equation from the system (\ref{stationaryconfigurationmainequ4}) above, it is easy to see that the degree of the system (\ref{stationaryconfigurationmainequ4}) is no more than $2^{11}$. We remark that by direct application of the following B\'{e}zout Theorem it follows that the number of central configurations  for the four-vortex problem is no more than $2^{10}=1024$.
\begin{lemma}\label{Bezout}\emph{(\cite{fulton2013intersection})}
Let $\mathcal{V}_1, \cdots,\mathcal{V}_m$ be  subvarieties of $\mathbb{P}^\mathcal{N}$. Let $\mathcal{U}_1, \cdots,\mathcal{U}_n$ be  the
irreducible components   of $\mathcal{V}_1\bigcap\cdots \bigcap\mathcal{V}_m$. Then
\begin{equation}\label{Bezoutform}
    \sum_{j=1}^{n}deg(\mathcal{U}_j)\leq \prod_{j=1}^{m}deg(\mathcal{V}_j).
\end{equation}

\end{lemma}

To obtain a better upper bound, we estimate the number of the irreducible components of the system (\ref{stationaryconfigurationmainequ4}) for $T=0$. This is a linear variety in  $\mathbb{P}^{11}_{\mathbb{C}}$:
\begin{equation}\label{stationaryconfigurationmainequ5}
\begin{array}{r}
 Z_{12} \left((\Gamma_1+\Gamma_2) W_{12}+\Gamma_3 W_{13}-\Gamma_3 W_{23}+\Gamma_4 W_{14}-\Gamma_4 W_{24}\right)=0 ,\\
 Z_{13} \left(\Gamma_2 W_{12}+\Gamma_2 W_{23}+(\Gamma_1+\Gamma_3) W_{13}+\Gamma_4 W_{14}-\Gamma_4 W_{34}\right)=0,\\
    Z_{14} \left(\Gamma_2 W_{12}+\Gamma_2 W_{24}+\Gamma_3 W_{13}+\Gamma_3 W_{34}+(\Gamma_1+\Gamma_4) W_{14}\right)=0,\\
    Z_{23} \left(\Gamma_1 (-W_{12})+\Gamma_1 W_{13}+\Gamma_2 W_{23}+\Gamma_3 W_{23}+\Gamma_4 W_{24}-\Gamma_4 W_{34}\right)=0,\\
    Z_{24} \left(\Gamma_1 (-W_{12})+\Gamma_1 W_{14}+\Gamma_2 W_{24}+\Gamma_3 W_{23}+\Gamma_3 W_{34}+\Gamma_4 W_{24}\right)=0,\\
   Z_{34} \left(\Gamma_1 (-W_{13})+\Gamma_1 W_{14}-\Gamma_2 W_{23}+\Gamma_2 W_{24}+\Gamma_3 W_{34}+\Gamma_4 W_{34}\right)=0,\\
{W_{12} \left((\Gamma_1+\Gamma_2) Z_{12}+\Gamma_3 Z_{13}-\Gamma_3 Z_{23}+\Gamma_4 Z_{14}-\Gamma_4 Z_{24}\right)}=0,\\
{W_{13} \left(\Gamma_2 Z_{12}+\Gamma_2 Z_{23}+(\Gamma_1+\Gamma_3) Z_{13}+\Gamma_4 Z_{14}-\Gamma_4 Z_{34}\right)}=0,\\
{W_{14} \left(\Gamma_2 Z_{12}+\Gamma_2 Z_{24}+\Gamma_3 Z_{13}+\Gamma_3 Z_{34}+(\Gamma_1+\Gamma_4) Z_{14}\right)}=0,\\
{W_{23} \left(\Gamma_1 (-Z_{12})+\Gamma_1 Z_{13}+\Gamma_2 Z_{23}+\Gamma_3 Z_{23}+\Gamma_4 Z_{24}-\Gamma_4 Z_{34}\right)}=0,\\
{W_{24} \left(\Gamma_1 (-Z_{12})+\Gamma_1 Z_{14}+\Gamma_2 Z_{24}+\Gamma_3 Z_{23}+\Gamma_3 Z_{34}+\Gamma_4 Z_{24}\right)}=0,\\
{W_{34} \left(\Gamma_1 (-Z_{13})+\Gamma_1 Z_{14}-\Gamma_2 Z_{23}+\Gamma_2 Z_{24}+\Gamma_3 Z_{34}+\Gamma_4 Z_{34}\right)}=0,\\
 \left((\Gamma_1+\Gamma_2) Z_{12}+\Gamma_3 Z_{13}-\Gamma_3 Z_{23}+\Gamma_4 Z_{14}-\Gamma_4 Z_{24}\right)\\
 ={\Lambda^2 }\left((\Gamma_1+\Gamma_2) W_{12}+\Gamma_3 W_{13}-\Gamma_3 W_{23}+\Gamma_4 W_{14}-\Gamma_4 W_{24}\right),\\
   Z_{12}=W_{12}.
\end{array}
\end{equation}
A straightforward computation shows that  the greatest lower bound  of the number of irreducible components of the linear variety (\ref{stationaryconfigurationmainequ5}) above is no more than  441
 in $\mathbb{P}^{11}_{\mathbb{C}}$. Therefore, the upper bound  of the number of central configurations  for the four-vortex problem is  at least $[\frac{2^{11}-441}{2}]=803$ by direct application of the  B\'{e}zout Theorem above.

\subsection{A new equivalent form of (\ref{stationaryconfiguration1}) and a better upper bound by  B\'{e}zout Theorem}
\indent\par
We use B\'{e}zout Theorem again, but transform the system  (\ref{stationaryconfiguration1}) or (\ref{stationaryconfiguration3}) into a new equivalent form.

Following O'Neil  \cite{o2007relative} we introduce the relations
\begin{equation}\label{stationaryconfigurationnew1}
\begin{array}{c}
  \frac{1}{2}\sum_{1\leq j, k\leq N, j\neq k}\frac{\Gamma_j\Gamma_k}{(\zeta-z_j)(\zeta-z_k)} = \overline{\Lambda}\sum_{1\leq k \leq N}\frac{\Gamma_k \overline{z}_k}{\zeta-z_k}\\
  \frac{1}{2}\sum_{1\leq j, k\leq N, j\neq k}\frac{\Gamma_j\Gamma_k}{(\zeta-\overline{z}_j)(\zeta-\overline{z}_k)} = {\Lambda}\sum_{1\leq k \leq N}\frac{\Gamma_k {z}_k}{\zeta-\overline{z}_k}
\end{array}
\end{equation}
by   the identity
\begin{equation}\label{identity1}
    \frac{1}{(\zeta-z_j)(\zeta-z_k)}=\frac{1}{z_j-z_k}(\frac{1}{\zeta-z_j}-\frac{1}{\zeta-z_k})
\end{equation}
and  (\ref{stationaryconfiguration1}).

It is easy to see that (\ref{stationaryconfiguration1}) is equivalent to (\ref{stationaryconfigurationnew1}) provided $z_j-z_k\neq 0$ for any $1\leq j< k\leq N$. Indeed, by (\ref{identity1}) it follows that
\begin{equation}
\begin{aligned}
&\frac{1}{2}\sum_{1\leq j, k\leq N, j\neq k}\frac{\Gamma_j\Gamma_k}{(\zeta-z_j)(\zeta-z_k)}=\sum_{1\leq j, k\leq N, j\neq k}\frac{\Gamma_j\Gamma_k}{(\zeta-z_k)(z_k-z_j)}\\
&=\sum_{1\leq  k\leq N}\frac{\Gamma_k}{\zeta-z_k}\sum_{1\leq j\leq N, j\neq k}\frac{\Gamma_j}{(z_k-z_j)}=\sum_{1\leq  k\leq N}\frac{\Gamma_k}{\zeta-z_k}\overline{V}_k.
\end{aligned}\nonumber
\end{equation}
Therefore, (\ref{stationaryconfigurationnew1}) holds if and only if ${V}_k={\Lambda}{z}_k$. That is to say, provided $z_j-z_k\neq 0$ for any $1\leq j< k\leq N$, $z$ is a central configuration if and only if  (\ref{stationaryconfigurationnew1}) holds for any $\zeta \in \mathbb{C}$.

Both sides of (\ref{stationaryconfigurationnew1}) are rational functions of $\zeta$;  by eliminating denominators of both sides  one gets two polynomial equations
in $\zeta$, with coefficients that are expressions with $z_k,\Lambda$  and their conjugations.  (\ref{stationaryconfigurationnew1}) holds   exactly when all coefficients of the two polynomials
are zero. According to these facts, we can transform the system  (\ref{stationaryconfiguration1}) into a new equivalent form by the relations of  their coefficients.

In a similar way,
 we can transform the system  (\ref{stationaryconfiguration3}) into a new equivalent form:
\begin{equation}\label{stationaryconfigurationnew2}
\begin{array}{c}
  \frac{1}{2}\sum_{1\leq j, k\leq N, j\neq k}\frac{\Gamma_j\Gamma_k}{(\zeta-z_j)(\zeta-z_k)} = \overline{\Lambda}\sum_{1\leq k \leq N}\frac{\Gamma_k w_k}{\zeta-z_k},\\
  \frac{1}{2}\sum_{1\leq j, k\leq N, j\neq k}\frac{\Gamma_j\Gamma_k}{(\zeta-w_j)(\zeta-w_k)} = {\Lambda}\sum_{1\leq k \leq N}\frac{\Gamma_k {z}_k}{\zeta-w_k};
\end{array}
\end{equation}
or the relations of  their coefficients (from now on we  consider only $N=4$):
\begin{equation}\label{stationaryconfigurationnew3}
\begin{array}{c}
 \Lambda M_z=0, ~~~~~~ \overline{\Lambda}M_w=0\\
  L-\overline{\Lambda}I+\overline{\Lambda}M_w \sum_{j=1}^4z_j=0,~~~~~~ L-{\Lambda}I+{\Lambda}M_z \sum_{j=1}^4w_j=0,\\
  \overline{\Lambda } (M_w \sum_{1\leq j< k\leq 4}z_j z_k+{F_z}-I \sum_{j=1}^4z_j)-\sum_{1\leq j< k\leq 4}\Gamma_j\Gamma_k(z_j+ z_k)+L \sum_{j=1}^4z_j=0,\\
{\Lambda } (M_z \sum_{1\leq j< k\leq N}w_j w_k+{F_w}-I \sum_{j=1}^4w_j)-\sum_{1\leq j< k\leq N}\Gamma_j\Gamma_k(w_j+ w_k)+L \sum_{j=1}^4w_j=0,\\
\sum_{1\leq j< k\leq 4,l<m,\{j,k,l,m\}=\{1,2,3,4\}}\Gamma_j \Gamma_k z_l z_m+\overline{\Lambda}  G_z=0,\\
\sum_{1\leq j< k\leq 4,l<m,\{j,k,l,m\}=\{1,2,3,4\}}\Gamma_j \Gamma_k w_l w_m+\Lambda  G_w=0,
\end{array}\nonumber
\end{equation}
or
\begin{equation}\label{stationaryconfigurationnew4}
\left\{
             \begin{array}{rr}
                M_z=0,& ~~~~~~ M_w=0,\\
             L-\overline{\Lambda}I=0,& ~~~~~~ L-{\Lambda}I=0, \\
             \overline{\Lambda } {F_z}-f_z=0,& ~~~~~~ {\Lambda } {F_w}-f_w=0,\\
\overline{\Lambda}  G_z +g_z=0,& ~~~~~~ \Lambda  G_w+g_w=0,
             \end{array}
\right.
\end{equation}
where
\begin{equation}
\begin{array}{c}
  M_z=\sum_{j=1}^4\Gamma_jz_j , ~~~~~~ M_w=\sum_{j=1}^4\Gamma_jw_j,\\
  I=\sum_{j=1}^4\Gamma_jz_jw_j,\\
   F_z=\sum_{j=1}^4\Gamma_jz_j^2 w_j, ~~~~~~f_z=\sum_{1\leq j< k\leq 4}\Gamma_j\Gamma_k(z_j+ z_k),\\
    F_w=\sum_{j=1}^4\Gamma_jz_j w_j^2,~~~~~~f_w=\sum_{1\leq j< k\leq 4}\Gamma_j\Gamma_k(w_j+ w_k),\\
G_z=\Gamma_1 w_1 z_2 z_3 z_4+\Gamma_2 w_2 z_1 z_3 z_4+\Gamma_3 w_3 z_1 z_2  z_4+\Gamma_4 w_4 z_1 z_2 z_3,\\
g_z=\sum_{1\leq j< k\leq 4,l<m,\{j,k,l,m\}=\{1,2,3,4\}}\Gamma_j \Gamma_k z_l z_m,\\
G_w=\Gamma_1 z_1 w_2 w_3 w_4+\Gamma_2 z_2 w_1 w_3 w_4+\Gamma_3 z_3 w_1 w_2  w_4+\Gamma_4 z_4 w_1 w_2 w_3,\\
g_w=\sum_{1\leq j< k\leq 4,l<m,\{j,k,l,m\}=\{1,2,3,4\}}\Gamma_j \Gamma_k w_l w_m.
\end{array}\nonumber
\end{equation}

It follows that, in the case of relative equilibria, provided $z_j-z_k\neq 0$ for any $1\leq j< k\leq 4$, normalized central configurations are characterized by
\begin{equation}\label{stationaryconfigurationnew5}
\left\{
             \begin{array}{rr}
             M_z=0,& ~~~~~~ M_w=0,\\
             L-\Lambda I=0, &~~~~~~ z_2-z_1= w_2-w_1, \\
             {\Lambda } {F_z}-f_z=0, &
             {\Lambda } {F_w}-f_w=0,\\
{\Lambda}  G_z+g_z=0,&\Lambda  G_w+g_w=0,
             \end{array}
\right.
\end{equation}
here, without loss of generality, one can further assume that $\Lambda=1$;
in the case of collapse configurations,  normalized central configurations are characterized by
\begin{equation}\label{stationaryconfigurationnew6}
\left\{
             \begin{array}{rr}
                M_z=0, &~~~~~~ M_w=0,\\
             L=0,~~~~~~ I=0, &z_2-z_1= w_2-w_1, \\
             {\overline{\Lambda} } {F_z}-f_z=0, &
             {\Lambda } {F_w}-f_w=0,\\
\overline{\Lambda} G_z+g_z=0,&\Lambda  G_w+g_w=0.
             \end{array}
\right.
\end{equation}

After embedding the system (\ref{stationaryconfigurationnew5}) or (\ref{stationaryconfigurationnew6}) above into a  system in the projective space $\mathbb{P}^{8}_{\mathbb{C}}$, it is easy to see that the degree of the systems  is no more than $2\times 3^{2}\times 4^{2}=288$. By direct application of the following B\'{e}zout Theorem it follows that the number of central configurations  for the four-vortex problem is no more than $144$.

We remark that
\begin{proposition}\label{isolatedtrivialsolution}
If $L=0$ and $z_j=z_k$ (or $w_j=w_k$) for some $ j\neq k$, then  there is only trivial solution in the system (\ref{stationaryconfigurationnew4}) above.
\end{proposition}
{\bf Proof.} %of Theorem \ref{asymptic2}:}

If  $z_j=z_k$  for some $ j\neq k$, assume that $\Xi$ is the subset  of the index set $\{1,2,3,4\}$   such
that $z_j=z_k\triangleq z_*$ for any $j,k\in\Xi$. According to  (\ref{stationaryconfigurationnew2}), it follows that
\begin{center}
$\sum_{j<k,j,k\in \Xi}\Gamma_j \Gamma_k=0$.
\end{center}
Otherwise, the right side of (\ref{stationaryconfigurationnew2}) has no double poles, while the left
side will have a double pole at $z=z_*$. By $L=0$ it is easy to see that $\Xi=\{1,2,3,4\}$.
Thus
\begin{equation}
\frac{\sum_{1\leq k \leq 4}\Gamma_k w_k}{\zeta-z_*}=0 ~~~~~~~~\text{or}~~~~~~~~\sum_{1\leq k \leq 4}\Gamma_k w_k=0.\nonumber
\end{equation}

Since $ w_2-w_1=z_2-z_1=0$, a similar argument shows that
\begin{equation}
w_1=w_2=w_3=w_4, ~~~~~~~~\text{and}~~~~~~~~\sum_{1\leq k \leq 4}\Gamma_k z_k=0.\nonumber
\end{equation}

Note that $\Gamma\neq 0$ by $L=0$.
As a result, there is only trivial solution in the system (\ref{stationaryconfigurationnew4}) above.

$~~~~~~~~~~~~~~~~~~~~~~~~~~~~~~~~~~~~~~~~~~~~~~~~~~~~~~~~~~~~~~~~~~~~~~~~~~~~~~~~~~~~~~~~~~~~~~~~~~~~~~~~~~~~~~~~~~~~~~~~~~~~~~~~~~~~~~~~~~~~~~~~~~~~\Box$\\

\subsubsection{An upper bound for collinear central configurations }
\indent\par
If $w_j=z_j$ ($j=1,2,3,4$), the systems (\ref{stationaryconfigurationnew5}) and (\ref{stationaryconfigurationnew6}) respectively reduce to
\begin{equation}\label{stationaryconfigurationnewcr1}
\left\{
             \begin{array}{rr}
             M_z=0,&L- I=0,\\
              {F_z}-f_z=0, &
   G_z+g_z=0;
             \end{array}
\right.
\end{equation}
and \begin{equation}\label{stationaryconfigurationnewcc1}
\left\{
             \begin{array}{rr}
                M_z=0, &
             L=0,~~~~~~ I=0,  \\
              {F_z}=f_z=0, &
             G_z=g_z=0.
             \end{array}
\right.
\end{equation}

It is easy to see that collinear normalized central configurations are characterized by the systems (\ref{stationaryconfigurationnew5}) or (\ref{stationaryconfigurationnew6}) respectively.  On the other hand, a straightforward computation shows that the system (\ref{stationaryconfigurationnewcc1}) has only trivial solution, thus there is no any collinear collapse configuration in the four-vortex problem.
In fact, it is also easy to see that there is no collinear collapse configuration for the general $N$-vortex problem.

\paragraph{An upper bound for collinear relative equilibria.}
In this case, we embed the system  (\ref{stationaryconfigurationnewcr1}) into a system in the projective space $\mathbb{P}^{4}_{\mathbb{C}}$:
\begin{equation}\label{stationaryconfigurationnewcr2}
\begin{array}{cccc}
             M_z=0,&
             L t^2-I =0, &
{F_z}-f_z t^2=0, &
   G_z+g_zt^2=0.
             \end{array}
\end{equation}
Then the system (\ref{stationaryconfigurationnewcr1}) is just an affine piece of the system (\ref{stationaryconfigurationnewcr2}) for $t\neq 0$. And the algebraic variety (\ref{stationaryconfigurationnewcr2}) is a disjoint union of  (\ref{stationaryconfigurationnewcr1}) and the  variety
\begin{equation}\label{stationaryconfigurationnewcr3}
\begin{array}{cccc}
             M_z=0,& I=0,&
             {F_z}=0, &
G_z=0.
             \end{array}
\end{equation}
Since  a straightforward computation shows that the system (\ref{stationaryconfigurationnewcr3}) has only trivial solution, it follows that the variety  (\ref{stationaryconfigurationnewcr2}) is equal to (\ref{stationaryconfigurationnewcr1}) and  has exactly $24$ points (counting with the appropriate mulitiplicity) in $\mathbb{P}^{4}_{\mathbb{C}}$. As a result,  there are at most $12$ collinear relative equilibria for the four-vortex problem.

\begin{remark}
The result, that there are at most $12$ collinear relative equilibria for the four-vortex problem, has been proved by Hampton and Moeckel  \cite{hampton2009finiteness}.
\end{remark}

\paragraph{Refined   B\'{e}zout Theorem.}

In the following we will frequently employ a refined version of B\'{e}zout Theorem
\begin{lemma}\label{Bezoutrefine}\emph{(\cite{patil1983remarks})}
Let $\mathcal{V}_1, \cdots,\mathcal{V}_m$ be pure dimensional  subvarieties of $\mathbb{P}^\mathcal{N}$.
 Let $\mathcal{U}_1, \cdots,\mathcal{U}_n$ be  the
irreducible components   of $\mathcal{X}\triangleq\mathcal{V}_1\bigcap\cdots \bigcap\mathcal{V}_m$. Then
\begin{equation}\label{Bezoutformnew}
    \sum_{j=1}^{n}l(\mathcal{X};\mathcal{U}_j)deg(\mathcal{U}_j)\leq \prod_{j=1}^{m}deg(\mathcal{V}_j),
\end{equation}
where $l(\mathcal{X};\mathcal{U}_j)$ is the length of well-defined primary ideals, i.e.,  the multiplicity  of $\mathcal{X}$ along $\mathcal{U}_j$.
\end{lemma}

Recall that
\begin{definition}\emph{(See \cite{eisenbud20163264})} The multiplicity $l(\mathcal{X};P)$ of $\mathcal{X}$ at a point $P\in \mathcal{X}$ is  the degree of the projectivized tangent cone $\mathbb{T}C_P\mathcal{X}$.
The multiplicity $l(\mathcal{X};\mathcal{U})$ of a scheme $\mathcal{X}$  along an
irreducible component $\mathcal{U}$,  is equal to the multiplicity of $\mathcal{X}$ at a
general point of $\mathcal{U}$.
\end{definition}
\begin{lemma}\label{Multiplicity}\emph{(\cite{fulton2013intersection})}
Let $\mathcal{V}_1, \cdots,\mathcal{V}_m$ be  pure-dimensional subschemes    of $\mathbb{P}^\mathcal{N}$, with
$$\sum_{j=1}^{m}dim(\mathcal{V}_j)=(m-1)\mathcal{N}.$$
Assume $P$ is an isolated point of $\mathcal{X}\triangleq\mathcal{V}_1\bigcap\cdots \bigcap\mathcal{V}_m$. Then
\begin{equation}\label{multiplicity}
   l(\mathcal{X};P)\geq \prod_{j=1}^{m}l(\mathcal{V}_j;\mathcal{U})+\sum_{j=1}^{n}deg(\mathcal{U}_j),
\end{equation}
where $\mathcal{U}_1, \cdots,\mathcal{U}_n$ are  the
irreducible components   of $\mathbb{T}C_P\mathcal{V}_1\bigcap\cdots \bigcap\mathbb{T}C_P\mathcal{V}_m$. In particular,
\begin{equation}\label{multiplicity1}
   l(\mathcal{X};P)\geq \prod_{j=1}^{m}l(\mathcal{V}_j;\mathcal{U})
\end{equation}
with equality if and only if $\mathbb{T}C_P\mathcal{V}_1\bigcap\cdots \bigcap\mathbb{T}C_P\mathcal{V}_m=\emptyset$.
\end{lemma}

\paragraph{An upper bound for collinear relative equilibria with $L=0$.}
In this case, the system  (\ref{stationaryconfigurationnewcr1}) becomes:
\begin{equation}\label{stationaryconfigurationnewcrL01}
\begin{array}{cccc}
             M_z=0,&
             I =0, &
{F_z}-f_z =0, &
   G_z+g_z=0.
             \end{array}
\end{equation}
The zero point $O$ is a trivial solution of system (\ref{stationaryconfigurationnewcrL01}), indeed, an isolated solution by  Proposition \ref{isolatedtrivialsolution}. Then by Lemma \ref{Multiplicity} it follows that $$l(\mathcal{X};O)= 4.$$
As a result, the system (\ref{stationaryconfigurationnewcrL01})  has exactly $20$ points (counting with the appropriate mulitiplicity) in $\mathbb{P}^{4}_{\mathbb{C}}$ such that $z_j-z_k\neq 0$ for any $1\leq j< k\leq 4$. Thus  there are at most $10$ collinear relative equilibria for the four-vortex problem with $L=0$.

\subsubsection{A  better upper bound for relative equilibria by  B\'{e}zout Theorem}
\indent\par

%We consider only the case with $L\neq 0$ below. For the case $L= 0$, one can use a similar method  to that of collapse configurations in next subsection.

We embed the system (\ref{stationaryconfigurationnew5}) above with $\Lambda =1$ into a polynomial system in the projective space $\mathbb{P}^{8}_{\mathbb{C}}$:
\begin{equation}\label{stationaryconfigurationnewr1}
\left\{
             \begin{array}{rr}
             M_z=0,& ~~~~~~ M_w=0,\\
             Lt^2- I=0, &~~~~~~ z_2-z_1= w_2-w_1, \\
              {F_z}-f_zt^2=0, &
              {F_w}-f_wt^2=0,\\
 G_z+g_zt^2=0,&  G_w+g_wt^2=0;
             \end{array}
\right.
\end{equation}

It is easy to see that the algebraic variety (\ref{stationaryconfigurationnewr1}) is a disjoint union of  (\ref{stationaryconfigurationnew5})  and the algebraic variety
\begin{equation}\label{stationaryconfigurationnewr2}
\left\{
             \begin{array}{rr}
             M_z=0,& ~~~~~~ M_w=0,\\
              I=0, &~~~~~~ z_2-z_1= w_2-w_1, \\
              {F_z}=0, &
              {F_w}=0,\\
  G_z=0,&  G_w=0.
             \end{array}
\right.
\end{equation}

First, we remark that, a straightforward computation shows that the algebraic variety (\ref{stationaryconfigurationnewr2}) is  one-dimensional. Indeed,  it is easy to see that (\ref{stationaryconfigurationnewr2}) at least contains two  one-dimensional irreducible components (i.e., two one-dimensional lines):
\begin{equation}\label{twoone-dimensionallines}
  \begin{array}{ccc}
     w_1=w_2=w_3=w_4=0, &M_z=0, &z_2-z_1=0; \\
    z_1=z_2=z_3=z_4=0, &M_w=0, &w_2-w_1=0;
  \end{array}
\end{equation}
and four isolated points:
\begin{equation}\label{fourisolatedpoints}
  \begin{array}{cc}
  z_2=0,\Gamma_1z_1+\Gamma_3z_3=0,z_4=0, &w_1=0, w_2+z_1=0,w_3=0,\Gamma_4w_4-\Gamma_1z_1=0;\\
     z_2=0,z_3=0,\Gamma_1z_1+\Gamma_4z_4=0, &w_1=0, w_2+z_1=0,\Gamma_3w_3-\Gamma_2z_1=0,w_4=0; \\
     z_1=0,\Gamma_2z_2+\Gamma_3z_3=0,z_4=0, &w_1+z_2=0, w_2=0,w_3=0,\Gamma_4w_4-\Gamma_1z_2=0;\\
      z_1=0,z_3=0,\Gamma_2z_2+\Gamma_4z_4=0, &w_1+z_2=0, w_2=0,\Gamma_3w_3-\Gamma_1z_2=0,w_4=0.
  \end{array}
\end{equation}

A straightforward computation shows that both  multiplicities of  two  one-dimensional irreducible components are at least $6$ and  all multiplicities of  four isolated points are at least $2$.
It follows that the degree of algebraic subset (\ref{stationaryconfigurationnew5}) is no more than $288-6\times2-4\times2=268$.

 \paragraph{An upper bound for strictly planar relative equilibria with $L\neq0$.}
 In this case, by the result for collinear central configurations above, it is easy to see that the degree of  (\ref{stationaryconfigurationnew5}) corresponding to  strictly planar relative equilibria is no more than $268-24=244$.
Therefore, the number of strictly planar relative equilibria  for the four-vortex problem with $L\neq 0$ is no more than $[\frac{244}{2}]=122$.

\paragraph{An upper bound for strictly planar relative equilibria with $L=0$.}
 In this case, firstly, by the result for collinear central configurations above, it is easy to see that the degree of  (\ref{stationaryconfigurationnew5}) corresponding to   collinear relative equilibria is  $20$.
Secondly, note that there is  an isolated  trivial solution of system (\ref{stationaryconfigurationnew5})  by  Proposition \ref{isolatedtrivialsolution}. And it is easy to see that the multiplicity of  this trivial solution is  $8$.

Therefore, the number of strictly planar relative equilibria  for the four-vortex problem with $L= 0$ is no more than $[\frac{268-8-20}{2}]=120$.

\paragraph{A better upper bound by computation of mixed volumes.}

On the other hand, by the method of computation of mixed volumes of  Newton polytope the system (\ref{stationaryconfigurationmainequ2}), we can obtain a better upper bound than the numbers above. The method is introduced by Hampton and Moeckel in \cite{hampton2006finiteness,hampton2009finiteness} for relative equilibria. We remark that the computation by Hampton and Moeckel in \cite{hampton2009finiteness} showed that the number  of  the strictly planar relative
equilibria  is no more than 74.

\subsubsection{A  better upper bound for collapse configurations by  B\'{e}zout Theorem}
\indent\par
%We focus on upper bounds of collapse configurations below.

In this case, we embed the system (\ref{stationaryconfigurationnew6}) above into a polynomial system in the projective space $\mathbb{P}^{8}_{\mathbb{C}}$:
\begin{equation}\label{stationaryconfigurationnewc1}
\left\{
             \begin{array}{rr}
                M_z=0, &~~~~~~ M_w=0,\\
             I=0, &z_2-z_1= w_2-w_1, \\
             {\overline{\Lambda} } {F_z}-t^2f_z=0, &
             {\Lambda } {F_w}-t^2f_w=0,\\
\overline{\Lambda} G_z+t^2g_z=0,&\Lambda  G_w+t^2g_w=0.
             \end{array}
\right.
\end{equation}

To obtain an upper bound of collapse configurations, we estimate the number of the irreducible components of the system (\ref{stationaryconfigurationnewc1}) for $t=0$. This is an algebraic variety in  $\mathbb{P}^{7}_{\mathbb{C}}$ which is the same as (\ref{stationaryconfigurationnewr2}) in form (note that $L=0$ here):
\begin{equation}\label{stationaryconfigurationnewc2}
\left\{
             \begin{array}{rr}
                M_z=0, &~~~~~~ M_w=0,\\
             I=0, &z_2-z_1= w_2-w_1, \\
              {F_z}=0, &
            {F_w}=0,\\
G_z=0,&  G_w=0.
             \end{array}
\right.
\end{equation}

A straightforward computation shows that  the  variety (\ref{stationaryconfigurationnewc2}) above is  one-dimensional and at least contains two  one-dimensional irreducible components (\ref{twoone-dimensionallines}) and four isolated points (\ref{fourisolatedpoints}).
Similarly, both  multiplicities of  two  one-dimensional irreducible components are at least $6$ and  all multiplicities of  four isolated points are at least $2$.
It follows that the degree of algebraic subset (\ref{stationaryconfigurationnew6}) is no more than $288-6\times2-4\times2=268$.

On the other hand, note that there is  an isolated  trivial solution of system (\ref{stationaryconfigurationnew5})  by  Proposition \ref{isolatedtrivialsolution}. And it is easy to see that the multiplicity of  this trivial solution is  $8$.

Therefore, the number of collapse configurations  for the four-vortex problem is no more than $[\frac{268-8}{2}]=130$.

\subsection{Conclusion on upper bounds }
\indent\par
To summarize, by the results  in this section and Theorem \ref{O'Neil} and  \ref{Hampton-Moeckel}, we get Corollary \ref{upperbounds}.

\section{Conclusion}
\indent\par

Inspired by the elegant method of Albouy and Kaloshin for celestial mechanics, which provides  an effective way to analyse of the singularities, we develop a  novel analysis of the singularities for a possible continuum of central configurations of the N-vortex problem.
We  proved that there are finitely many complex  central configurations in the planar four-vortex problem.  As a result, there are finitely many  stationary configurations consisting of equilibria, rigidly translating configurations, relative equilibria (uniformly rotating configurations)
 and collapse configurations.

  Once the finiteness is proved, an explicit upper bound on the number of
relative equilibria
 and collapse configurations is obtained by direct application of B\'{e}zout Theorems. However, to obtain good   upper bounds, it is necessary to transform the system  (\ref{stationaryconfiguration1})  into an available  equivalent form,  which  is based on an   observation of O'Neil for the planar $N$-vortex problem, and to estimate the  multiplicity of the irreducible components not corresponding to central configurations. To obtain better upper bounds,  the method  of mixed volumes for the system (\ref{stationaryconfigurationmainequ2})  is  employed.   The method is introduced by Hampton and Moeckel. Unfortunately, the   computation is not easy, thus we  simply apply the data of  Hampton and Moeckel in \cite{hampton2009finiteness} to show that the number  of  the strictly planar relative
equilibria  is no more  than 74.

%\section*{Acknowledgements}

\newpage

\bibliographystyle{plain}

%\bibliography{C:/Users/YuXiang/Desktop/YuReference}

\begin{thebibliography}{10}

\bibitem{Albouy2012Finiteness}
Alain Albouy and Vadim Kaloshin.
\newblock Finiteness of central configurations of five bodies in the plane.
\newblock {\em Annals of Mathematics}, 176(1):535--588, 2012.

\bibitem{aref1979motion}
H.~Aref.
\newblock Motion of three vortices.
\newblock {\em The Physics of Fluids}, 22(3):393--400, 1979.

\bibitem{aref1983integrable}
H.~Aref.
\newblock Integrable, chaotic, and turbulent vortex motion in two-dimensional
  flows.
\newblock {\em Annual Review of Fluid Mechanics}, 15:345--389, 1983.

\bibitem{aref2003vortex}
H.~Aref, P.~K. Newton, M.~A. Stremler, T.~Tokieda, and D.~L Vainchtein.
\newblock Vortex crystals.
\newblock {\em Advances in applied Mechanics}, 39:2--81, 2003.

\bibitem{aref2005vortex}
H.~Aref and M.~van Buren.
\newblock Vortex triple rings.
\newblock {\em Physics of fluids}, 17(5):057104, 2005.

\bibitem{eisenbud20163264}
David Eisenbud and Joe Harris.
\newblock {\em 3264 and all that: A second course in algebraic geometry}.
\newblock Cambridge University Press, 2016.

\bibitem{fulton2013intersection}
W.~Fulton.
\newblock {\em Intersection theory}, volume~2.
\newblock Springer Science \& Business Media, 2013.

\bibitem{grobli1877specielle}
W.~Gr{\"o}bli.
\newblock {\em Specielle Probleme {\"u}ber die Bewegung geradliniger paralleler
  Wirbelf{\"a}den}, volume~8.
\newblock Druck von Z{\"u}rcher und Furrer, 1877.

\bibitem{hampton2009finiteness}
M.~Hampton and R.~Moeckel.
\newblock Finiteness of stationary configurations of the four-vortex problem.
\newblock {\em Transactions of the American Mathematical Society},
  361(3):1317--1332, 2009.

\bibitem{hampton2006finiteness}
Marshall Hampton and Richard Moeckel.
\newblock Finiteness of relative equilibria of the four-body problem.
\newblock {\em Inventiones mathematicae}, 163(2):289--312, 2006.

\bibitem{helmholtz1858integrale}
H.~Helmholtz.
\newblock {\"U}ber integrale der hydrodynamischen gleichungen, welche den
  wirbelbewegungen entsprechen.
\newblock {\em Journal f{\"u}r die reine und angewandte Mathematik},
  1858(55):25--55, 1858.

\bibitem{kirchoff1877vorlesungen}
G.~Kirchhoff.
\newblock Vorlesungen {\"u}ber mathematische physik: Mechanik, 1876.

\bibitem{mumford1995algebraic}
D.~Mumford.
\newblock {\em Algebraic geometry I: complex projective varieties}.
\newblock Springer Science \& Business Media, 1995.

\bibitem{novikov1975dynamics}
E.~A. Novikov.
\newblock Dynamics and statistics of a system of vortices.
\newblock {\em Zh. Eksp. Teor. Fiz}, 68(1868-188):2, 1975.

\bibitem{novikov1979vortex}
E.~A. Novikov and Yu.~B. Sedov.
\newblock Vortex collapse.
\newblock {\em Zhurnal Eksperimentalnoi i Teoreticheskoi Fiziki}, 77:588--597,
  1979.

\bibitem{o1987stationary}
K.~A. O'Neil.
\newblock Stationary configurations of point vortices.
\newblock {\em Transactions of the American Mathematical Society},
  302(2):383--425, 1987.

\bibitem{o2007relative}
K.~A. O'Neil.
\newblock Relative equilibrium and collapse configurations of four point
  vortices.
\newblock {\em Regular and Chaotic Dynamics}, 12(2):117--126, 2007.

\bibitem{palmore1982relative}
J.~I. Palmore.
\newblock Relative equilibria of vortices in two dimensions.
\newblock {\em Proceedings of the National Academy of Sciences},
  79(2):716--718, 1982.

\bibitem{patil1983remarks}
Dilip~P Patil and Wolfgang Vogel.
\newblock Remarks on the algebraic approach to intersection theory.
\newblock {\em Monatshefte f{\"u}r Mathematik}, 96(3):233--250, 1983.

\bibitem{roberts1999continuum}
G.~E. Roberts.
\newblock A continuum of relative equilibria in the five-body problem.
\newblock {\em Physica D: Nonlinear Phenomena}, 127(3-4):141--145, 1999.

\bibitem{synge1949motion}
J.~L. Synge.
\newblock On the motion of three vortices.
\newblock {\em Canadian Journal of Mathematics}, 1(3):257--270, 1949.

\end{thebibliography}

\end{document}